\documentclass[aps,physrev,preprint,groupedaddress,showkeys]{revtex4-2}

\usepackage{lmodern}
\usepackage{amsmath}
\usepackage{array}
\usepackage{adjustbox}
\usepackage{multirow}
\usepackage{import}
\usepackage{tikz}
\usepackage{diagbox}
\usepackage{makecell}
\usepackage{calc}
\usepackage{tabularx}
\usepackage{float}
\usepackage{xcolor}
\usepackage{booktabs}

\usepackage{dirtytalk}
\usepackage[inkscapelatex=false]{svg}
\usepackage{subcaption}
\usepackage{graphicx}
\usepackage{tikz}
\usetikzlibrary{calc}

\newsavebox{\panelbox}

\begin{document}


\title{High Resolution and High-Speed Live Optical Flow Velocimetry}


\author{Juan Pimienta}
\email{ju-pimienta@photonlines.com}
\affiliation{Photon Lines, 10 avenue des Touches, Pacé, 35740, France.}
\author{Jean-Luc Aider}
\email{jean-luc.aider@espci.psl.eu}
\affiliation{Laboratoire PMMH, ESPCI-PSL, CNRS, 7-9 Quai Saint Bernard, Paris, 75005, France.}


\begin{abstract}
Particle Image Velocimetry (PIV) is the most widely used optical technique for measuring two-dimensional velocity fields in fluids. However, with the standard cross-correlation (CC) algorithm, improving the spatial resolution of instantaneous velocity fields and obtaining dense velocity fields in real time and at high frequencies remains challenging. Optical Flow Velocimetry (OFV) provides an alternative route to per-pixel velocimetry. In this study, we show that dense velocity fields (one vector per pixel) with high effective spatial resolution can be obtained in real time at frequencies up to kHz using an optical flow approach. First, using synthetic images on two cases (a well-defined Rankine vortex and a homogeneous isotropic turbulence DNS dataset), we show that resolving large displacement gradients down to small scales is feasible when the particle seeding is appropriately tuned. Second, we show that real-time processing rates can be achieved through algorithmic and GPU-oriented optimizations, together with appropriate choices of OF parameters. With this implementation, $21$ Mp velocity fields are measured in real-time (live) at $90$ Hz, while $4$ Mp velocity fields can be measured up to  $460$ Hz and $1$ Mp velocity fields can be measured up to $1400$ Hz. The approach is further validated on experimental measurements of the flow past a circular cylinder, where dense instantaneous fields are obtained and used to compute derived quantities in real time over long durations. These processing rates enable real-time computation of derived flow quantities during experiments, including long-duration monitoring, low-frequency dynamics captured from sustained high-rate acquisition, and closed-loop flow-control strategies based on OFV measurements. In addition to enabling live operation, the approach accelerates standard off-line post-processing, reducing turnaround time and computational cost.
\end{abstract}


\maketitle

\section{Introduction}
\textcolor{black}{Particle Image Velocimetry (PIV) is a non-intrusive optical technique that involves illuminating a flow with a laser beam and seeding it with light-reflecting particles. Two successive images of the seeded flow are recorded, and velocities are typically obtained by calculating cross-correlation within Interrogation Windows (IW) \citep{raffel2018particle}. Although widely used in fluid mechanics to study complex flows \citep{westerweel2013particle}, this approach is limited by the trade-off between spatial resolution imposed by the size of the IW, dynamic range of displacement gradients and computational cost, especially for large images, long acquisitions and real-time operation.}

\textcolor{black}{The performance of cross-correlation PIV (CC-PIV) has been improved through numerous developments such as adaptive and multi-pass schemes with window deformation, as well as super-resolution and hybrid PIV/PTV approaches \citep{theunissen2006adaptive,scarano2000advances,keane1995super,susset2006novel,stitou2001extension}. More recently, Machine Learning (ML) refinements have been introduced to improve accuracy and spatial resolution \citep{zhang2020unsupervised,choi2023deep,ilg2017flownet} but increasing the post-processing time. In parallel, efforts in real-time processing have reduced latency, but necessarily at the cost of compromises on spatial resolution or specialized hardware \citep{Maruyama2001AnAT,Fujiwara2003ARV, Siegel2003RealTimePI,Shchapov2018SupercomputerRE,Willert2010RealtimePI,McCormick2024ReactiveCO}.}

\textcolor{black}{An alternative to CC-PIV for calculating velocity fields from the displacement of particles transported by a flow is the Optical Flow (OF) velocimetry (OFV or OF-PIV). OF estimates the apparent motion of the pixels between two image frames by imposing constant brightness and spatial regularity between two successive images and is not limited to seeded flows. Indeed, this method has been successfully implemented for Real-Time (RT) processing in machine vision to measure the velocity of objects or for UAV control  \citep{Piga2021ROFTRO,Wang2011RealTimeVS,Lim2017RealtimeOF,Brebion2021RealTimeOF}. Notably  \citet{romera2023optical} reported dense velocity fields of 1 vector per pixel at 20~Hz for $2048\times2048~px$ (4.2 Mp) images. In fluid mechanics, OF's accuracy has been assessed in detail for various methods. \citet{Mendes2021ACS} benchmarked several OF approaches using synthetic PIV images and found that OF can achieve high accuracy over a broader range of image parameters than CC-PIV, while still requiring adequate seeding and good image quality. Particularly, the optimal particle seeding needed to obtain a dense velocity field seems to be higher than for CC-PIV \cite{Jassal2025ARO}.}    

\textcolor{black}{Deep-learning (DL) OF methods have also been adapted for PIV including RAFT-based (Recurrent All-pairs Field Transforms) approaches and RAFT-PIV \citep{Cai2019DenseME,fischer2015flownetlearningopticalflow,Yu2023DeepDR,Teed2020RAFTRA,hui2020liteflownet3resolvingcorrespondenceambiguity,Lagemann2021DeepRO}. However, these methods typically rely on trained models meaning that their performance and computational cost depend on network design and training data. In contrast, the variational approach presented in this study does not require any training and is optimized for RT velocity fields computation \citep{Gautier2013RealtimePF,Gautier2013ControlOT,Gautier2015FrequencylockRC,Varon2017ReactiveCO}.}

\textcolor{black}{Nevertheless, computing dense velocity field in real time, keeping high spatial resolution at high frequencies, remains a challenge. In this study we present advances in the GPU implementation of Optical Flow Velocimetry (OFV) that produces dense, per-pixel velocity vectors and is designed for both fast post-processing and RT high frequency processing. The accuracy of the algorithm is quantified on two synthetic benchmarks: a Rankine vortex with controlled displacement gradients and seeding conditions, and homogeneous isotropic turbulence (HIT) using the Madrid DNS database \cite{Cardesa2017TheTC}, spanning particle sizes and densities. In both cases, the ability to measure high gradients even at small scales is demonstrated, as long as the proper experimental and OF parameters are chosen.We also report processing speeds that, to the best of our knowledge, have not been previously reported for dense per-pixel OFV at these image sizes (up to 21 Mp) and processing parameters for both \emph{offline} (post-processing) and \emph{online} (Live measurements). Finally, we present the application of the OFV method to the flow around a cylinder at $Re_D\approx6482$ in a hydrodynamic channel, which allows us to obtain highly detailed instantaneous velocity fields. We also demonstrate the unique ability of the RT-OFV method to measure various quantities derived from these instantaneous velocity fields over a virtually unlimited time period. }

\section{Live-Optical Flow Velocimetry\label{sec:LOFV}}

\textcolor{black}{The concept of Optical Flow can be defined as \textit{the apparent velocities from intensity changes in a scene}, as first introduced by Gibson \citep{Gibson1967TheSC} in 1967. Although the notion of OF was formulated, there was no mathematical solution that allowed for the estimation of velocity vectors from intensity changes between successive images until 1981, when Horn--Schunk \citep{HORN1981185} proposed the first solution.   }

\textcolor{black}{Mathematically, OF algorithms rely} on the brightness-constancy assumption: the image intensity attached to a moving scene is approximately conserved between two successive snapshots. Linearizing this assumption yields the optical flow constraint equation:

\begin{equation}
    I_x\,u + I_y\,v + I_t = 0,
    \label{eq:intensity_gradient}
\end{equation}
where $I_x$, $I_y$, and $I_t$ are the spatial and temporal derivatives of the instantaneous intensity field $I(x, y, t)$ and $\mathbf{V}(x,y) = (u(x,y),v(x,y))^{T}$ is the inter-frame displacement field at each pixel\textcolor{black}{; the corresponding velocity is $\mathbf{V}/\delta t$.}

\textcolor{black}{However, Eq.~\ref{eq:intensity_gradient} is insufficient to resolve the OF problem since it provides one constraint for two unknowns. Horn--Schunck \citep{HORN1981185} closed the problem by adding a global smoothness regularization. Here, we adopt the local Lucas--Kanade (LK) approach \citep{Lucas1981AnII}, which assumes that the displacement is approximately constant within a small  neighborhood (kernel) $N(m)$ centered at pixel location $m$. The kernel size is controlled by the kernel radius $KR$ (in pixels), which defines the extent of $N(m)$ around each pixel.
The local displacement $\mathbf{V}(m)$ is estimated by minimizing a symmetric sum-of-squared-differences (SSD) objective, in which the two images are evaluated at midpoint-warped coordinates $k \mp \mathbf{V}/2$ for all pixels $k \in N(m)$:
\begin{equation}
\mathbf{V}^*(m)
=
\arg\min_{\mathbf{V}}
\sum_{k \in N(m)}
\left[
I^{t}\!\left(k - \frac{\mathbf{V}}{2}\right)
-
I^{t'}\!\left(k + \frac{\mathbf{V}}{2}\right)
\right]^2 ,
\label{eq:of_symmetric}
\end{equation}
where $t' = t+\delta t.$
This symmetric additive formulation differs from the original approach in the LK algorithm that uses a forward additive method. It yields a favorable balance between computational cost and accuracy \citep{Champagnat2011FastAA,varonthesis,Pan2015EvaluatingTA}. 
}

\textcolor{black}{In their basic differential form, optical flow algorithms are accurate for small inter-frame displacements (1 to 2 pixels). To handle larger displacements between two successive snapshots, a Gaussian-pyramid (coarse--to--fine) scheme \citep{Bouguet1999PyramidalIO} is used to successively reduce the image size using a Burt--Adelson Gaussian pyramid \citep{Burt1983TheLP}, thereby sub--sampling large displacements. This introduces a second key parameter: the number of \emph{pyramid sub--levels} (PSL). At each PSL, the image size is divided by 2 in both directions.}

A third parameter is the number of Gauss--Newton iterations used to reduce the SSD objective at each pyramid sub--level. In addition, we apply a pre-processing step consisting of local (pixel-centered) intensity normalization across the image to improve robustness to illumination variations.

The overall velocity-field estimation pipeline comprises six steps:
\begin{enumerate}
    \item Local intensity normalization.
    \item Gaussian-pyramid sub--sampling (construction of coarse--to--fine levels).
    \item Per--pixel displacement estimation at the kernel scale (LK with Gauss--Newton).
    \item Up--sampling and projection of the displacement to the next finer level.
    \item Apply steps 3--4 from coarsest to the finest pyramid level.
    \item Final dense velocity-field computation at full resolution.
\end{enumerate}

\begin{figure}[h]
    \centering
    \includegraphics[width=\linewidth]{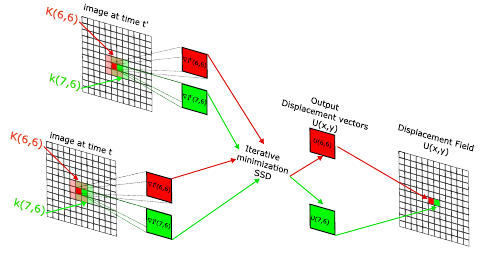}
    \caption{Diagram showing the main steps used to compute the displacement vectors at the kernel scale.}
    \label{fig:KR}
\end{figure}

The kernel-scale estimation is sketched in Fig.~\ref{fig:KR}. Two sub--images of size $12\times12~px$ at times $t$ and $t' = t+\delta t$ are considered. In this example, two neighboring pixels at $(x,y)=(6,6)$ (red) and $(7,6)$ (green) are evaluated with a kernel radius $KR=1$~px. The neighborhood associated with each pixel is indicated by the lighter shade (left of Fig.~\ref{fig:KR}). At this scale, spatial intensity gradients are computed and the Gauss--Newton iterations minimize the SSD objective in Eq.~\ref{eq:of_symmetric} to recover the displacement at each pixel. Repeating this over the image pair yields a dense displacement field \textcolor{black}{(one vector per pixel)}.
In what follows, \emph{eyePIV}\texttrademark, a dedicated OFV plugin is used. It has been developed jointly by the two teams (PMMH and Photon Lines) for real--time measurements. \emph{eyePIV} is the result of a continuous collaboration, leading to various optimizations and studies performed over the years \cite{giannopoulos:tel-03364421,pimientaalvernia:tel-05105277}.

\section{Workstation and Live-OFV setup\label{sec:setup}}

\textcolor{black}{The off-line benchmarks and experimental measurements presented below have been carried out on a custom-made workstation designed to optimize data acquisition and transmissions between the CPU, GPUs, hard drive and RAM. A dedicated motherboard is equipped with an AMD Ryzen Threadripper PRO 3955 WX processor (16 cores, 3.90~GHz) and 128~GB of RAM at 3.2~GHz. The workstation is designed to accommodate multiple Graphics Processing Units (GPUs), providing flexibility and modularity This is an important constraint which requires a dedicated power supply to deal with the energy consumption of large GPUs.  For the tests reported in this study, it was equipped with a single 5th-generation NVIDIA RTX~5090 GPU (Graphics Processing Unit).}

\textcolor{black}{For the live experimental measurements, a Mikroton 21CXP12 camera was used to stream the image sequences directly to the workstation through a dedicated CoaXPress frame grabber card. This camera can acquire images up to $5120 \times 4096$~px ($20.97$~Mp) at a maximum frame rate of $f=230$~frames per second (fps) that can be streamed in real-time to the workstation through the CoaXPress card.}

\section{Synthetic image generation\label{sec:Imagegen}}

Synthetic images were used to rigorously evaluate the accuracy and measurement capabilities of displacement gradients at different scales. Two test cases were implemented. First, a Rankine vortex, whose images were generated by a third-party tool, enabled the accurate evaluation of well-defined displacement gradients. Second, a HIT-type direct numerical simulation (DNS), whose images were generated by an in-house tool, allowed the algorithm to be tested on a complex and realistic turbulent flow.

\subsection{Rankine vortex}

The synthetic PIV images for the Rankine vortex were generated using the \textit{PIV-image-generator} of \citet{MENDES_piv_image_gen}. This MATLAB-based tool creates synthetic particle fields which are moved by the chosen analytical velocity fields (e.g., simple shears or a Rankine vortex). The specified analytical flow displaces randomly distributed particles between successive snapshots, thereby producing image pairs similar to the ones obtained by PIV measurements, but on a known displacement field.

In what follows, we evaluate the accuracy and precision of the OFV algorithm for a single Rankine vortex of well-defined size and intensity, with different image parameters.

The Rankine vortex is an insightful benchmark because it exhibits spatially localized, high-displacement gradients. This configuration allows independent control of the core radius and rotational speed (i.e., the displacement gradient), both of which are known to be challenging for CC-PIV \citep{Beresh_velocity_gradient_PIV_turbulence, westerweel_piv_gradients}. Following \citet{MENDES_piv_image_gen}, the analytical description of the Rankine vortex is defined as follows. \\

\noindent\fbox{%
    \parbox{\textwidth}{%
    The azimuthal displacement is:
    \begin{center}
        $u_\theta(r) = \frac{\Gamma}{2\pi M R} \times \Biggl\{ \begin{array}{l}
    \frac{r}{R}~~r\leq R \\
    \\
    \frac{R}{r}~~r > R,
  \end{array}$
    \end{center}
    where $r$ is the radial position, $R$ the vortex core radius, $\Gamma$ the vortex core circulation, and $M$ a scale factor. The particle displacement over a time interval $ \Delta t$ is in Cartesian coordinates:
    
     \begin{center}
     
    $\Delta x(r,\theta) ~=~r\cos{\left(\frac{u_\theta}{r}\Delta t + \theta \right)} $,
    \\
    \vspace{0.5cm}
    $
    \Delta y(r,\theta) ~=~r\sin{\left(\frac{u_\theta}{r}\Delta t + \theta \right)}
    $,

     \end{center}   
     where $(r,\theta)$ are the initial polar coordinates of the seeding particle.
    }
    \label{tab:flows}
}
\\

\begin{table}[h!]
    \centering
    \begin{tabular}{|p{4cm}|p{5cm}|p{4cm}|}
    \hline
     \textbf{Maximum \newline displacement [pixels]} & \textbf{Particles concentration \newline [particles / IW]} & \textbf{Rankine core \newline radius [pixels]}\\
     \hline
     8, 16, 24, 32 & 5, 10, 15 & 12, 25, 50, 75, 100, 150\\
     \hline
    \end{tabular}
    \caption{Parameters used for the generation of the synthetic images. \textit{IW} stands for interrogation window.}
    \label{tab:images}
\end{table}

\begin{table}[h!]
    \centering
    \begin{tabular}{|c|c|c|c|}
        \hline
         \textbf{Parameter} & \textbf{Minimum}  & \textbf{Maximum} & \textbf{Step}\\
         \hline
         \textbf{Kernel Radius} & 2 & 7 & 1 \\
         \hline
         \textbf{Pyramid sub-level} & 2 & 4 & 1 \\
         \hline
         \textbf{Iterations} & 1 & 4 & 1 \\
         \hline
         \textbf{Normalization Radius} & 1 & 4 & 1 \\
         \hline
    \end{tabular}
    \caption{Main OFV parameters and their ranges used in this study.}
    \label{tab:OFPIV}
\end{table}

To generate the synthetic images several parameters defining the virtual optical setup and the seeding particles must be defined. To reduce the space parameters, the following parameters were kept constant: image size ($1024 \times 1024$~px), particle radius ($r_p=1.5$~px), laser-sheet thickness ($2.0$~mm), standard deviation of the out-of-plane motion ($\sigma=0.025$~mm), and no added noise. The variable parameters were the maximum expected displacement, the particle concentration (expressed as particles per interrogation window, [part/IW]), and the vortex core radius $R$. The generator defines the concentration relative to a $16\times16$ px interrogation window. We have retained this definition based on the IW, which is convenient for CC-PIV (calculation of the CC in the IW) and well-know by the PIV users. Nevertheless, it should be noted that this definition has no real significance for OFV, as we will see later.

\begin{figure}[h!]
    \centering
    \includegraphics[width = \textwidth]{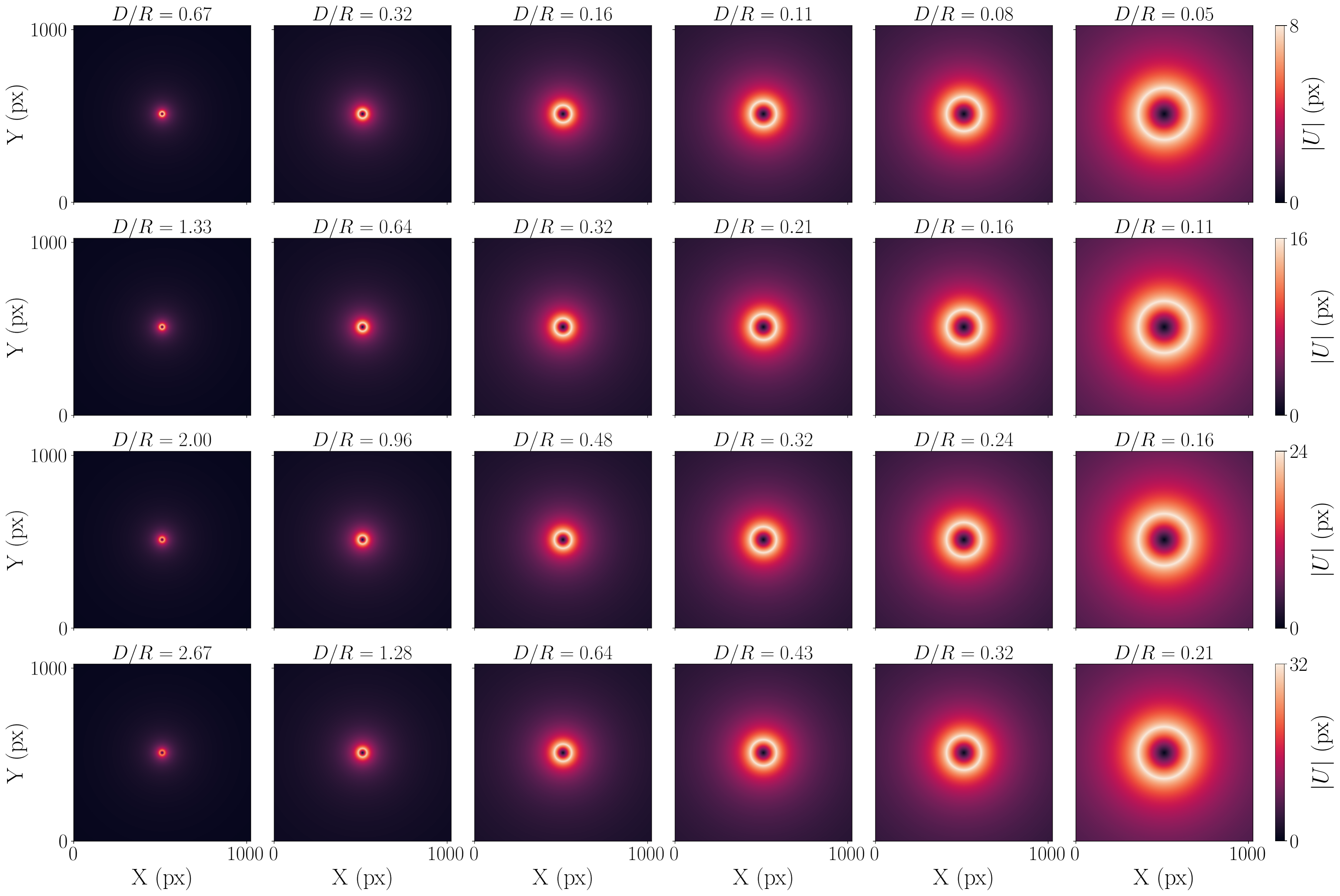}
    \caption{Theoretical displacement magnitude of the Rankine vortex for various radii and velocities. The core radius increases from left to right (from $r = 12$ to $r=150$ ~px) while the maximum displacement of the field increases from 8 to 32~px from top to bottom.}
    \label{fig:th_displacement}
\end{figure}

Table~\ref{tab:images} lists the variable parameters used to create the synthetic images. For statistical robustness, ten pairs of images were created for each combination of parameters, resulting in 216 sets of 10 pairs of images.

The OFV parameter space has been kept reasonably constrained, based on an empirically observed ``convergence zone": moving outside this region leads to degraded results, especially for high gradient flows. The OFV parameters are reported in the table~\ref{tab:OFPIV} with their minimum, maximum and incremental values, leading to $288$ unique combinations. For each image--parameter set across all OFV combinations, there were $2\,880$ data points, totaling $207\,360$ data points from $414\,720$ image pairs for the entire study.

\begin{figure}[h!]
    \begin{subfigure}{0.48\linewidth}
        \centering
        \includegraphics[width=\linewidth]{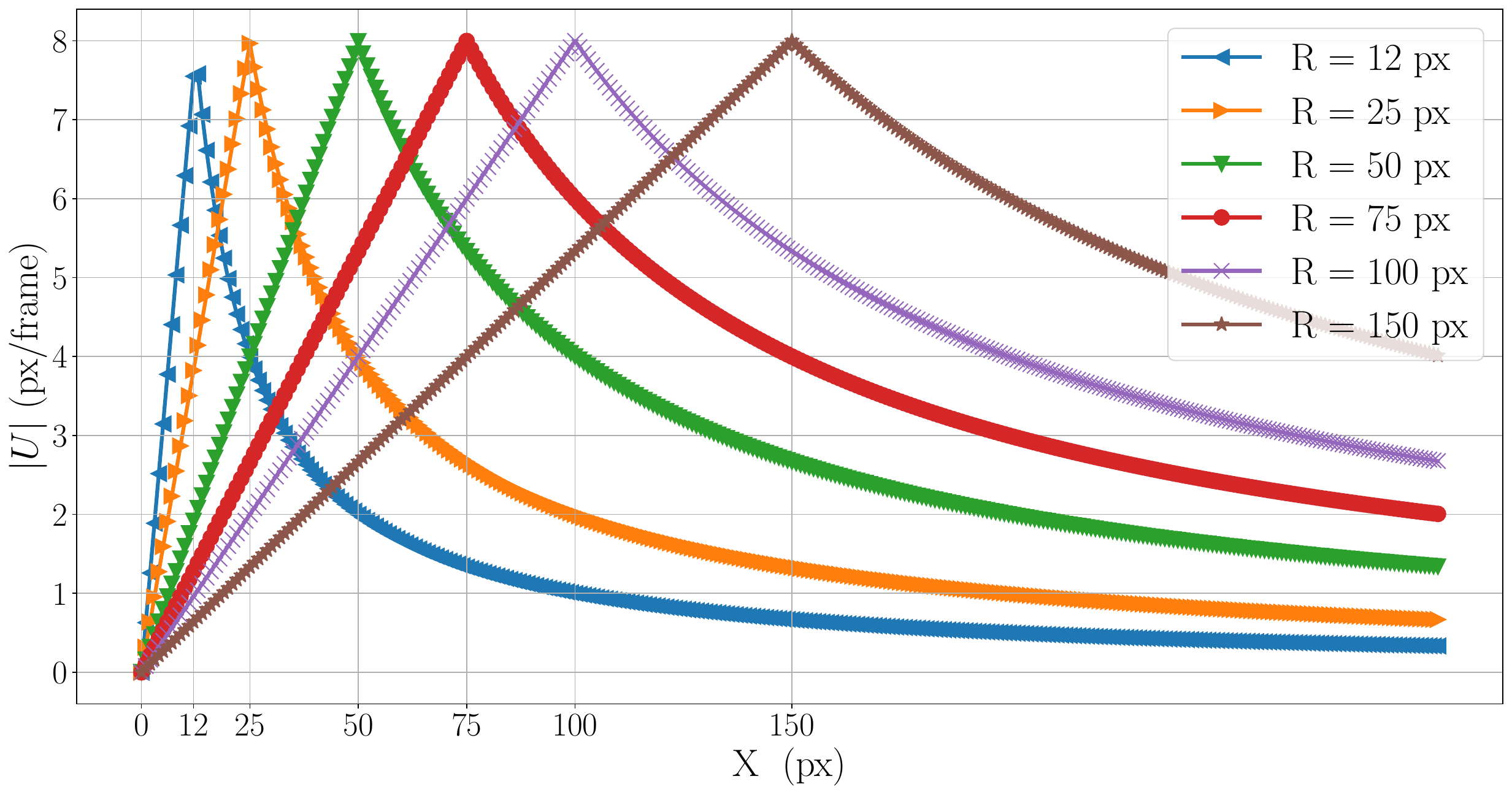}
        \caption{}
        \label{fig:grad_prof}    
    \end{subfigure}
    \begin{subfigure}{0.48\linewidth}
        \centering
        \includegraphics[width=\linewidth]{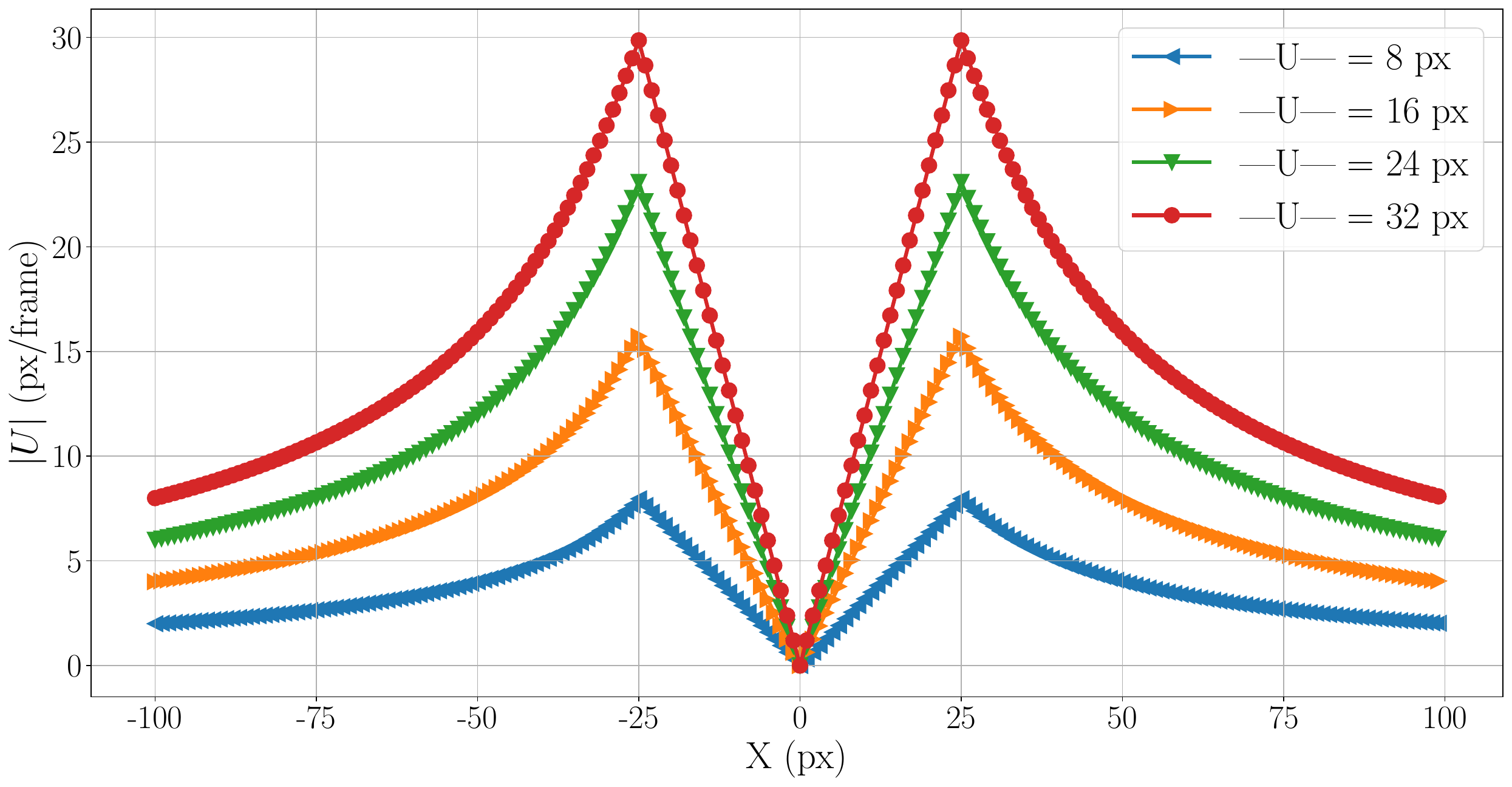}
        \caption{}
        \label{fig:rad_prof}    
    \end{subfigure}
    \caption{Theoretical displacement profiles used to move the particles in the images. (a) Displacement magnitude radial profiles with a given magnitude ($|U|=8$ px/frame) and different core radii leading to variation of the spatial displacement gradient. (b) Displacement profiles obtained for a given fixed radius ($r=25$ px) of the Rankine vortex and for different  amplitudes. It also leads to variations of the displacement gradients. Both cases show zoomed-in profiles around the core's center. }
    \label{fig:Rankine_th_profiles}
    
\end{figure}

Figure~\ref{fig:th_displacement} shows typical examples of contours of displacement fields obtained with the analytical definition for the Rankine vortex, computed for various set of parameters. Due to the fixed image size (1 MP), all cases clearly illustrate large and small vortices as well as low and high rotational speeds, which is relevant for evaluating performance under near-turbulent conditions. Indeed, it simulates the diversity of scales and displacement gradients characteristic of turbulent flows.

To illustrate the influence of maximum displacement and core radius on the displacement gradient, the Fig.~\ref{fig:Rankine_th_profiles} presents various theoretical displacement profiles as a function of amplitude, along the radial direction, for different amplitudes and core radii. The modification of the displacement gradients is clearly visible, which can lead to extreme cases of strong gradients at small scales. This phenomenon represents both a challenge for PIV and a major objective for accurate analysis of turbulent flows.

\subsection{Homogeneous Isotropic Turbulence}

\begin{figure}[h]
    \centering
    \includegraphics[width=0.65\linewidth]{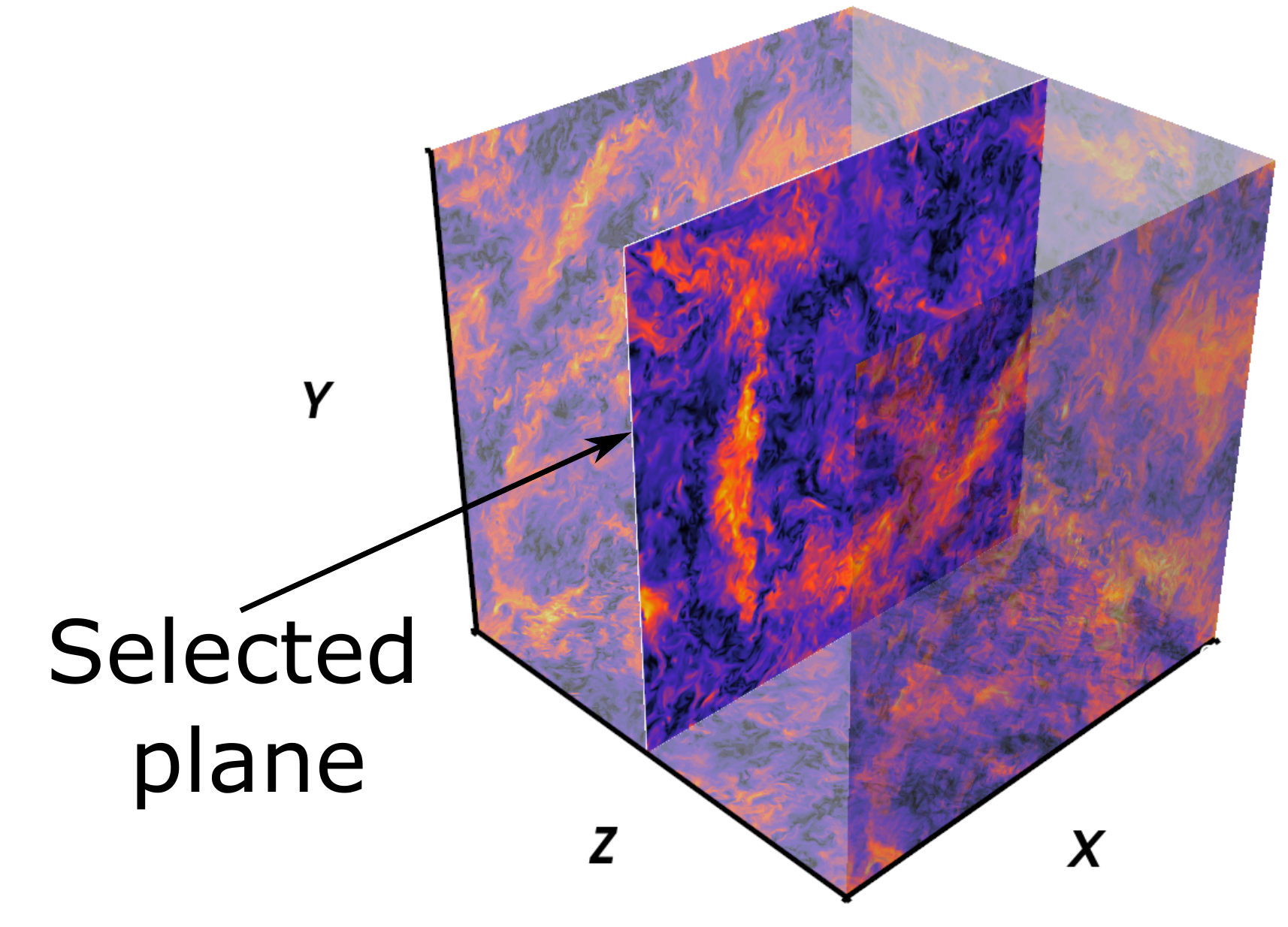}
    \caption{3D representation of a displacement datacube snapshot from the DNS HIT dataset. The selected plane used for the study is in the middle of the cube.  }
    \label{fig:datacube}
\end{figure}

\begin{figure}
    \centering
    \begin{subfigure}{0.48\linewidth}
        \centering
        \includegraphics[width=\linewidth]{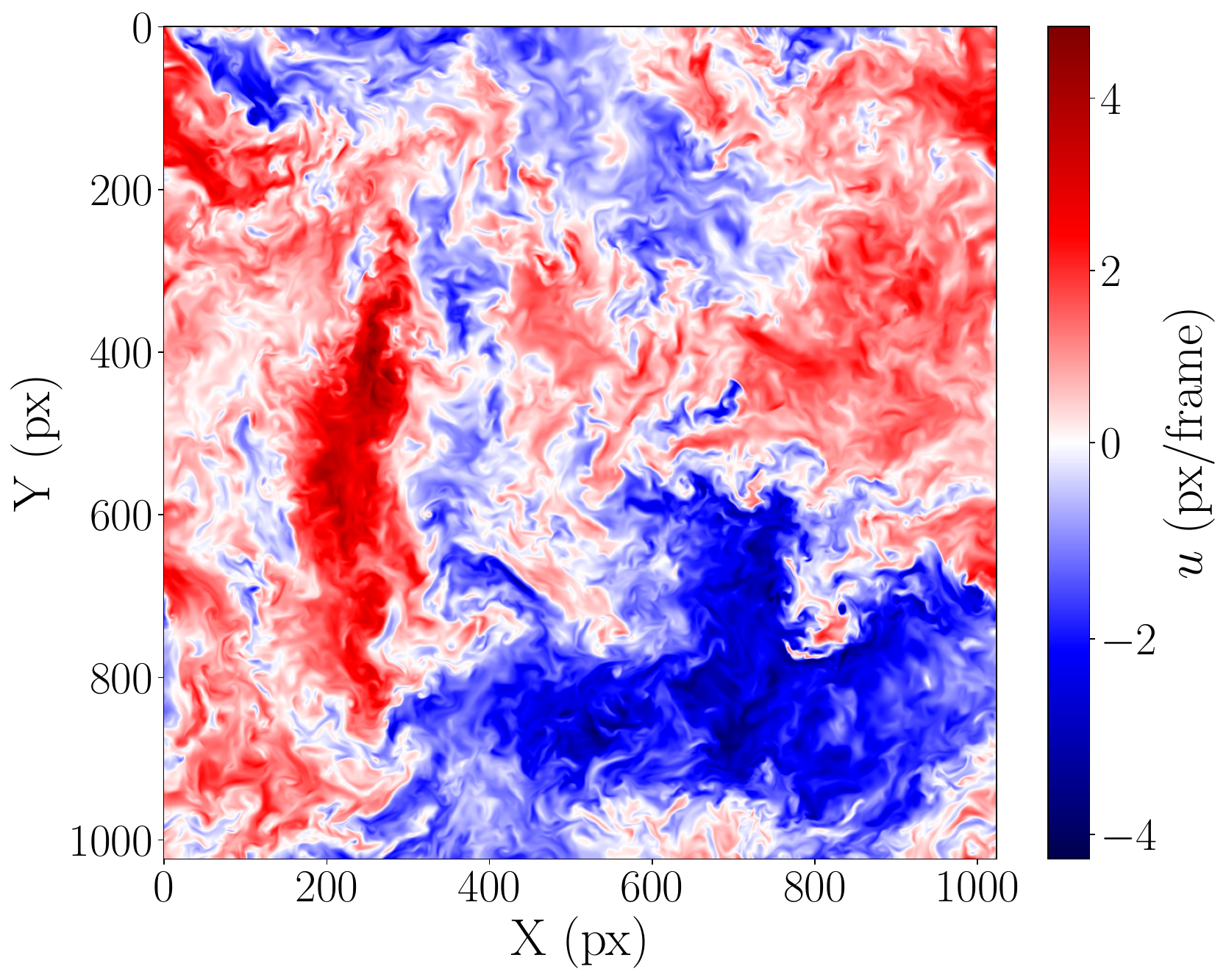}
        \caption{}
        \label{fig:u_dns}
    \end{subfigure}
    \begin{subfigure}{0.48\linewidth}
        \centering
        \includegraphics[width=\linewidth]{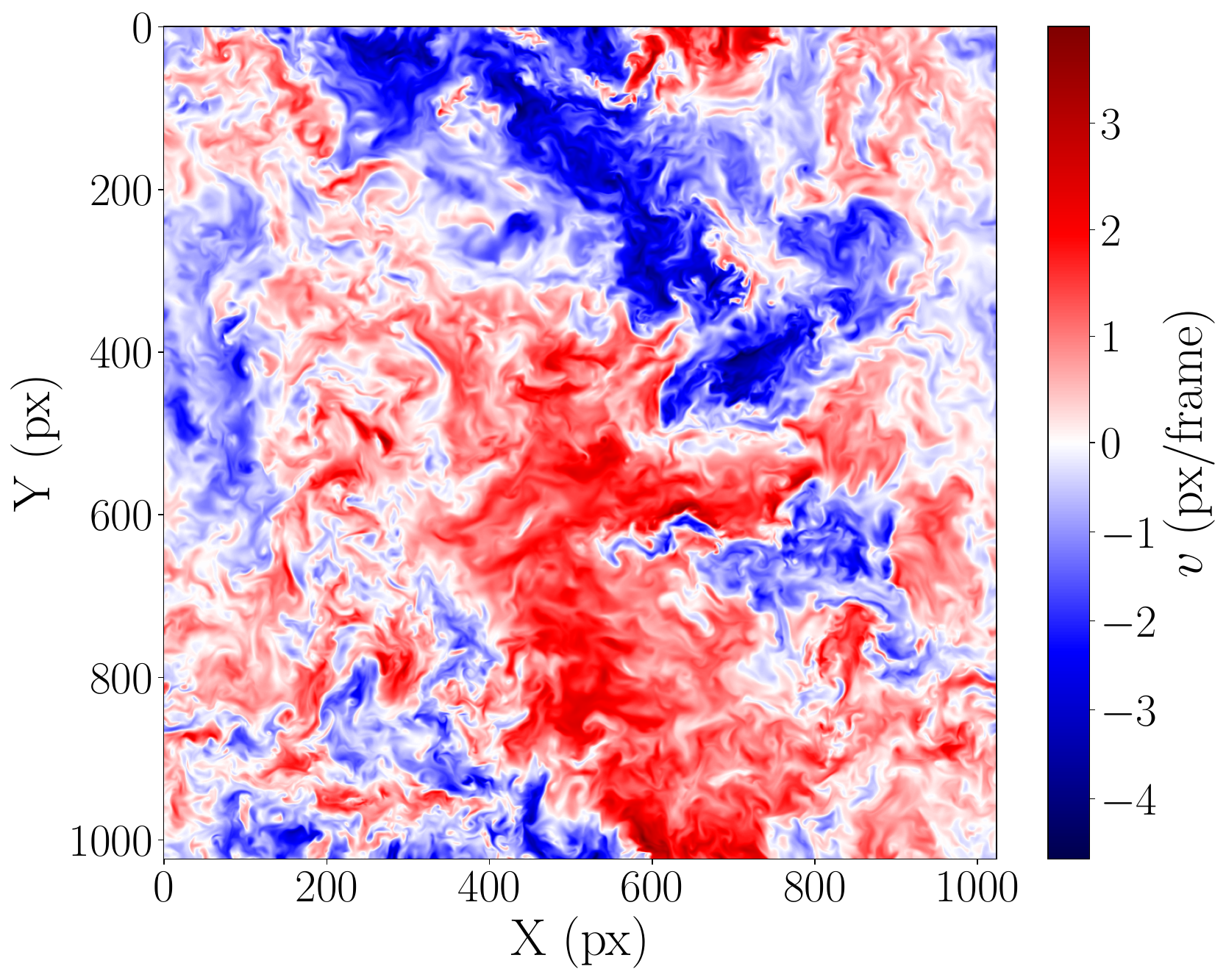}
        \caption{}
        \label{fig:v_dns}
    \end{subfigure}
    \caption{Example of the two components of an instantaneous displacement fields in the selected plane of the HIT dataset. (a) $u(x,y)$. (b) $v(x,y)$. The particles will be moved by this instantaneous complex flow at each time step. }
    \label{fig:DNS_fields}
\end{figure}

To evaluate OFV with a more complex flow, with a wide variety of displacement gradients at different scales at the same time, we used 199 snapshots from the Homogeneous Isotropic Turbulence (HIT) dataset of the Polytechnic University of Madrid \cite{Cardesa2017TheTC}. The dataset comprises instantaneous velocity fields in a cube on a $1024^3$ grid at $Re_\lambda=315$ with a maximum resolution $k\eta=2$. From each instantaneous 3D displacement field, the displacement components in the plane ($u,v$) were extracted in the median plane $z$ of the data cube. For the sake of simplicity, we consider the units to be pixels per image, since the extracted fields are expressed on a grid of pixels. 
Figure~\ref{fig:datacube} indicates the selected plane in the volume, while Fig.~\ref{fig:DNS_fields} shows an example of the 2D DNS instantaneous displacement fields in the selected plane. It clearly displays rich small-scale structures and sharp gradients all over the selected plane, typical from a turbulent flow.\\

\begin{figure}[h!]
    \centering
    \includegraphics[width=0.98\linewidth]{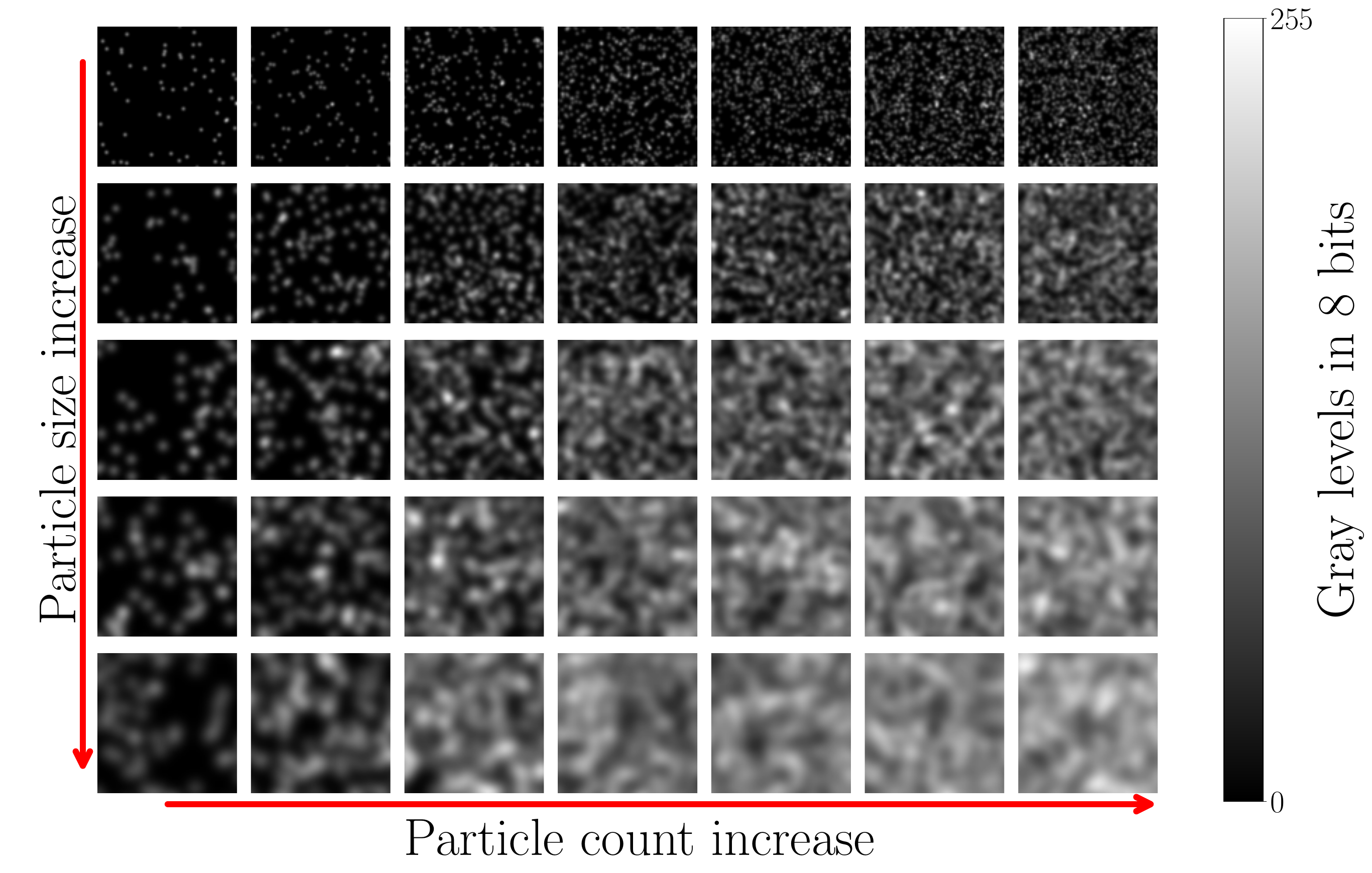}
    \caption{Synthetic particle images for the HIT dataset. Left to right: increasing particle size; top to bottom: increasing particle count. Panel shows $100\times100$ px sub--samples.}
    \label{fig:particle_images_HIT}
\end{figure}

In the absence of open-source software to generate pairs of synthetic particle images from existing displacement fields, a synthetic particle image generator was developed in--house to create images from displacement fields derived from the DNS. Specifically, particles are randomly placed within $1024\times1024$ px images and then displaced using the DNS displacement fields, which serve as displacement vectors per pixel. The particles are represented as Gaussian blobs whose size is defined by the full width at half maximum (FWHM) of a Gaussian with a standard deviation $\sigma$. These two parameters are related by the following equation:
\[
\mathrm{FWHM} = 2\sqrt{2\ln 2}\;\sigma \approx 2.35482\,\sigma.
\]
Here, $\sigma$ is varied to control the particle size.
Particles are initialized in a finite measurement volume with random depth coordinate $z$ and rendered  into a 2D plane. 

To study the effect of the concentration of particles and of their diameter, we varied $\sigma$ from $1$ to $5$ (effective diameters $\approx 2.4$ to $\approx 11.8$~px) and the total particle count from $5{,}000$ to $125{,}000$ per image. Figure~\ref{fig:particle_images_HIT} shows $100\times100$ px sub--samples that illustrate how both the number and size of particles alter the image texture.

\begin{figure}[h!]
    
    \begin{subfigure}{0.6\linewidth}
        \centering
        \includegraphics[width=\textwidth]{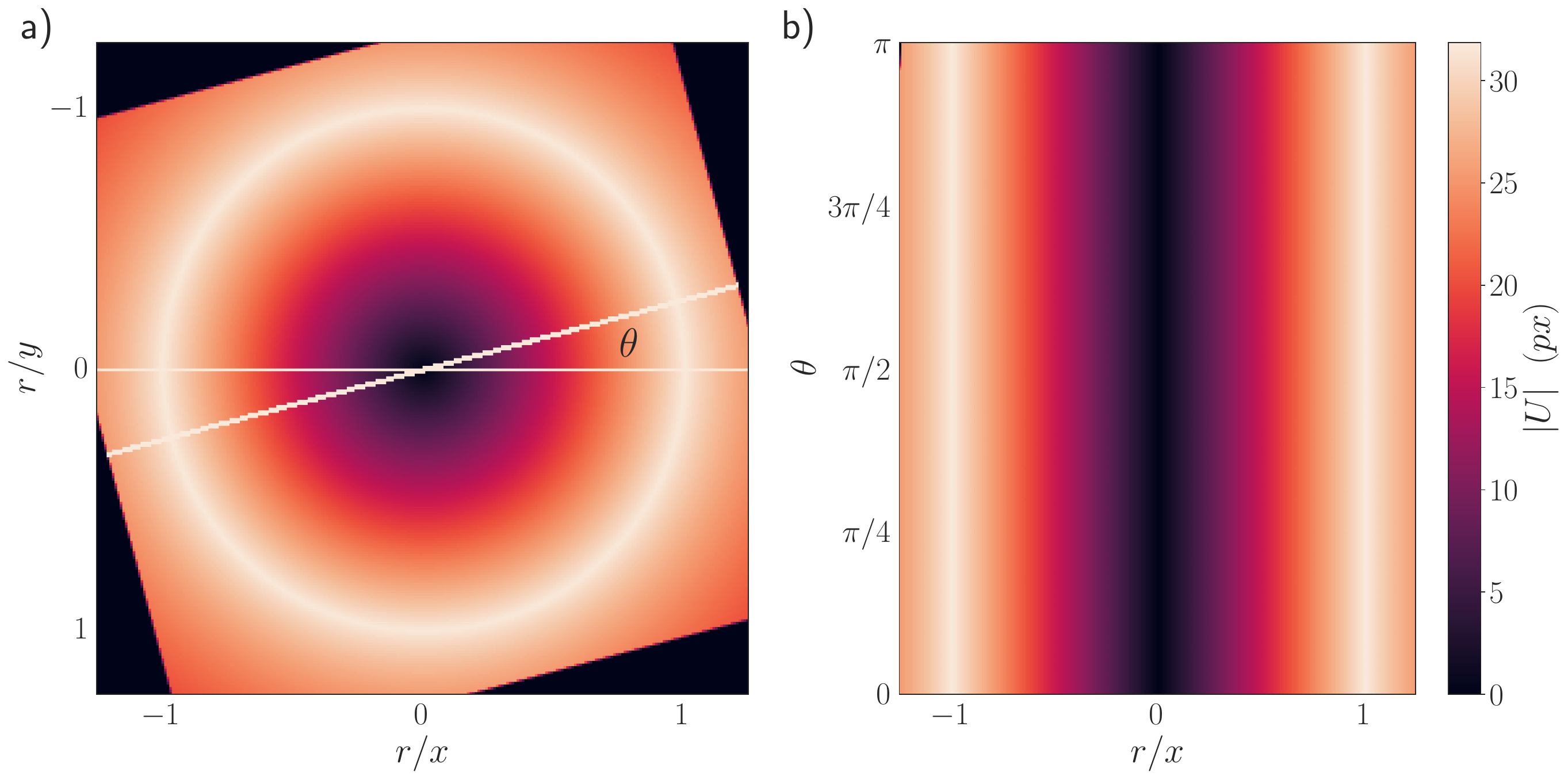}    
    \end{subfigure}
    \begin{subfigure}{0.6\linewidth}
        \includegraphics[width=\textwidth]{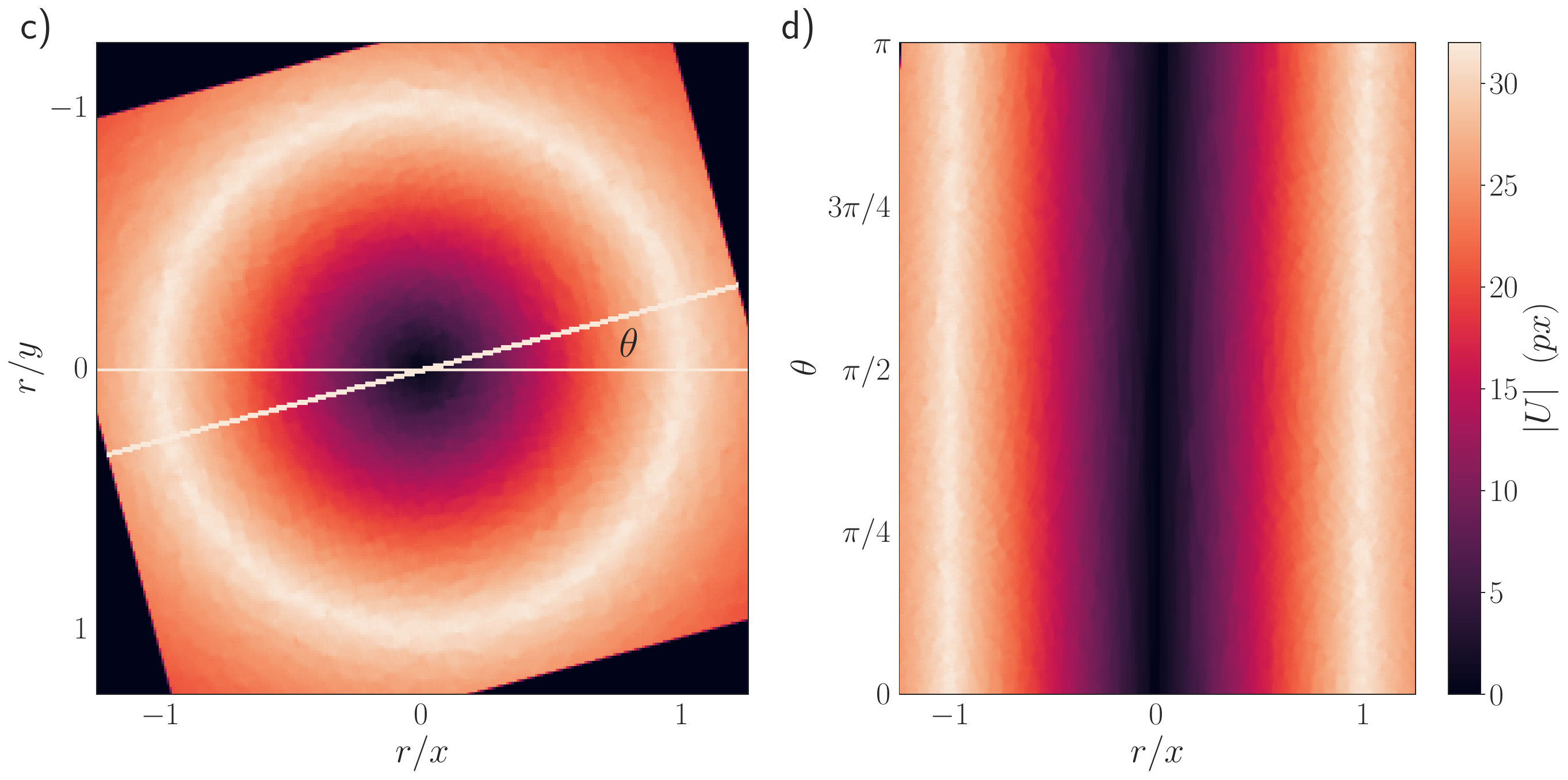}
    \end{subfigure}
    \begin{subfigure}{0.6\linewidth}
        \includegraphics[width=\textwidth]{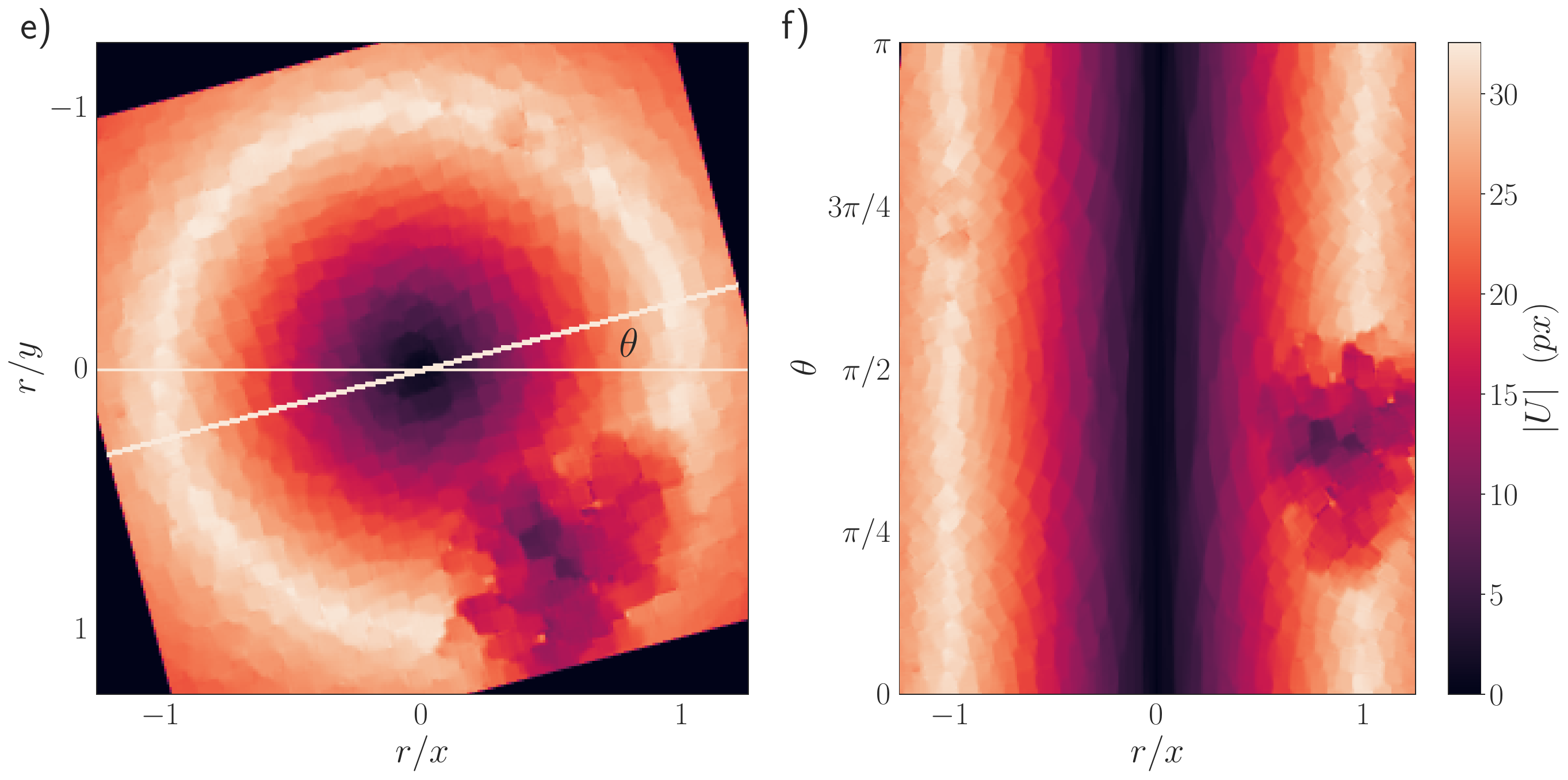}
    \end{subfigure}
    \caption{a) Principle used to obtain all theoretical displacement profiles for the Rankine vortex via rotation of the displacement field. b) Corresponding contour plot of the theoretical profiles versus rotation angle $\theta$ for $R=100$~px and maximum displacement $32$~px. As the theoretical Rankine vortex is invariant by rotation, it leads to a band-like contour in the ($\theta$, $r/x$) plot. c, d) Same as a, b) but instead of the theoretical displacement field, the displacement field based on the displacement of particles and computed by OFV is shown. The comparison is accurate in this case, but one can see that the displacement field is slightly blurred, compared to the theoretical field. \textcolor{black}{e, f) Same as c, d) but portraying a sub-optimal OFV result from a different parameter configuration. This comparison shows the localized errors that could be smoothed by global error metrics. } }
    \label{fig:rankine_stretch}
\end{figure}

\subsection{Error metrics\label{sec:error_estimation}}

We used two metrics to quantify OFV accuracy: (i) a point--wise absolute displacement error applicable to both the Rankine and HIT cases and (ii) a Rankine-specific profile comparison based on an angular “unfolding” of the vortex.

\paragraph{Absolute displacement error (Rankine and HIT).}
The pointwise absolute displacement error is
\begin{equation}
    Err(x,y) \;=\; \sqrt{\big(u_{\text{th}} - u_{\text{OFV}}\big)^2 + \big(v_{\text{th}} - v_{\text{OFV}}\big)^2},
    \label{eq:absolute_error}
\end{equation}

where subscripts “th” and “OFV” denote respectively the``theoretical" and the components of the displacement estimated by OFV. This metric is reported directly and via spatial averages (e.g., $\langle Err \rangle_{x,y}$) as appropriate.

\begin{figure}[!h]
    \centering
    \includegraphics[width = 0.6\textwidth]{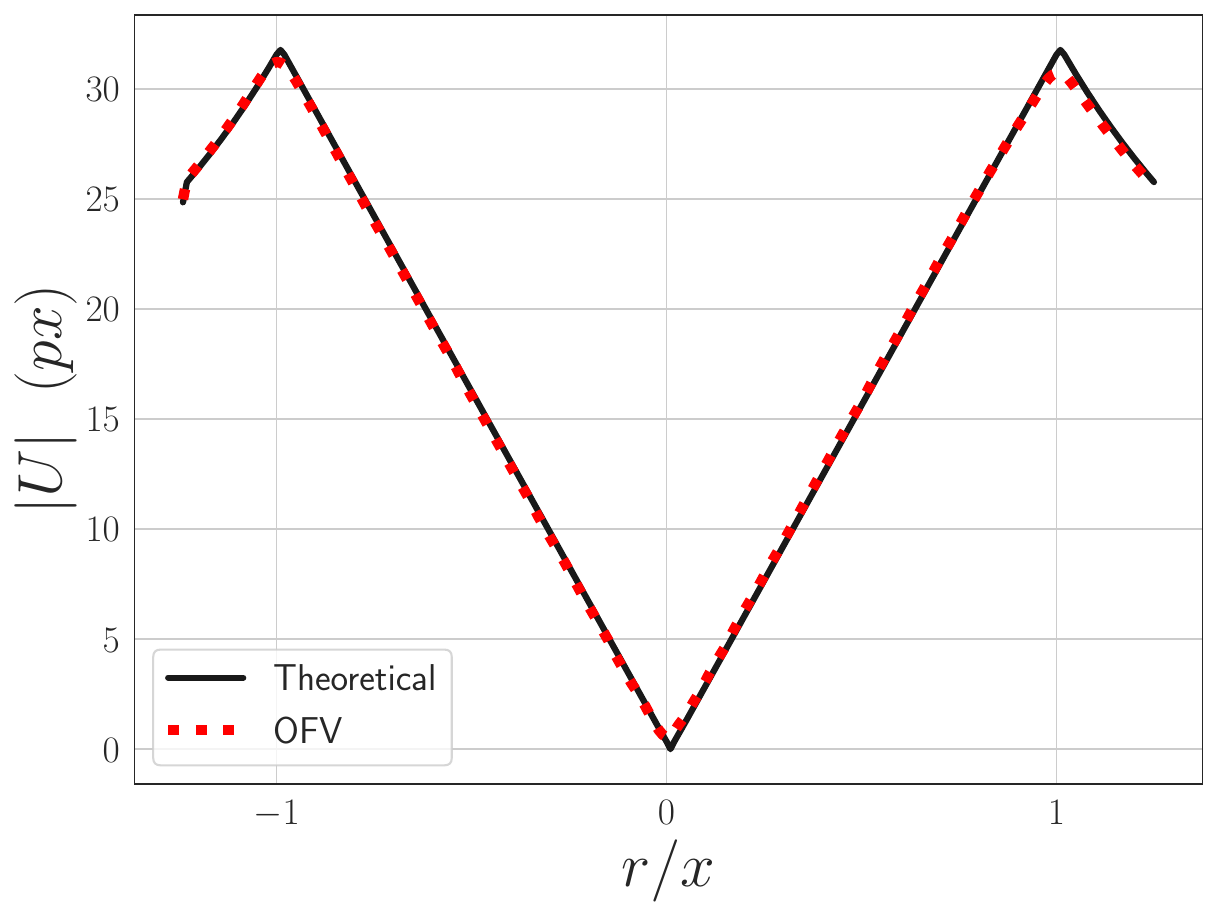}
    \caption{Theoretical vs OFV displacement profiles for a Rankine vortex ($R=100$~px, maximum displacement $32$~px). The radial profile is the $\theta$-averaged from Fig.~\ref{fig:rankine_stretch}b and d), showing close agreement in amplitude and in the location of extrema.}
    \label{fig:profile_comparison_rank}
\end{figure}

\paragraph{Rankine profile–unfolding metric.}
For the Rankine vortex, we compare one-pixel displacement-magnitude profiles across the core by rotating the displacement field and aggregating profiles over angle (Fig.~\ref{fig:rankine_stretch}). The procedure is the following:
(i) define a region of interest (ROI) centered on the vortex;
(ii) extract a one-pixel profile of $U(x,y)=\sqrt{u^2+v^2}$ along the mid-height ($y=Y/2$) and store it in an array $R$;
(iii) rotate the ROI by an angle $\Delta\theta$ chosen such that the arc length at radius $r$ corresponds to approximately one-pixel shift ($r\,\Delta\theta \approx 1$~px);
(iv) extract the new mid-height profile and append it to $R$;
(v) repeat steps (iii)–(iv) until a total rotation of $180^{\circ}$ is covered (Fig.~\ref{fig:rankine_stretch}a–b).
Applying this procedure to both the theoretical and the OFV fields produces $R_{\text{th}}$ and $R_{\text{OFV}}$. It is then possible to obtain the  mean radial displacement profiles for each configuration. Fig.~\ref{fig:profile_comparison_rank} shows the mean displacement profile computed by OFV with a given set of parameter (corresponding to Fig.~\ref{fig:rankine_stretch}c--d), compared to the theoretical profile. In this case, the OF computation recovers accurately the displacement amplitude and gradient. 

We then define the scalar discrepancy $\Psi$ as:

\begin{equation}
    \Psi \;=\; \frac{1}{A}\,\sum_{i=1}^{N_x}\sum_{j=1}^{N_\theta} \bigl|\,R_{\text{th}}(i,j) - R_{\text{OFV}}(i,j)\,\bigr|,
    \label{eq:psi_metric}
\end{equation}
where $i$ indexes the radial position, $j$ the rotation angle samples, and $A$ is the normalization by the unfolded vortex area. This criterion will be used in the following to quantify the accuracy of OF computation compared to the theoretical profiles at the local scale. 

\textcolor{black}{In comparison, Fig.~\ref{fig:rankine_stretch} c--d) show an OFV result in good agreement with the theoretical one, while panels e--f) display a suboptimal OFV result with a rougher texture and a localized region where the displacements are not correctly estimated. This comparison shows the relevance of this metric that is focused in the high displacement gradient region, hence providing a local quality criterion of the displacement vectors computation.}

\section{Accuracy benchmarks on synthetic flows}
\subsection{Rankine vortex - Influence of particle concentration\label{subsec:RK_results}}

\begin{figure}[h!]
    \centering
    \includegraphics[width=0.9\linewidth]{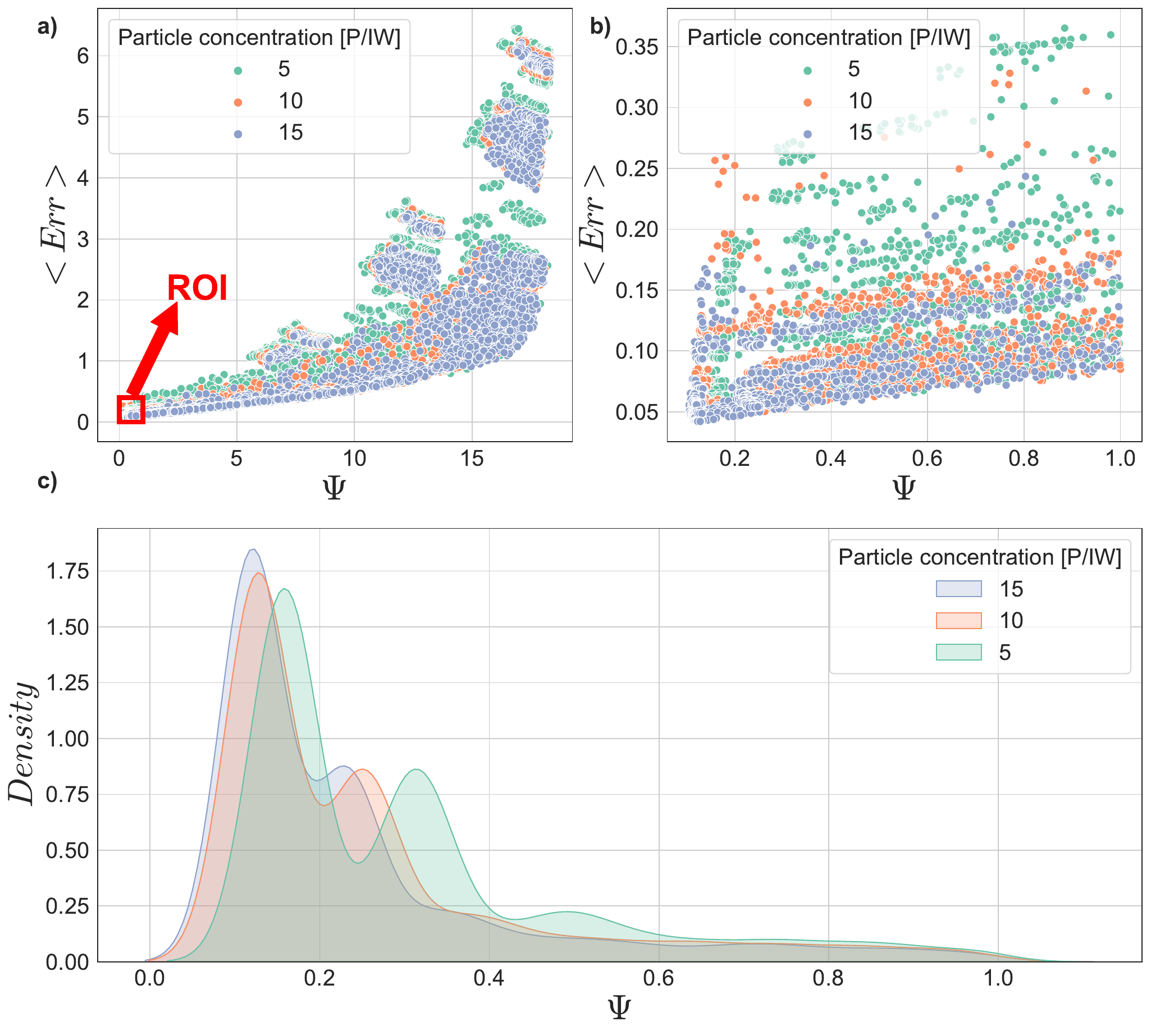}
    \caption{Scatter--plot  error analysis for a Rankine vortex with core radius $r=75$~px. \textbf{a)} Scatter plot of both error metrics for all displacements and concentrations. \textbf{b)} Zoomed-in view of the ROI. \textbf{c)} Probability density function of $\Psi$ within the ROI.}
    \label{fig:rankine_core_r75}
\end{figure}

\begin{table}[!ht]
    \centering
    \begin{adjustbox}{width=1\textwidth}
        \begin{tabular}{|>{\centering\arraybackslash}m{.15\linewidth}|>{\centering\arraybackslash}m{.15\linewidth}|>{\centering\arraybackslash}m{.2\linewidth}|>{\centering\arraybackslash}m{.1\linewidth}|>{\centering\arraybackslash}m{.1\linewidth}|
        >{\centering\arraybackslash}m{.1\linewidth}|
        >{\centering\arraybackslash}m{.1\linewidth}|
        >{\centering\arraybackslash}m{.1\linewidth}|}
        \hline
             \textbf{Rankine \newline core radius\newline (px)}&\textbf{Maximum\newline displacement (px)}&\textbf{Particle\newline concentration (p/IW)} & \textbf{KR\newline(px)}&\textbf{NR\newline(px)}&\textbf{PSL}&\textbf{IT}&\textbf{D/R} \\
             \hline
             \multirow{4}{*}{12}& \multicolumn{1}{c|}{8}&\multicolumn{1}{c|}{15}&\multicolumn{1}{c|}{2}&\multicolumn{1}{c|}{1}&\multicolumn{1}{c|}{2}&\multicolumn{1}{c|}{4}&\multicolumn{1}{c|}{0.67}\\ 
            &16&5&2&3&4&4&1.33\\
            &24&15&2&2&4&4&2.00\\
            &32&10&2&1&4&4&2.67\\
             \hline
             \multirow{4}{*}{25}& \multicolumn{1}{c|}{8}&\multicolumn{1}{c|}{15}&\multicolumn{1}{c|}{3}&\multicolumn{1}{c|}{1}&\multicolumn{1}{c|}{3}&\multicolumn{1}{c|}{4}&\multicolumn{1}{c|}{0.32}\\
             &16&15&2&2&4&4&0.64\\
             &24&5&2&1&4&3&0.96\\
             &32&10&2&1&4&3&1.28\\
             \hline
             \multirow{4}{*}{50}& \multicolumn{1}{c|}{8}&\multicolumn{1}{c|}{15}&\multicolumn{1}{c|}{5}&\multicolumn{1}{c|}{1}&\multicolumn{1}{c|}{4}&\multicolumn{1}{c|}{3}&\multicolumn{1}{c|}{0.16}\\
             &16&15&4&1&3&4&0.32\\
             &24&10&2&3&4&4&0.48\\
             &32&5&3&1&4&4&0.64\\
             \hline
             \multirow{4}{*}{75}& \multicolumn{1}{c|}{8}&\multicolumn{1}{c|}{15}&\multicolumn{1}{c|}{4}&\multicolumn{1}{c|}{1}&\multicolumn{1}{c|}{4}&\multicolumn{1}{c|}{2}&\multicolumn{1}{c|}{0.11}\\
             &16&15&4&1&4&4&0.21\\
             &24&15&3&1&4&4&0.32\\
             &32&15&2&2&4&4&0.43\\
             \hline
             \multirow{4}{*}{100}& \multicolumn{1}{c|}{8}&\multicolumn{1}{c|}{15}&\multicolumn{1}{c|}{6}&\multicolumn{1}{c|}{1}&\multicolumn{1}{c|}{3}&\multicolumn{1}{c|}{4}&\multicolumn{1}{c|}{0.08}\\
             &16&15&4&1&3&4&0.16\\
             &24&15&3&1&4&4&0.24\\
             &32&15&2&1&4&4&0.32\\
             \hline
             \multirow{4}{*}{150}& \multicolumn{1}{c|}{8}&\multicolumn{1}{c|}{15}&\multicolumn{1}{c|}{6}&\multicolumn{1}{c|}{1}&\multicolumn{1}{c|}{4}&\multicolumn{1}{c|}{2}&\multicolumn{1}{c|}{0.05}\\
             &16&15&5&1&4&3&0.11\\
             &24&15&4&1&4&3&0.16\\
             &32&15&4&1&4&4&0.21\\
             \hline
        \end{tabular}
    \end{adjustbox}
    \caption{OFV parameter configuration yielding the best results for each case. \textbf{KR}: kernel radius; \textbf{NR}: normalization radius; \textbf{PSL}: pyramid sub-levels; \textbf{IT}: iterations; \textbf{D/R}: ratio of maximum displacement to core radius.}
    \label{tab:of_best}
\end{table}

Figure~\ref{fig:rankine_core_r75}a) shows the spatially averaged displacement error over the entire instantaneous velocity fields, $\langle Err\rangle_{x,y}$, together with the unfolded profile criterion $\Psi$ for all maximum displacements and concentrations of particles for a Rankine vortex with a given core radius $r=75$~px. Both metrics here are combined to show the local to global ratio of errors. Figure~\ref{fig:rankine_core_r75}b) shows a zoomed ROI where both metrics indicate minimal error, with $\Psi<1$. Since $\Psi$ is normalized by the area of the unfolded vortex, $\Psi=1$ can be interpreted as an average difference of one pixel of displacement between the theoretical and OFV fields. Figure~\ref{fig:rankine_core_r75}c presents the probability density function (PDF) of $\Psi$ within the ROI for the three particle concentrations. It shows that the particle concentration significantly influences the error margins across all displacement magnitudes. For the given set of OFV parameters (see Tab.~\ref{tab:OFPIV}), the highest particle concentrations correspond to the lowest error levels.
\def\w{0.3\linewidth}
\textcolor{black}{
\begin{figure}
    \includegraphics[width=0.8\linewidth]{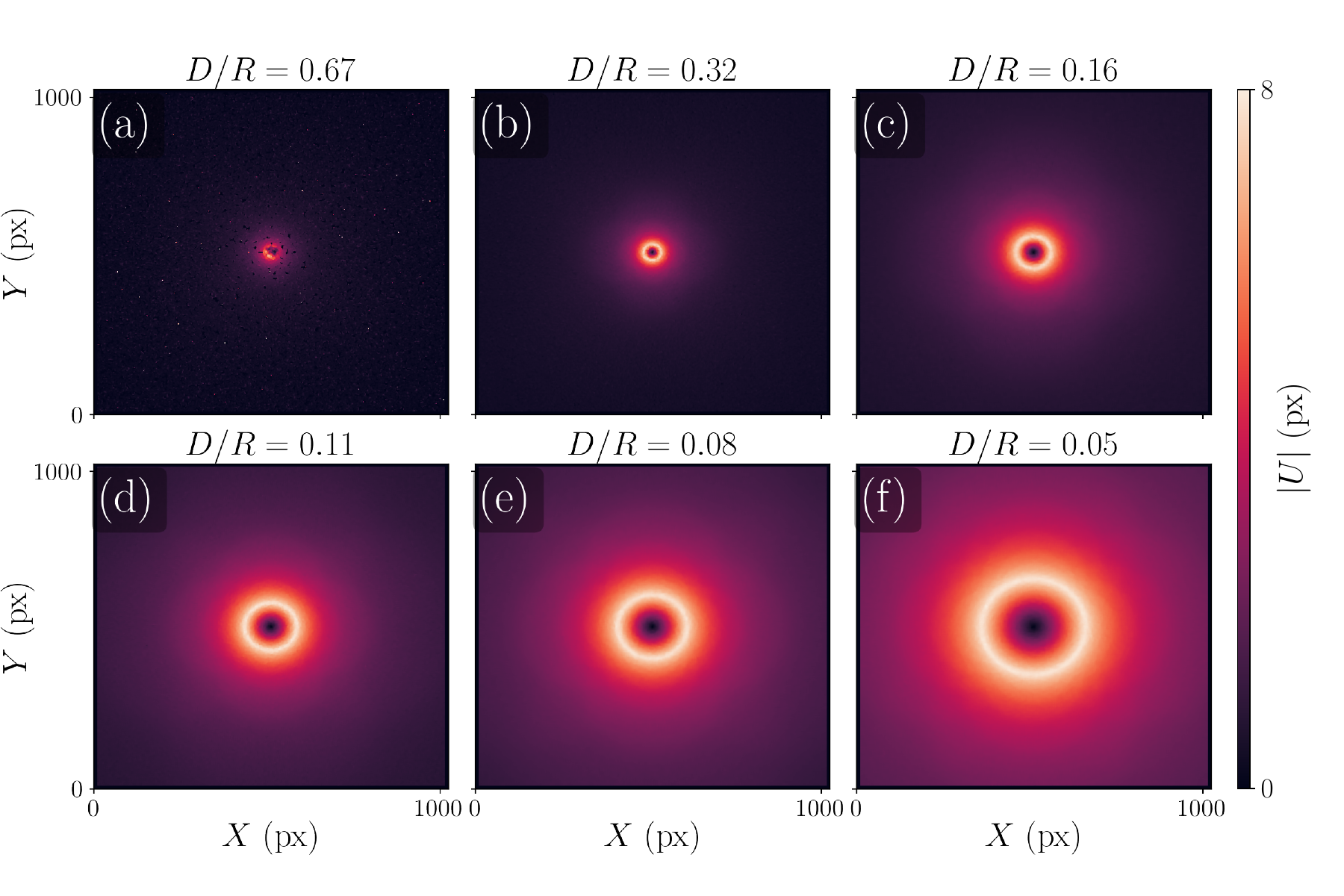}
    \caption{\textcolor{black}{Best OFV results for maximum displacement $|U|=8$ px, and different vortex core radius. Panels (a--f) correspond to $r=12,~25,~50,~75,~100,~150$ px (\textbf{D/R} indicated in each panel). Displacement magnitude maps.}}
    \label{fig:Best_U_8}
\end{figure}}
\textcolor{black}{
\begin{figure}
    \includegraphics[width=0.8\linewidth]{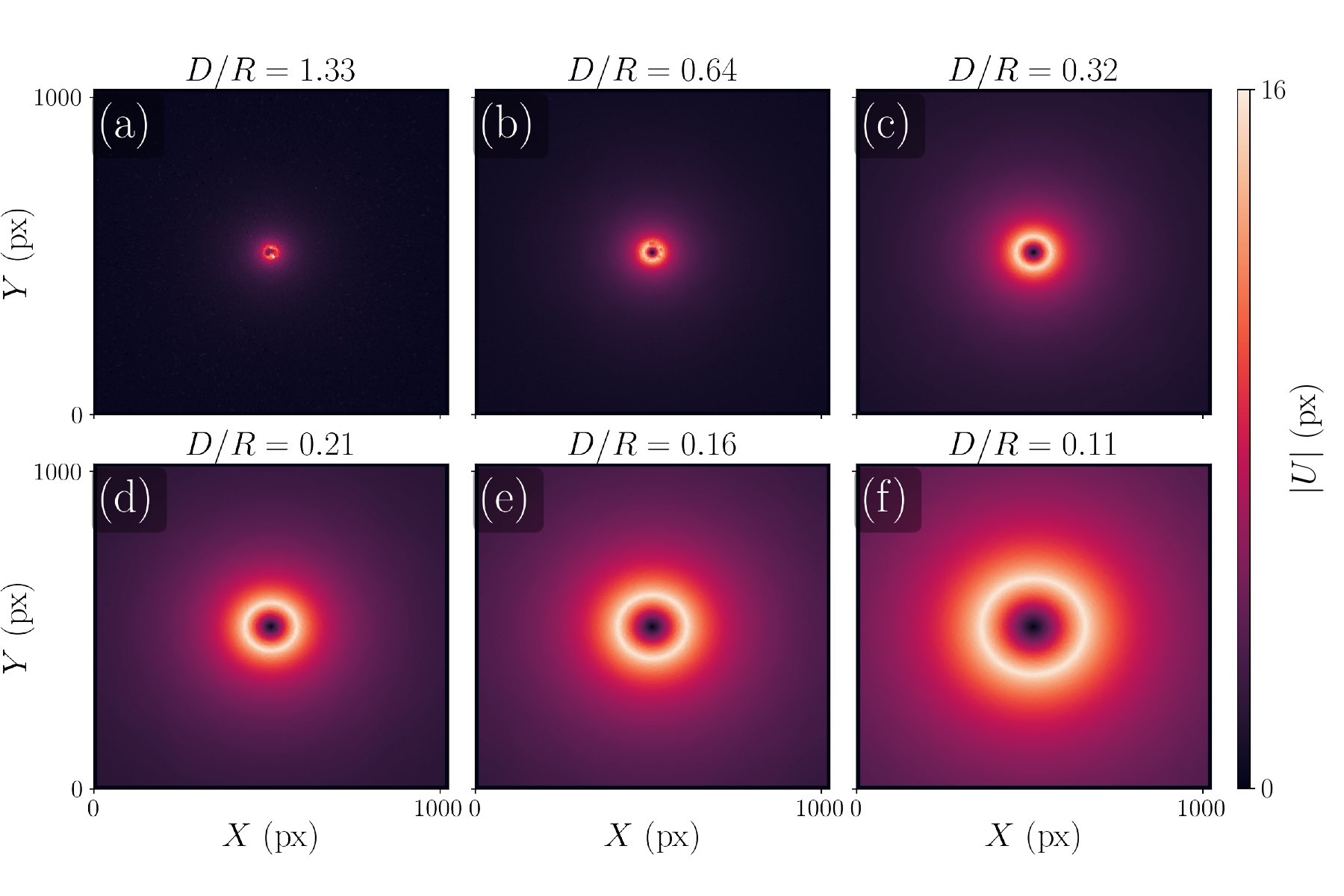}
    \caption{\textcolor{black}{Best OFV results for maximum displacement $|U|=16$ px, and  different vortex core radius. Panels (a--f) correspond to $r=12,~25,~50,~75,~100,~150$ px (\textbf{D/R} indicated in each panel). Displacement magnitude maps.}}
    \label{fig:Best_U_16}
\end{figure}}
\textcolor{black}{
\begin{figure}
    \includegraphics[width=0.8\linewidth]{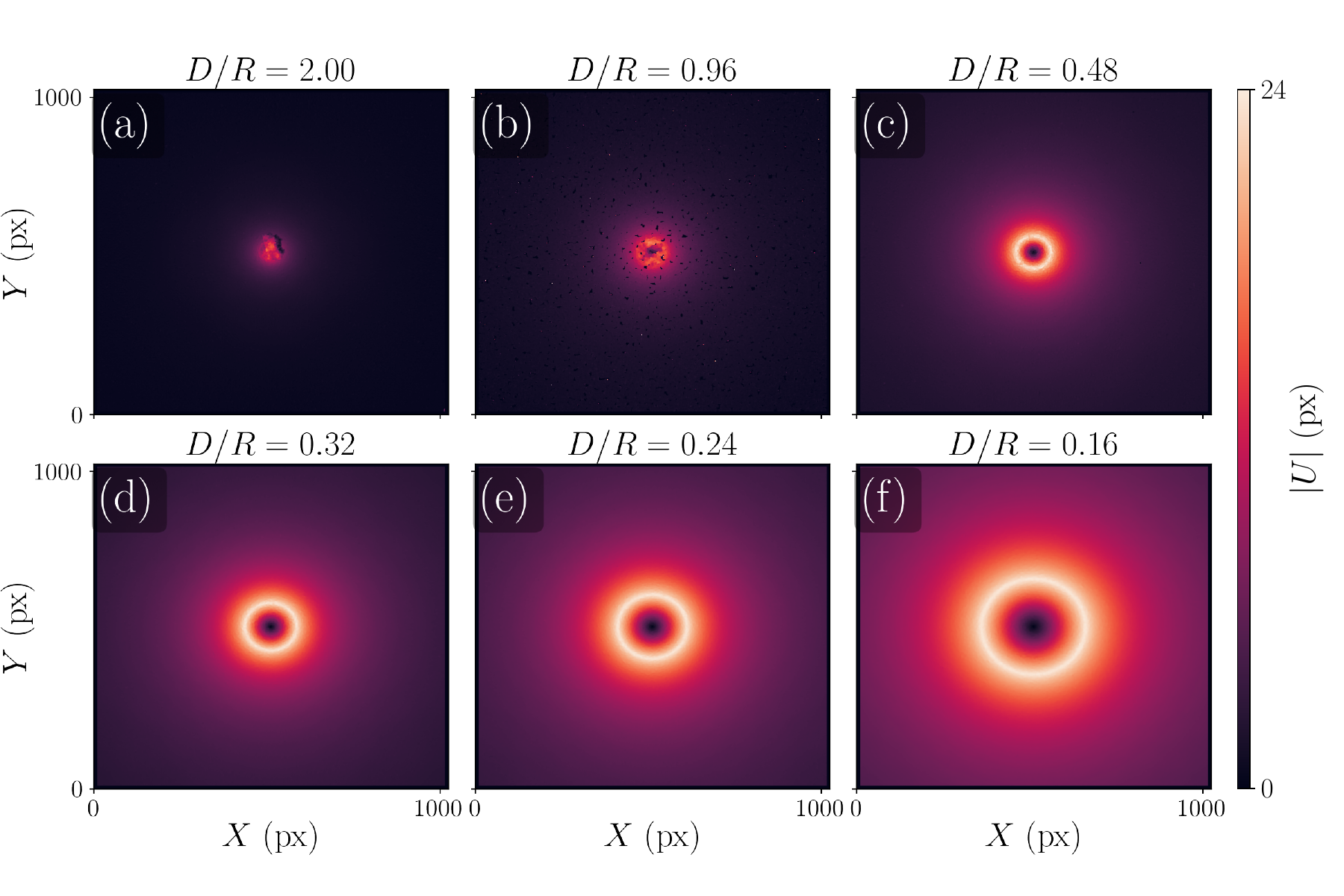}
    \caption{\textcolor{black}{Best OFV results for maximum displacement $|U|=24$ px, and different vortex core radius. Panels (a--f) correspond to $r=12,~25,~50,~75,~100,~150$ px (\textbf{D/R} indicated in each panel). Displacement magnitude maps.}}
    \label{fig:Best_U_24}
\end{figure}}
\textcolor{black}{
\begin{figure}[!ht]
    \includegraphics[width=0.8\linewidth]{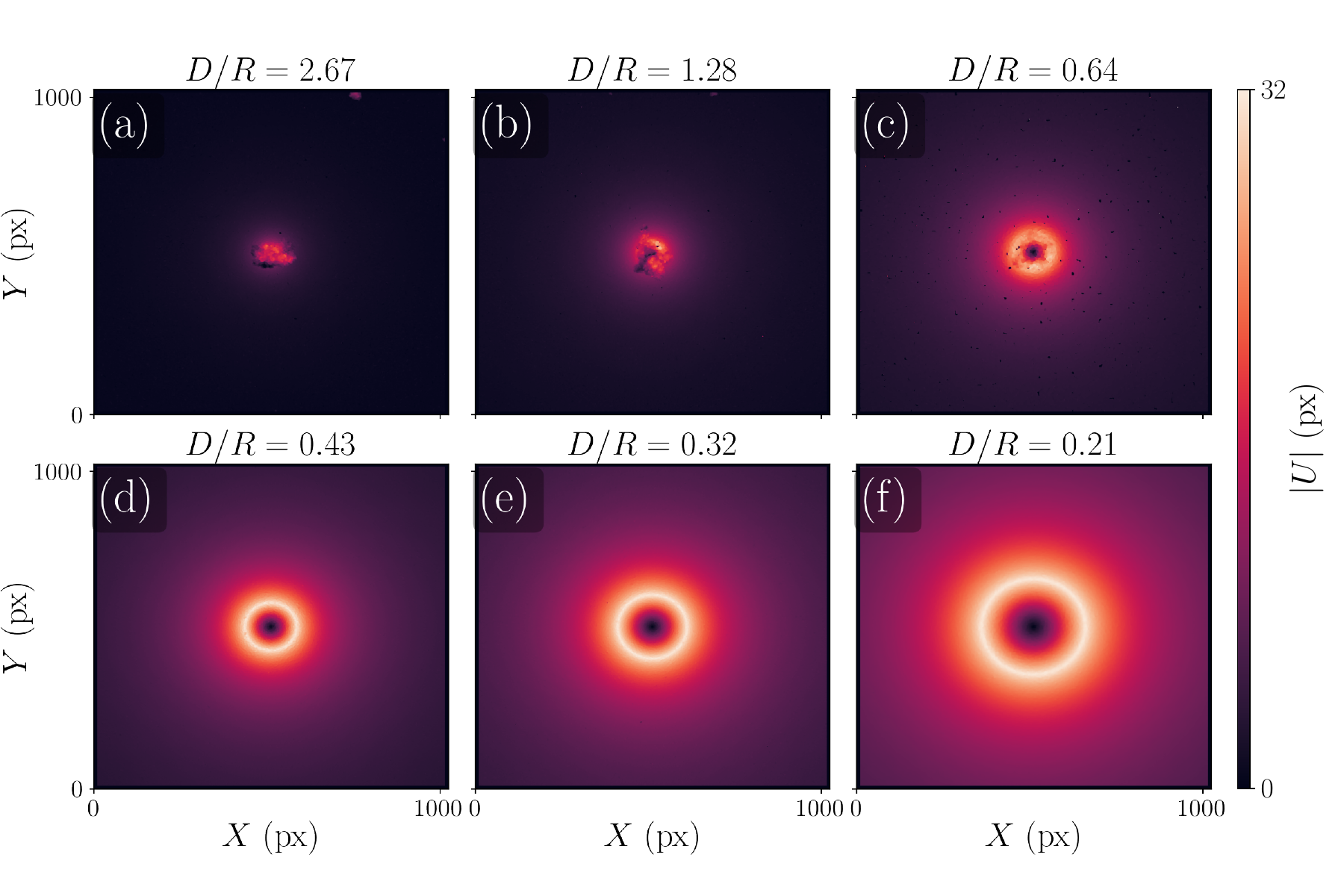}
    \caption{\textcolor{black}{Best OFV results for maximum displacement $|U|=32$ px, and  different vortex core radius. Panels (a--f) correspond to $r=12,~25,~50,~75,~100,~150$ px (\textbf{D/R} indicated in each panel). Displacement magnitude maps.}}
    \label{fig:Best_U_32}
\end{figure}}
The sets of parameters minimizing the $\Psi$ error were then sought. Table~\ref{tab:of_best} lists the parameter sets leading to the best OFV results. Figs.~\ref{fig:Best_U_8}--\ref{fig:Best_U_32} show the corresponding displacement--magnitude fields. As mentioned earlier, the best results are generally obtained with images showing a higher concentration of particles. For concentrations below 15~p/IW, although Figs.~\ref{fig:Best_U_8}--\ref{fig:Best_U_32} and Table~\ref{tab:of_best} report the best available results, algorithmic limits may be approached or exceeded.

There is a clear link between the maximum displacement $D$ and the smallest resolvable structure (here, the core radius $R$). We therefore use $D/R$ as an indicator of displacement gradient: larger $D/R$ corresponds to smaller, faster vortices; smaller $D/R$ to larger, slower vortices. In Figs.~\ref{fig:Best_U_8}--\ref{fig:Best_U_32}, a resolution limit appears around $D/R \approx 0.7$, consistent with the expectation that small, rapidly rotating vortices are the most challenging to measure.

\begin{figure}
    \centering
    \includegraphics[width=0.8\linewidth]{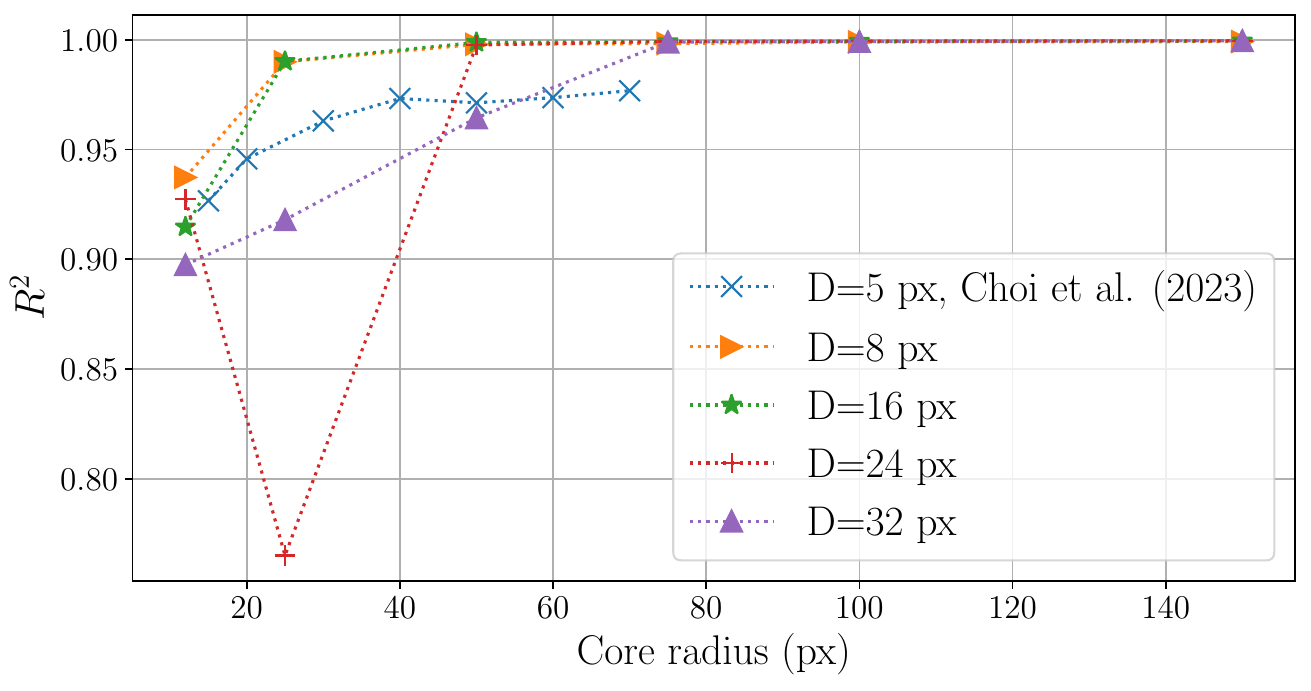}
    \caption{$R^2$ criterion for evaluating the computed displacements vs the theoretical data for different core radius and maximum displacements. Blue cross from \citet{choi2023deep}.}
    \label{fig:R2}
\end{figure}

In \citet{Lagemann2022GeneralizationOD}, RAFT-PIV applied to a turbulent boundary layer showed similar trends: larger displacement gradients increased errors in both displacement components, while higher seeding density reduced them. \citet{choi2023deep} also validated a CNN-based OF method with Rankine-vortex images (various $R$; $D=5$~px). Their $R^2$ ($R^2 = 1 - \frac{\Sigma (V - \hat{V})^2}{\Sigma (V - \bar{V})^2}$, where $V$ is the theoretical data, $\bar{V}$ is the mean of the theoretical data, and $\hat{V}$ is the result from their code) comparison against reference data is reproduced for context. Figure~\ref{fig:R2} compares our $R^2$ values to \citet{choi2023deep}. When the limits are not exceeded, OFV solves displacement gradients as well, or even better.

Another observation from Table~\ref{tab:of_best} and Figs.~\ref{fig:Best_U_8} -- \ref{fig:Best_U_32} is that \textbf{KR} relates more with displacement \emph{gradient} than with displacement magnitude. For example, with $r=100$~px and $D=32$~px \textcolor{black}{(Fig.~\ref{fig:Best_U_32})}, the optimal \textbf{KR} is only $2$~px. Ranking results by $D/R$ reveals a trend: higher $D/R$ (stronger gradients) selects smaller \textbf{KR}, and vice versa. Algorithmically, larger kernels smooth the field, whereas smaller kernels better capture localized variations (i.e., high gradients). This is evident, for example,  in \textcolor{black}{Fig.~\ref{fig:Best_U_24},  where the displacement is the same, but the size of the structure changes, and the \textbf{KR} reported in Tab.~\ref{tab:of_best} decreases inversely to $D/R$. This shows an inverse correlation between \textbf{KR} and displacement \emph{gradient}.}

Comments on functional ranges of particle concentration for OF algorithms have been made by \citet{Mendes2021ACS}. After a comprehensive evaluation of different OF methods for PIV, they reported a broader domain of high-accuracy operation for OF than for CC-PIV, and highlighted the need for a stronger theoretical basis. We add a practical perspective and went further: defining OFV seeding in terms of ``particles per IW'' or ``particles per pixel'', similarly to the CC-PIV, is \textit{not adapted to OFV}. Instead, the operative concept should be the \textit{\textbf{image texture}}. Indeed, OF methods infer motion from time evolution of the spatial distribution of intensity $I(x, y, t)$. It does not depend on a number of particles in an IW, which has no meaning in OF. OFV rather depends on meaningful \textit{local variations of intensity}, meaning that enriching the texture of the snapshots directly enhances OFV spatial resolution. This aligns with the classical concepts of Gibson \cite{Gibson1967TheSC,Gibson1979TheEA,Gibon1950,gibson2002theory} (at the origin of the concept of OF) and with computer-vision practice, where texture manipulation has been shown to improve OF results \citep{ANDALIBI20151}. Thus, for OFV, even for seeding-based approach like Particle Image Velocimetry, the texture of the images should be the experimentalist’s primary criterion. This means that the seeding should be chosen to optimize the texture of the snapshots. This criterion naturally leads to larger concentrations of particles for OFV than for CC-PIV to optimize the spatial resolution.\\ 

\subsection{Rankine vortex - Resolution of small scales}

\begin{figure}
    \centering
    \begin{subfigure}{0.35\linewidth}
        \centering
        \includegraphics[width=\linewidth]{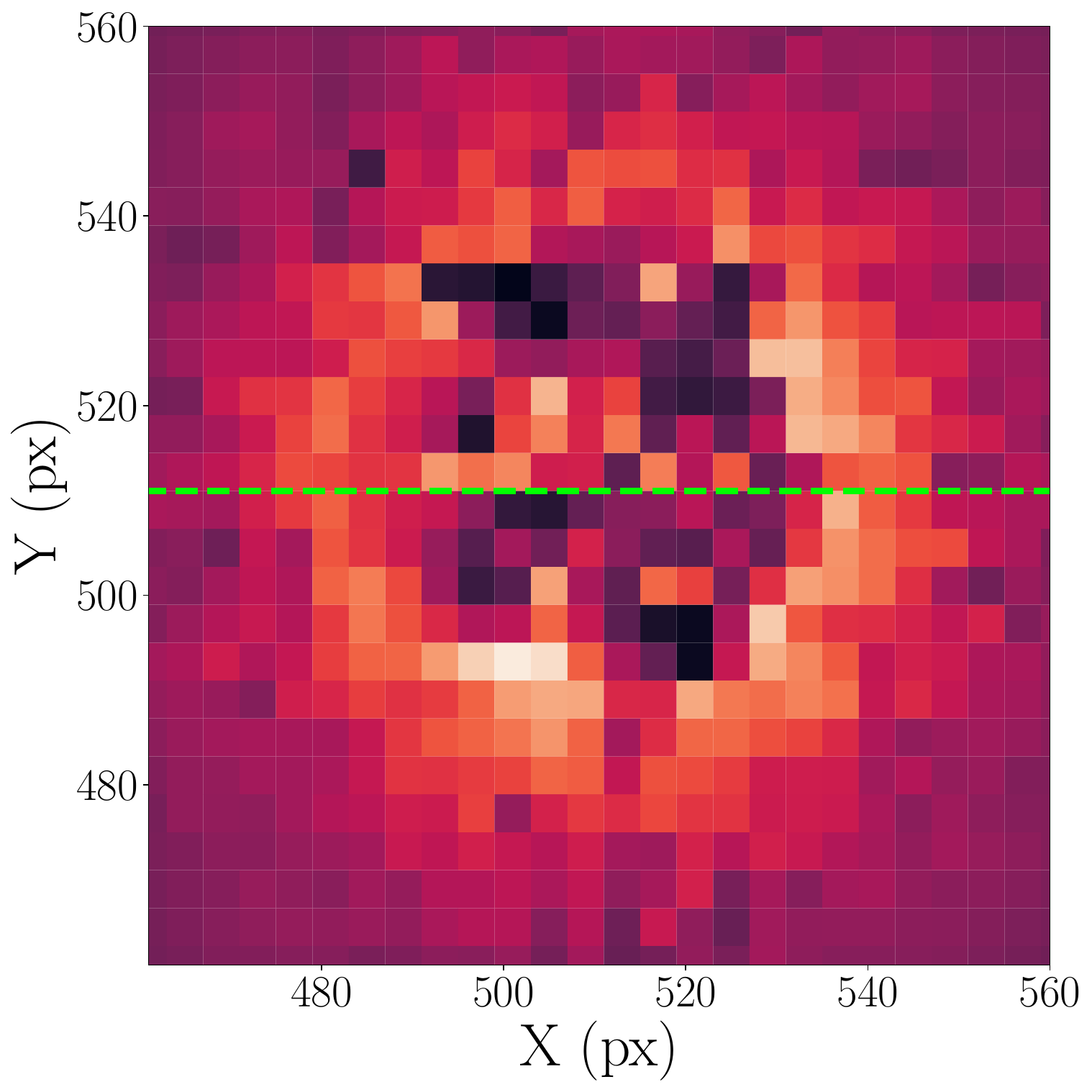}
        \caption{}
        \label{fig:cc_r12}
    \end{subfigure}
    \centering
    \begin{subfigure}{0.35\linewidth}
        \centering
        \includegraphics[width=\linewidth]{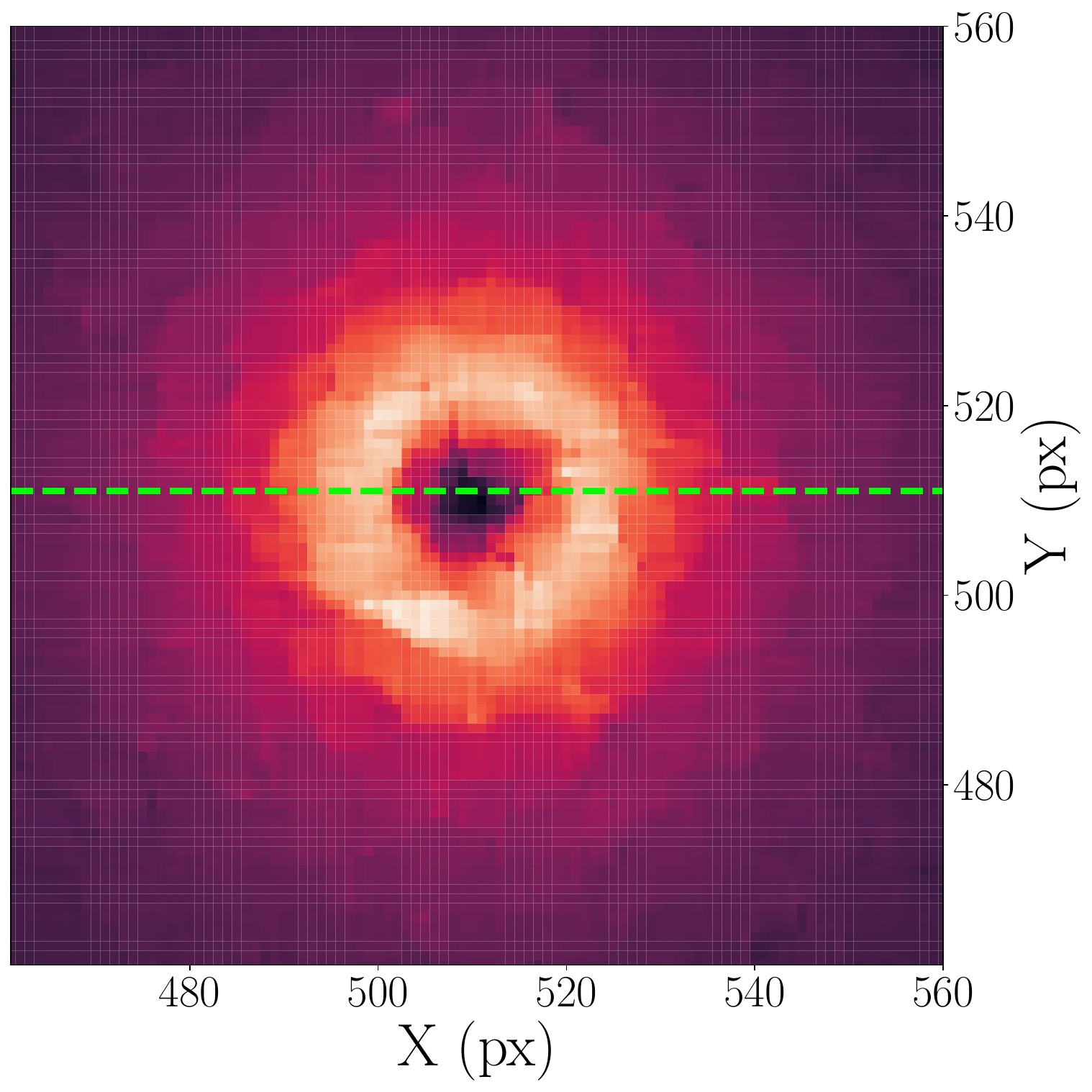}
        \caption{}
        \label{fig:of_r12}
    \end{subfigure}
    
    \begin{subfigure}{0.8\linewidth}
        \includegraphics[width=\linewidth]{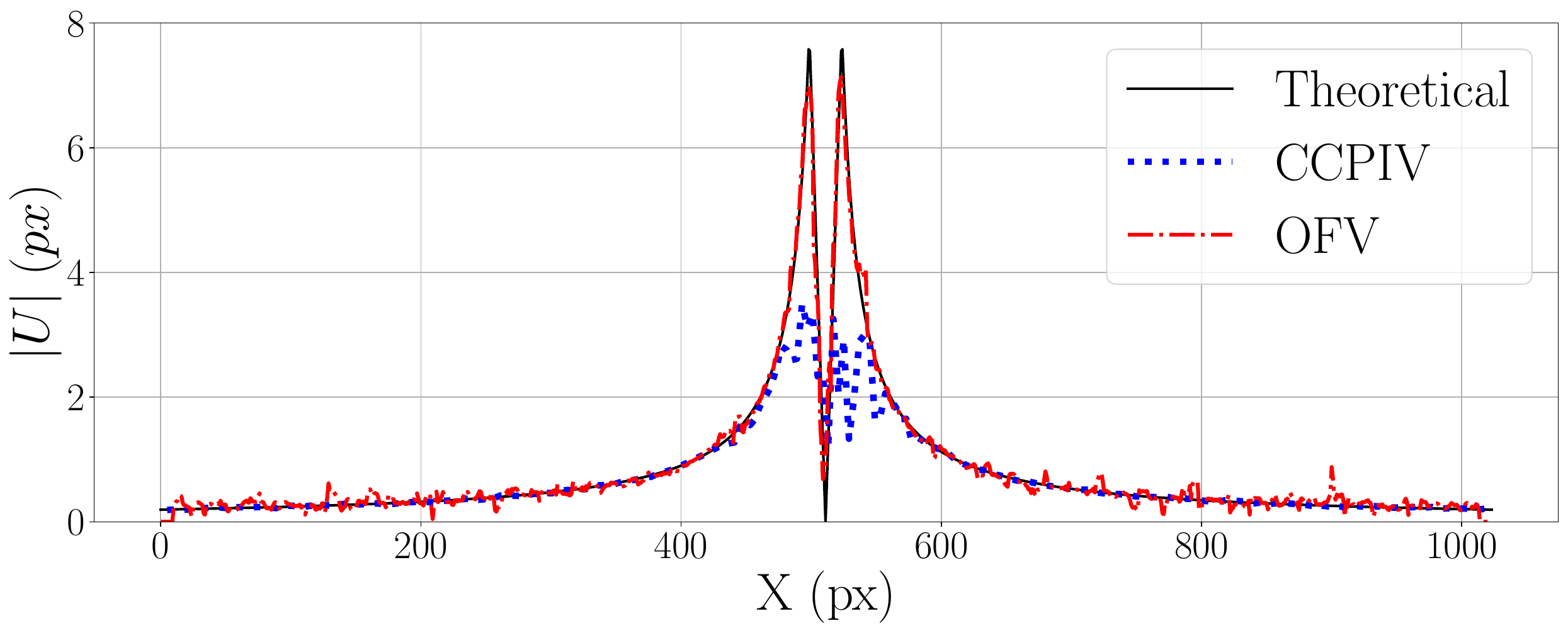}
        \caption{}
        \label{fig:r12_profiles}
    \end{subfigure}
    \caption{Rankine vortex with $r=12$~px and $D=8$~px. \textbf{(a)} CC-PIV zoom. \textbf{(b)} OFV zoom. \textbf{(c)} Mid-height displacement-magnitude profile (dotted green line). OFV from an image pair with 15~p/IW.}
    \label{fig:rk_comparison_12}
\end{figure}

\begin{figure}
    \centering
    \begin{subfigure}{0.35\linewidth}
        \centering
        \includegraphics[width=\linewidth]{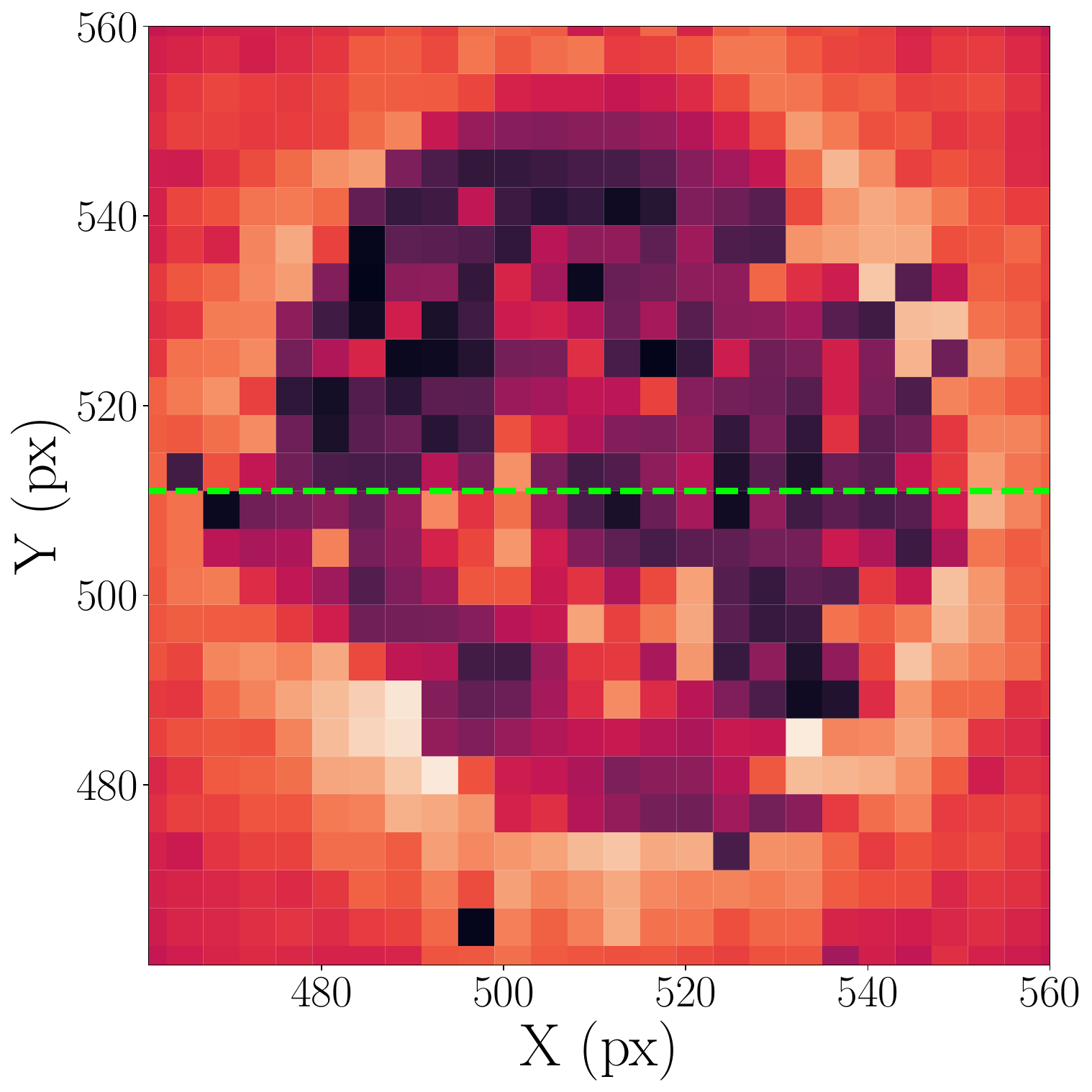}
        \caption{}
        \label{fig:cc_r25}
    \end{subfigure}
    \centering
    \begin{subfigure}{0.35\linewidth}
        \centering
        \includegraphics[width=\linewidth]{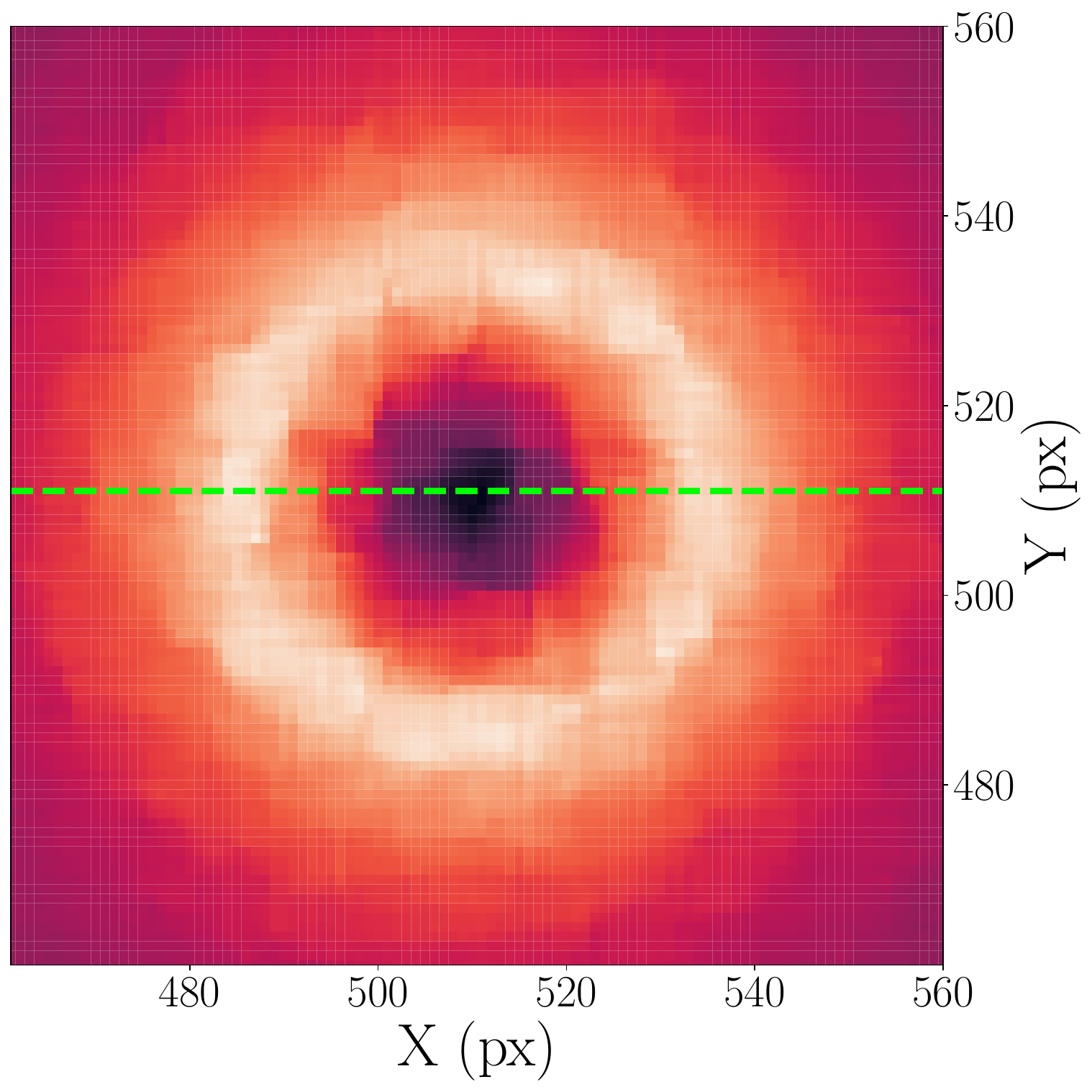}
        \caption{}
        \label{fig:of_r25}
    \end{subfigure}
    
    \begin{subfigure}{0.8\linewidth}
        \includegraphics[width=\linewidth]{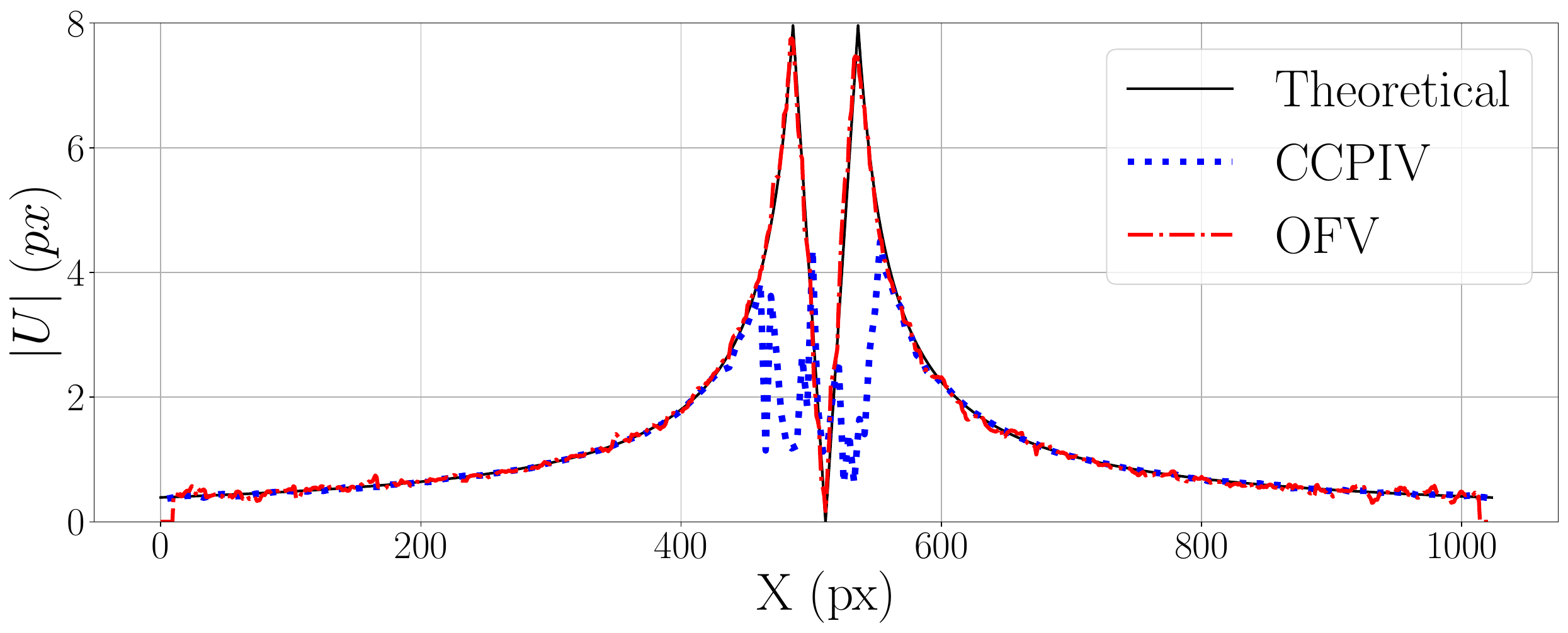}
        \caption{}
        \label{fig:r25_profiles}
    \end{subfigure}
    \caption{Rankine vortex with $r=25$~px and $D=8$~px. \textbf{(a)} CC-PIV zoom. \textbf{(b)} OFV zoom. \textbf{(c)} Mid-height displacement-magnitude profile (dotted green line). OFV from an image pair with 15~p/IW.}
    \label{fig:rk_comparison_25}
\end{figure}

A key question for OFV users is its ability to resolve small structures with precision, even with relatively large displacements, as can be encountered in a turbulent flow. Figures~\ref{fig:rk_comparison_12} and \ref{fig:rk_comparison_25} compare the best results obtained with OFV and CC-PIV for two very small vortices ($r=12$ and $r=25$~px), and a large displacement ($D=8$~px, comparable to the size of the vortex). The CC-PIV velocity fields were obtained with a multi-pass FFT implementation \citep{Thielicke_2021} with four passes, $IW=[64,32,16,8]$~px, and 50\% overlap, to optimize the spatial resolution. The Figures~\ref{fig:rk_comparison_12}--\ref{fig:rk_comparison_25}~(a–b) show zoomed views of the vortex core for respectively CC-PIV and OFV, while the Figures~\ref{fig:rk_comparison_12}--\ref{fig:rk_comparison_25}~(c) compares a one-pixel radial profile along the mid-height of the vortex ($y=Y/2$).

One can clearly see in Figs.~\ref{fig:rk_comparison_12}–-\ref{fig:rk_comparison_25} that the vortex structure is accurately recovered with OFV, despite its small size. For $r=12$~px (Fig.~\ref{fig:rk_comparison_12}), the global maxima are slightly underestimated (Fig.~\ref{fig:r12_profiles}), but the peak discrepancies are within 5\% of the theoretical profile. The peaks locations and gradients are well captured, even along a single pixel line (no spatial averaging). Similar conclusions hold for $r=25$~px (Fig.~\ref{fig:rk_comparison_25}): OFV recovers the structure of the vortex more clearly and resolves the extrema better than CC-PIV. These cases also confirm that OFV limitations are tied more to displacement \emph{gradient} than magnitude.

\textcolor{black}{Figs.~\ref{fig:error_U_8} to \ref{fig:error_U_32}  present the error maps $Err(x,y)$ for all Rankine configurations with their best OFV settings. The influence of $D/R$ is clear: the error level decreases as the gradient diminishes. Errors appears concentrated around the vortex core, which is expected since this is the region with the highest displacement gradients, which are the most challenging to resolve. However, it is worth noting that when $D/R$ does not exceed $0.7$, absolute error levels remain sub--pixel. }

\textcolor{black}{
\begin{figure}[h!]
    \includegraphics[width=\linewidth]{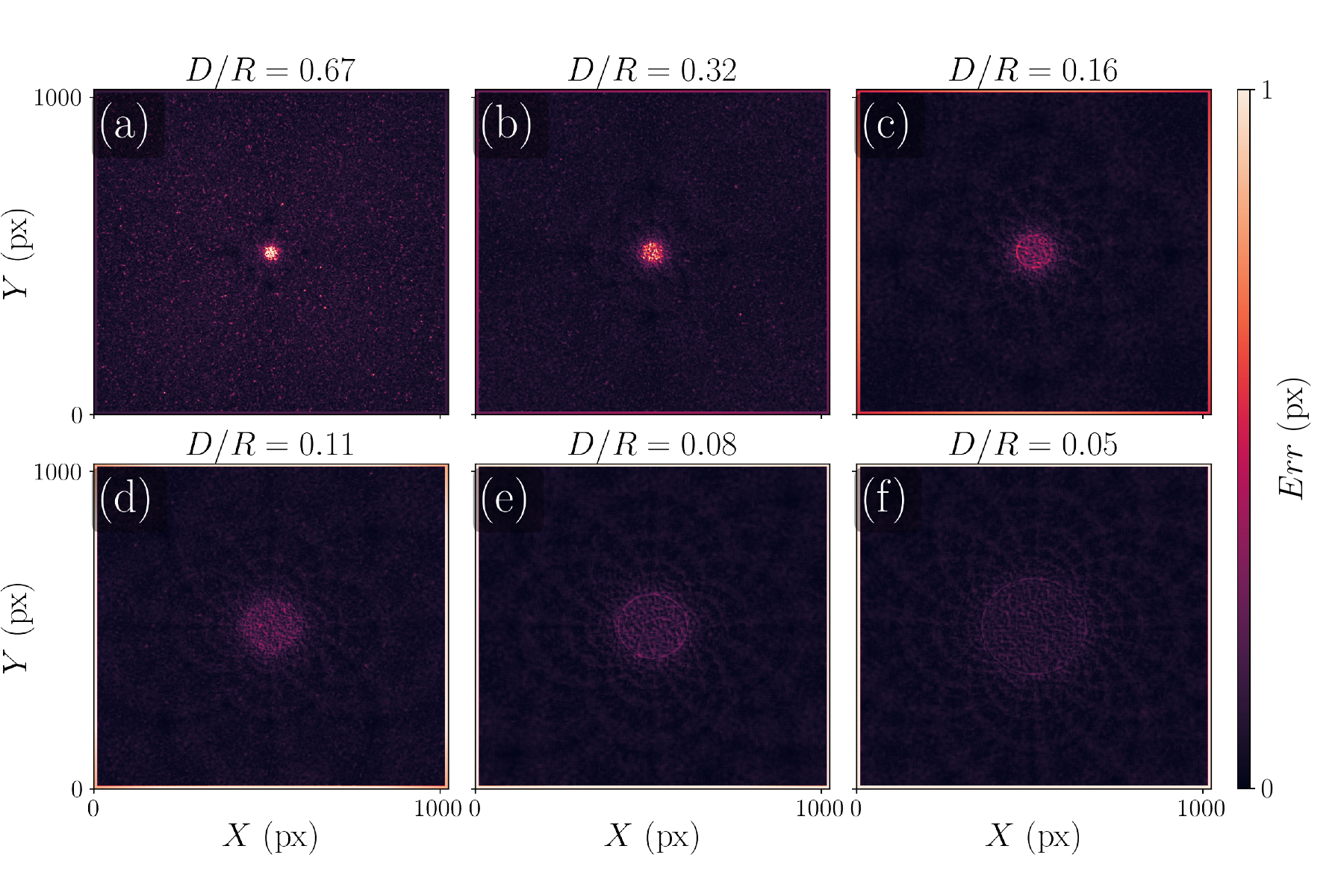}
    \caption{\textcolor{black}{Absolute displacement-error maps $Err(x,y)$ for the best OFV result for maximum displacement $|U|=8$ px, through the different vortex core radius sizes. Panels (a--f) correspond to $r=12,~25,~50,~75,~100,~150$ px (\textbf{D/R} indicated in each panel).}}
    \label{fig:error_U_8}
\end{figure}}

\textcolor{black}{
\begin{figure}[h!]
    \includegraphics[width=\linewidth]{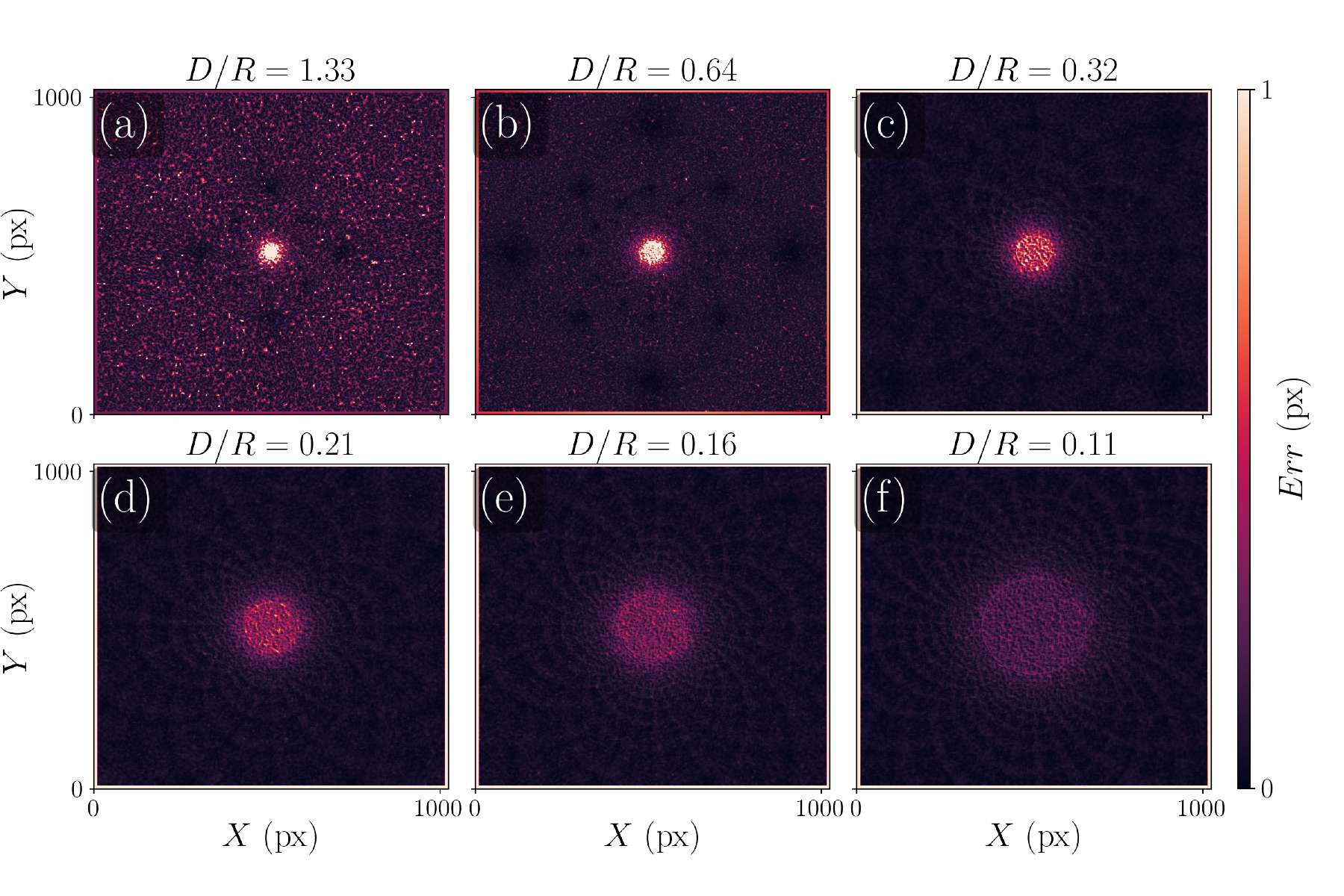}
    \caption{\textcolor{black}{Absolute displacement-error maps $Err(x,y)$ for the best OFV result for maximum displacement $|U|=16$ px, through the different vortex core radius sizes. Panels (a--f) correspond to $r=12,~25,~50,~75,~100,~150$ px (\textbf{D/R} indicated in each panel).}}
    \label{fig:error_U_16}
\end{figure}}


\subsection{Homogeneous Isotropic Turbulence (HIT)}

\begin{figure}[h!]
    \centering
    \includegraphics[width=0.8\linewidth]{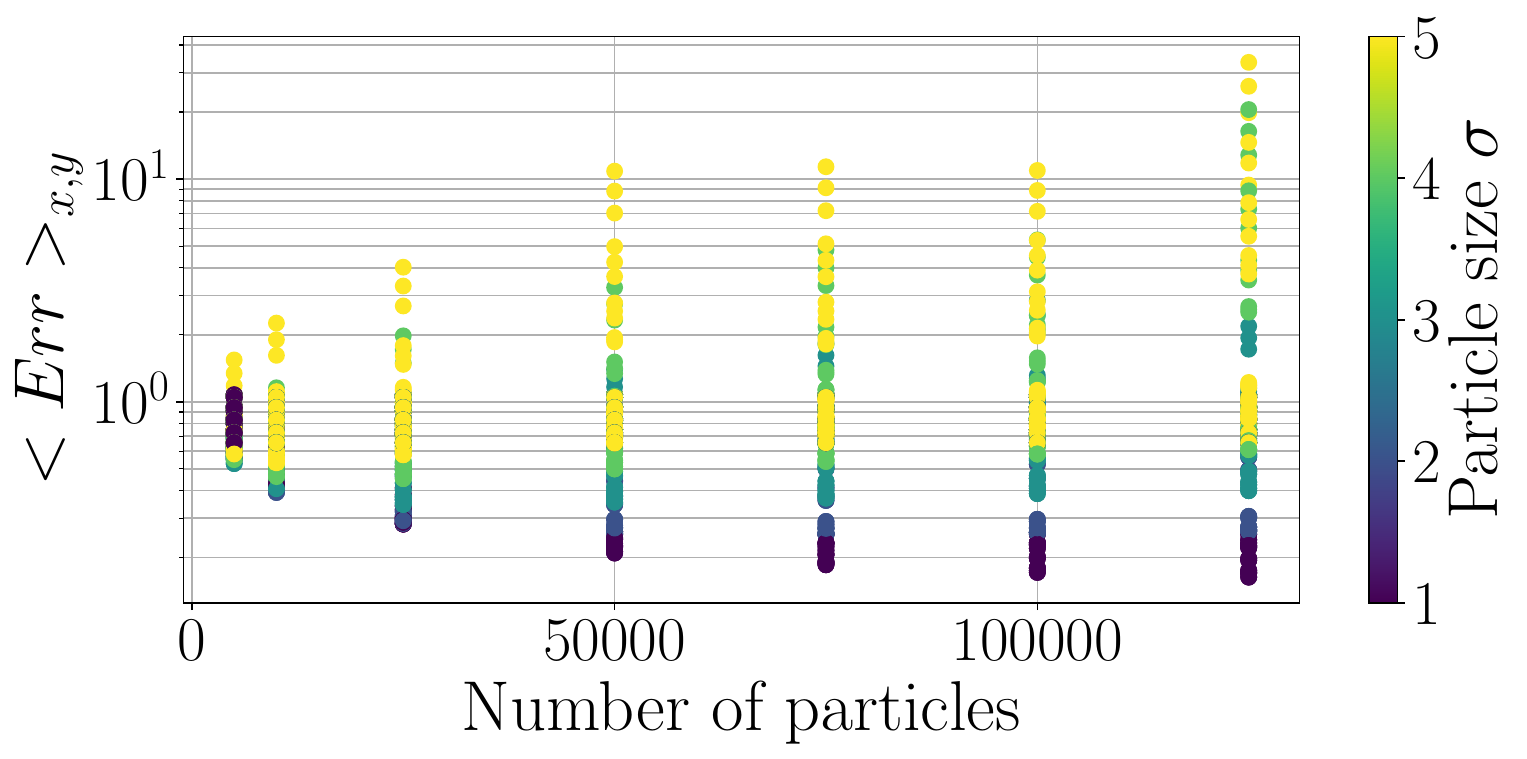}
    \caption{Spatially averaged absolute displacement error $\langle Err\rangle_{x,y}$ for HIT, discriminated by particle size and particle count.}
    \label{fig:error_dns}
\end{figure}

The HIT evaluation follows the Rankine analysis but extends the study of the influence of particle size and concentration on a more complex flow. Figure~\ref{fig:error_dns} shows $\langle Err\rangle_{x,y}$ as a function of the number of particles, for various particle sizes (color bar) and for all OFV parameter combinations used. One can see that OFV performs better with smaller particle size. Although many configurations yield good agreement ($\langle Err\rangle_{x,y}<1$), the best results are obtained for $\sigma=1$ with $125{,}000$ particles. In CC-PIV terms, this corresponds to a particle diameter $D_p\approx 2.35$~px and an average concentration (particles per image pixel) $C_p=N/(X\cdot Y)=0.119$~ppp, where $N$ is the number of particles, and $X,Y$ are the image dimensions. This combination produces rich image texture and  spatial intensity gradients, enhancing OFV’s ability to extract displacements from every pixel. The best OFV parameters follow the Rankine guidelines, with small kernels preferred: $KR=2$~px, $NR=1$~px, $PSL=2$, $IT=5$.

\textcolor{black}{
\begin{figure}[h!]
    \includegraphics[width=\linewidth]{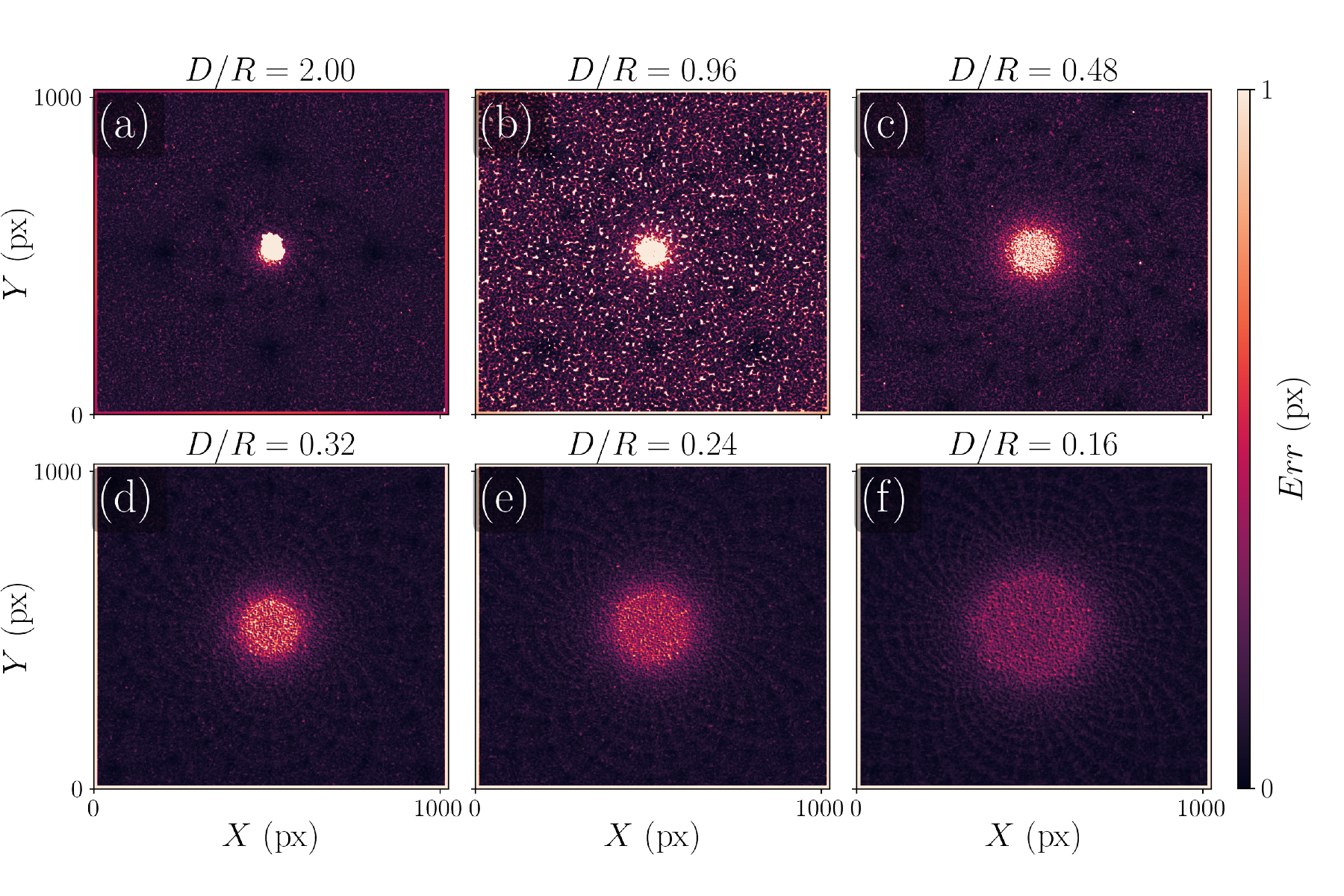}
    \caption{\textcolor{black}{Absolute displacement-error maps $Err(x,y)$ for the best OFV result for maximum displacement $|U|=24$ px and different vortex core radius. Panels (a--f) correspond to $r=12,~25,~50,~75,~100,~150$ px (\textbf{D/R} indicated in each panel).}}
    \label{fig:error_U_24}
\end{figure}}

\textcolor{black}{
\begin{figure}[h!]
    \includegraphics[width=\linewidth]{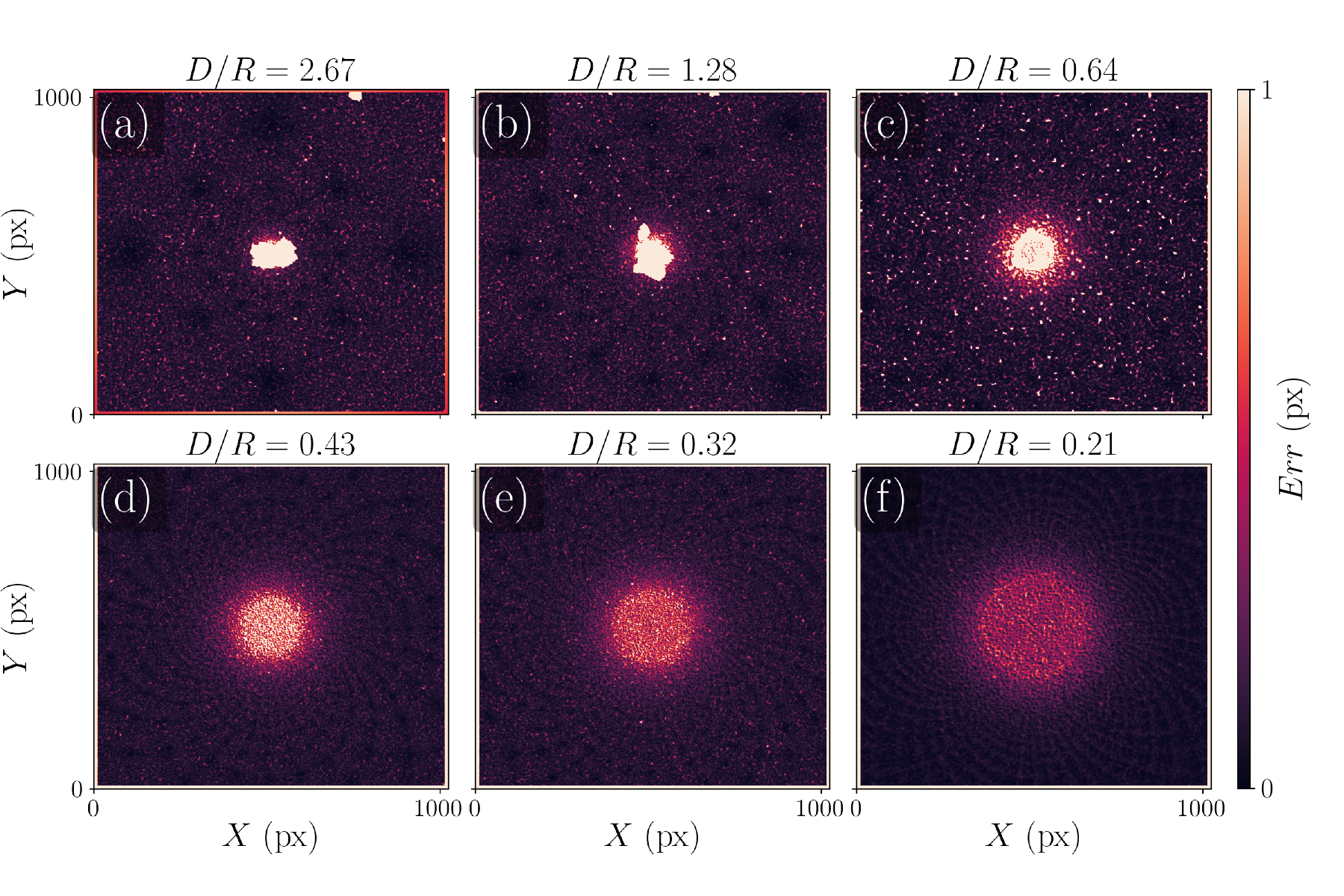}
    \caption{\textcolor{black}{Absolute displacement-error maps $Err(x,y)$ for the best OFV result for maximum displacement $|U|=32$ px and different vortex core radius. Panels (a--f) correspond to $r=12,~25,~50,~75,~100,~150$ px (\textbf{D/R} indicated in each panel).}}
    \label{fig:error_U_32}
\end{figure}}

\begin{figure}[h!]
    \begin{subfigure}{0.48\linewidth}
        \centering
        \caption*{\textbf{DNS}}
        \includegraphics[width=\linewidth]{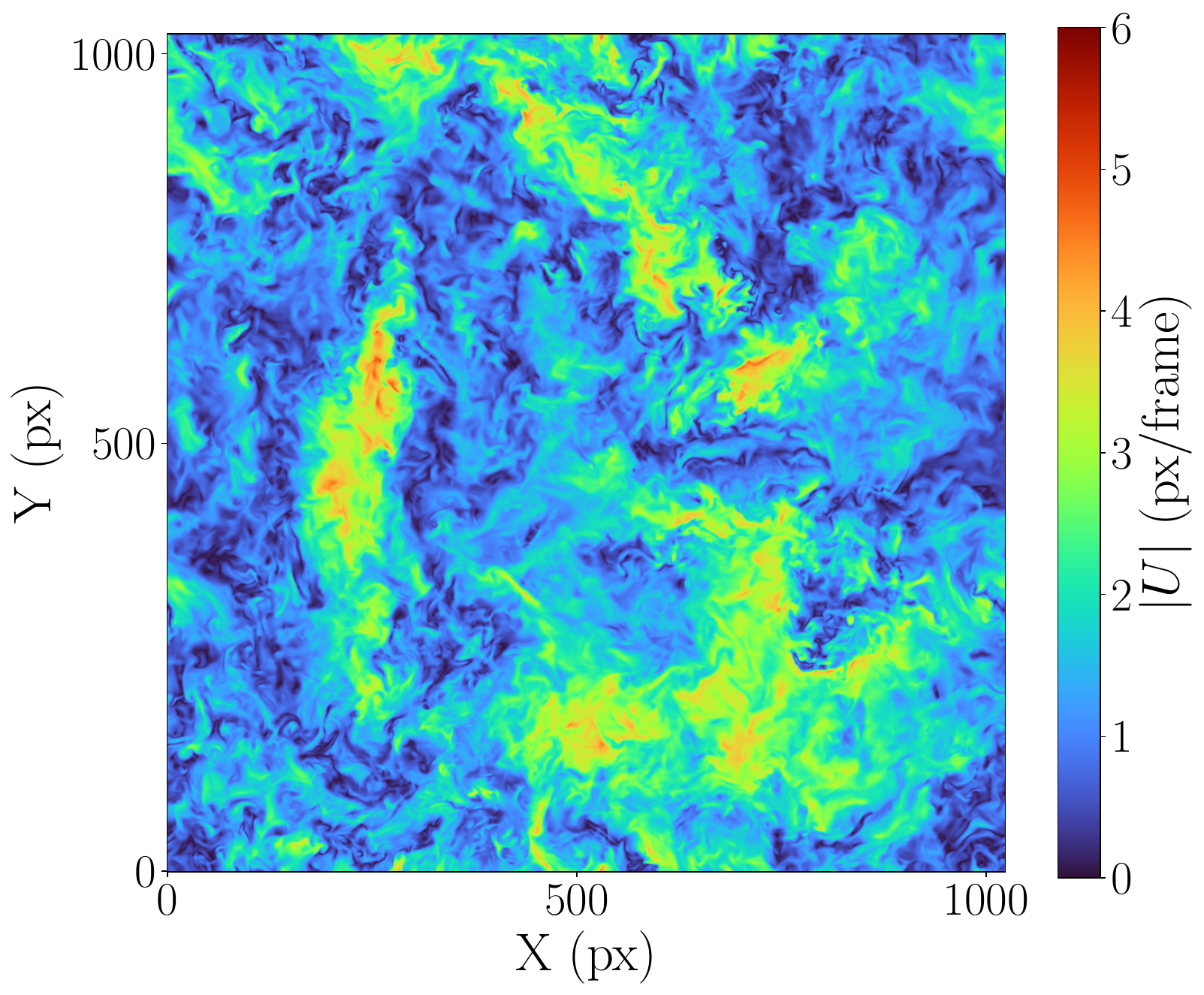}
        \caption{}
        \label{fig:mag_dns}
    \end{subfigure}
    \begin{subfigure}{0.48\linewidth}
        \centering
        \caption*{\textbf{OFV}}
        \includegraphics[width=\linewidth]{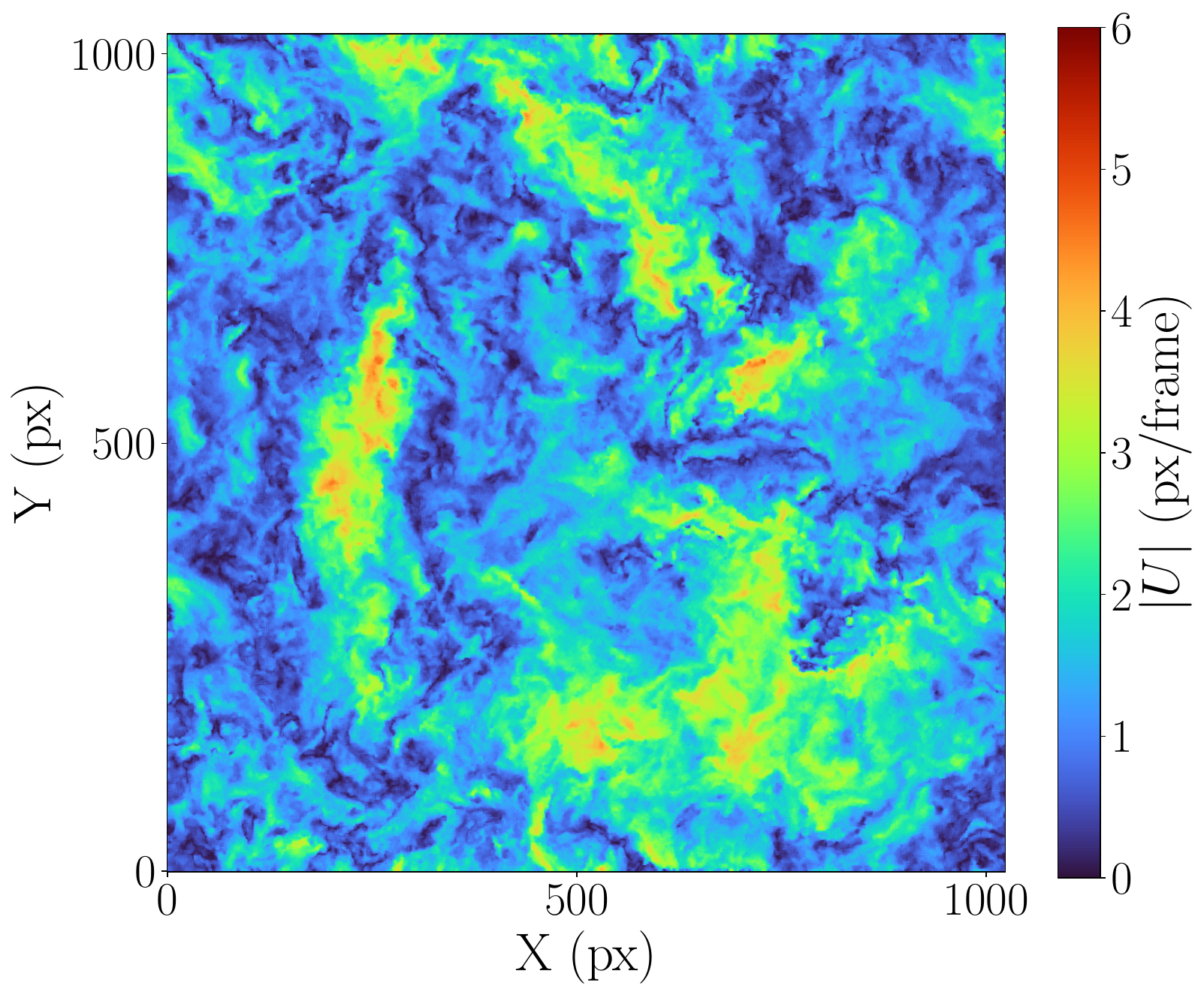}
        \caption{}
        \label{fig:mag_of}
    \end{subfigure}
    \centering
    \begin{subfigure}{0.5\linewidth}
        \centering
        \caption*{\textbf{CC-PIV}}
        \includegraphics[width=\linewidth]{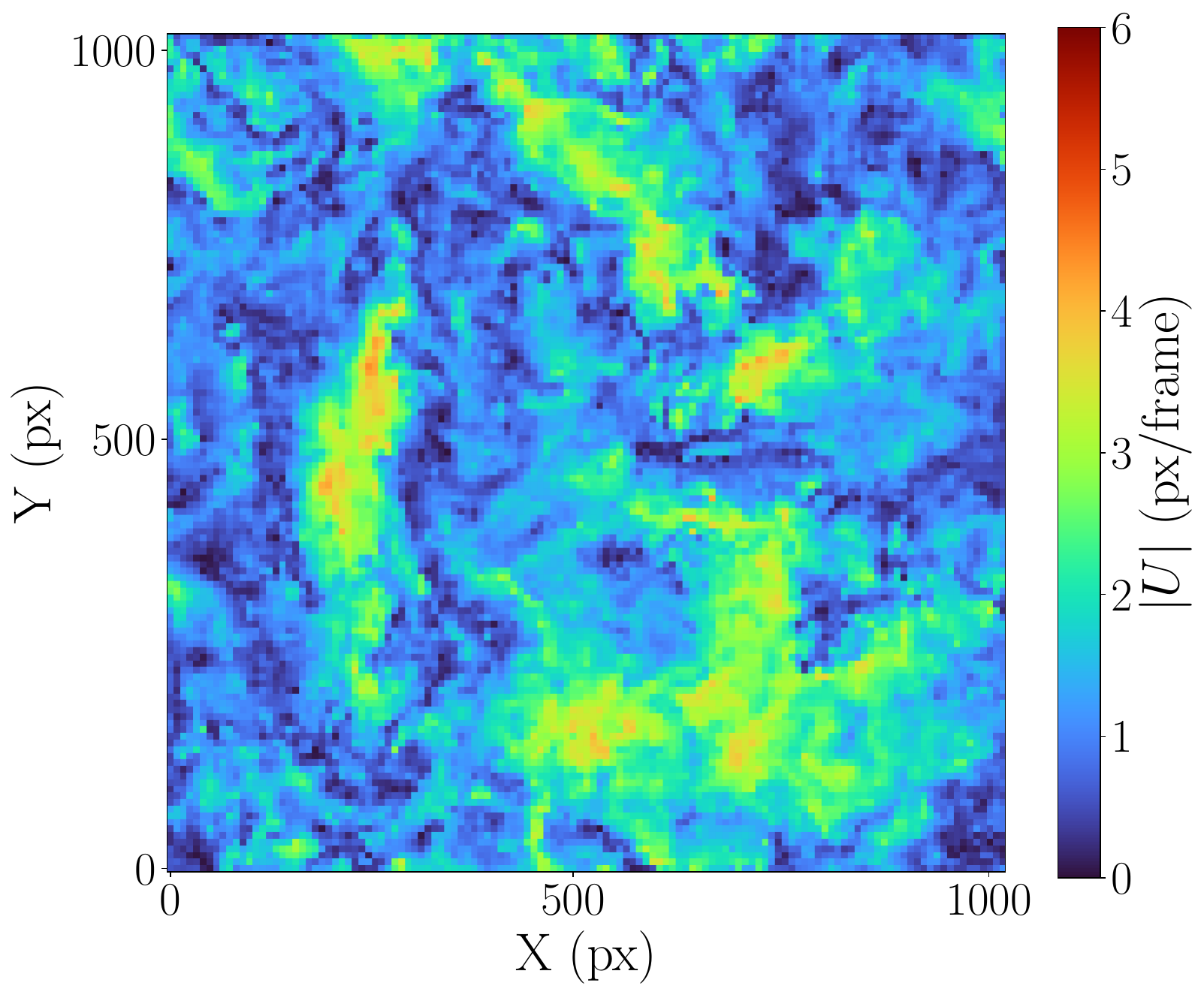}
        \caption{}
        \label{fig:mag_cc}
    \end{subfigure}
    
    \caption{Snapshots of instantaneous displacement magnitude in the mid-section of the HIT dataset: \textbf{(a)} DNS, \textbf{(b)} OFV, \textbf{(c)} CC-PIV.}
    \label{fig:mag_comparison}
\end{figure}

As in the Rankine section, the results obtained with OFV are compared to the ones obtained with a CC-PIV algorithm \cite{Thielicke_2021}, using three IW passes  ($64\times64$, $32\times32$, $16\times16$~px) with 50\% overlap. Figure~\ref{fig:mag_comparison} shows that OFV produces a dense  displacement field, on par with the original DNS displacement field, whereas CC-PIV yields a blurred displacement field (one vector per IW).

\begin{figure}[h!]
    \begin{subfigure}{\linewidth}
        \centering
        \includegraphics[width=\linewidth]{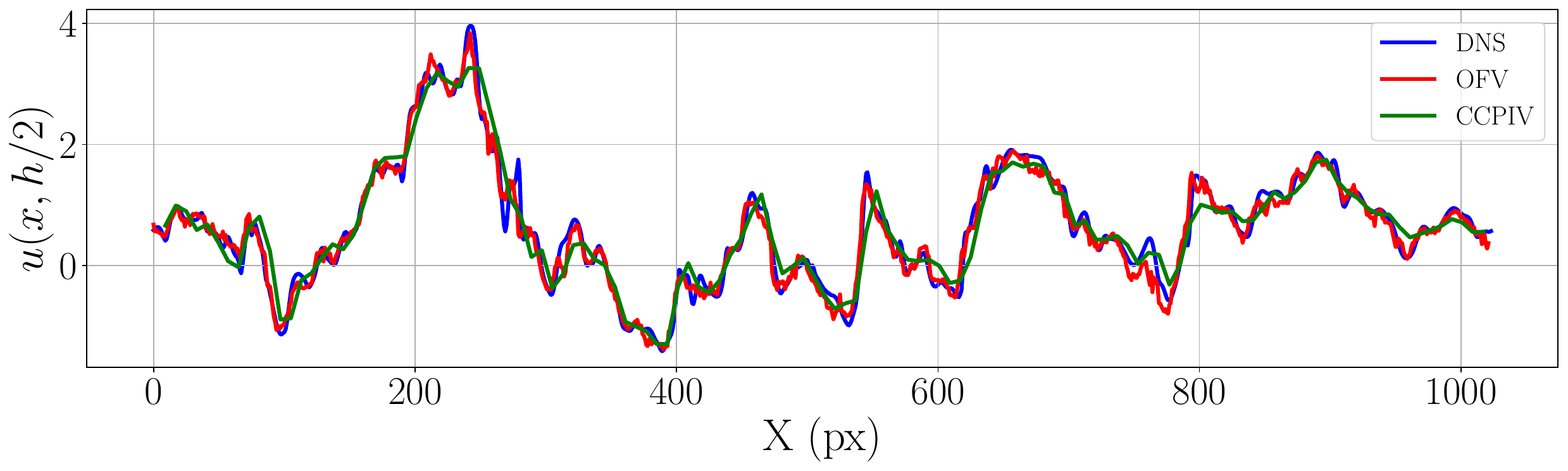}
        \caption{}
        \label{fig:profiles_stream}
    \end{subfigure}
    \begin{subfigure}{\linewidth}
        \centering
        \includegraphics[width=\linewidth]{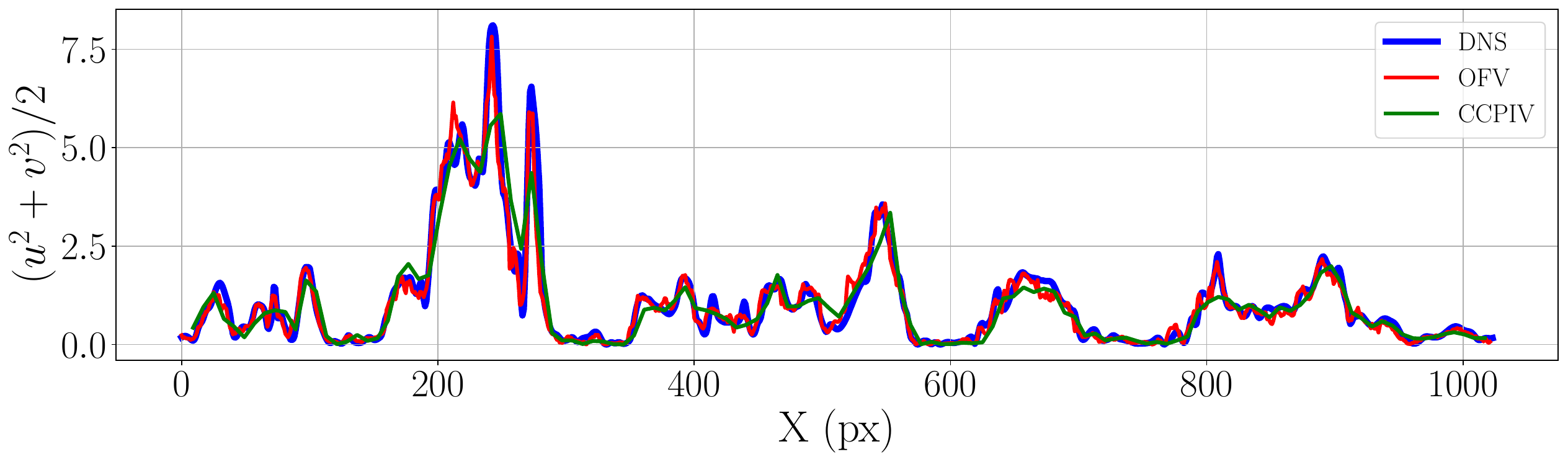}
        \caption{}
        \label{fig:profiles_energy}
    \end{subfigure}
    \caption{Comparison between DNS, OFV, and CC-PIV along a mid-height line along the $X$ direction. \textbf{(a)} Profile of instantaneous $u$-displacement. \textbf{(b)} Profile of instantaneous Kinetic energy profile.}
    \label{fig:profile_comparison}
\end{figure}

Fig.~\ref{fig:profile_comparison} compares the $u$ component and kinetic energy profiles along the mid-height line ($y=Y/2$) in the HIT dataset obtained with OFV, CC-PIV and the DNS.  It shows that OFV better captures the sharp gradients present in the DNS velocity fields. This supports the claim that OFV recovers more accurately the smaller structures and higher gradients.

\begin{figure}
    \centering
    \includegraphics[width=\linewidth]{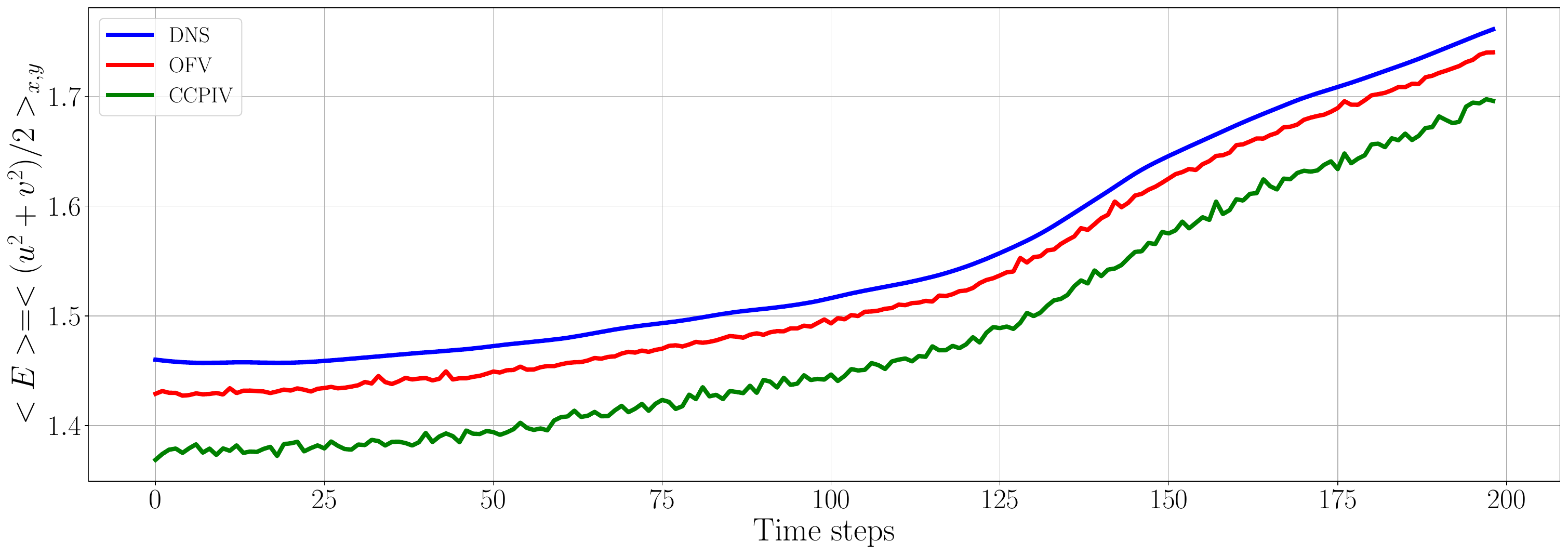}
    \caption{Spatially averaged kinetic energy over the 199 time steps. Comparison between the DNS, OFV and CCPIV}
    \label{fig:kinetic_energy}
\end{figure}

While Fig.~\ref{fig:profile_comparison} shows instantaneous velocity profiles taken along a single line in a single snapshot of the HIT dataset, Fig.~\ref{fig:kinetic_energy} aggregates the time evolution over 199 time steps of the spatially averaged kinetic energy $\langle E\rangle_{x,y}(t)$ . OFV follows closely the evolution of the DNS data (underestimation of the mean kinetic energy lower than $<2\%$), while CC-PIV data is more noisy with a constant under--estimation of $\sim 5\%$.

\begin{figure}[!h]
    \centering
    \includegraphics[width=\linewidth]{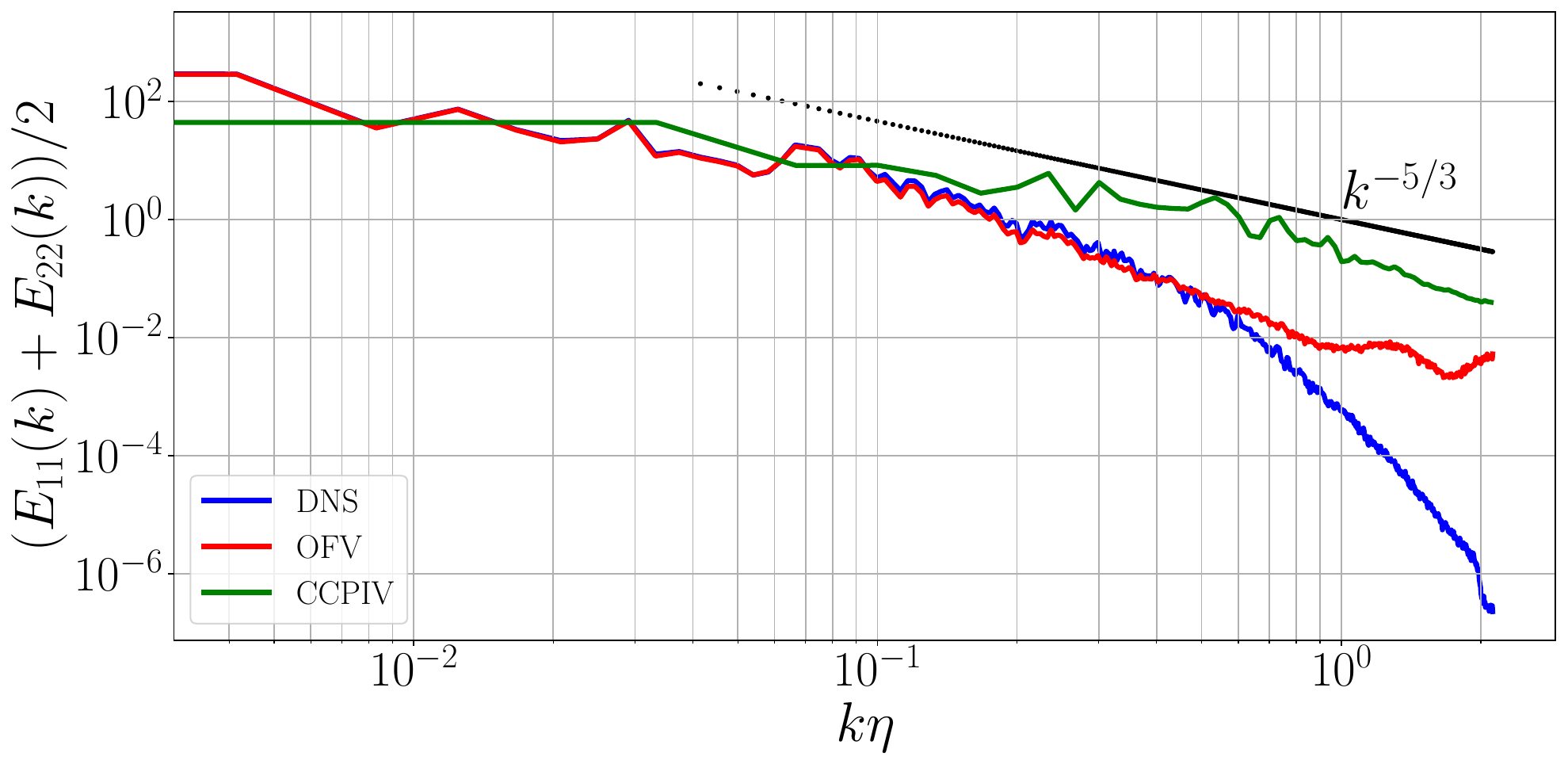}
    \caption{Spatial spectra of kinetic energy averaged over 199 time steps, computed along the mid-height  profiles.}
    \label{fig:spectra}
\end{figure}

Guided by these observations and to go further into the analysis of the data, the spatial spectra of the instantaneous kinetic energy has been computed along the mid--height profile shown in Fig.~\ref{fig:profiles_energy}. It has then been time--averaged over the 199 time steps. For display, we account for the DNS mapping $1516\,\eta=1024$~px, where $\eta$ is the Kolmogorov microscale, to present spectra across scales. Fig.~\ref{fig:spectra} shows that OFV follows the DNS closely at large scales and through the mid-range; divergence begins near $k\eta\approx 0.5$, with a cutoff near $k\eta\approx 1$. If OFV does not reach the DNS resolution at the smallest scales, it clearly outperforms CC-PIV, which fails to track the DNS spectrum.

\begin{figure}
\begin{subfigure}{0.48\linewidth}
    \centering
    \includegraphics[width=\linewidth]{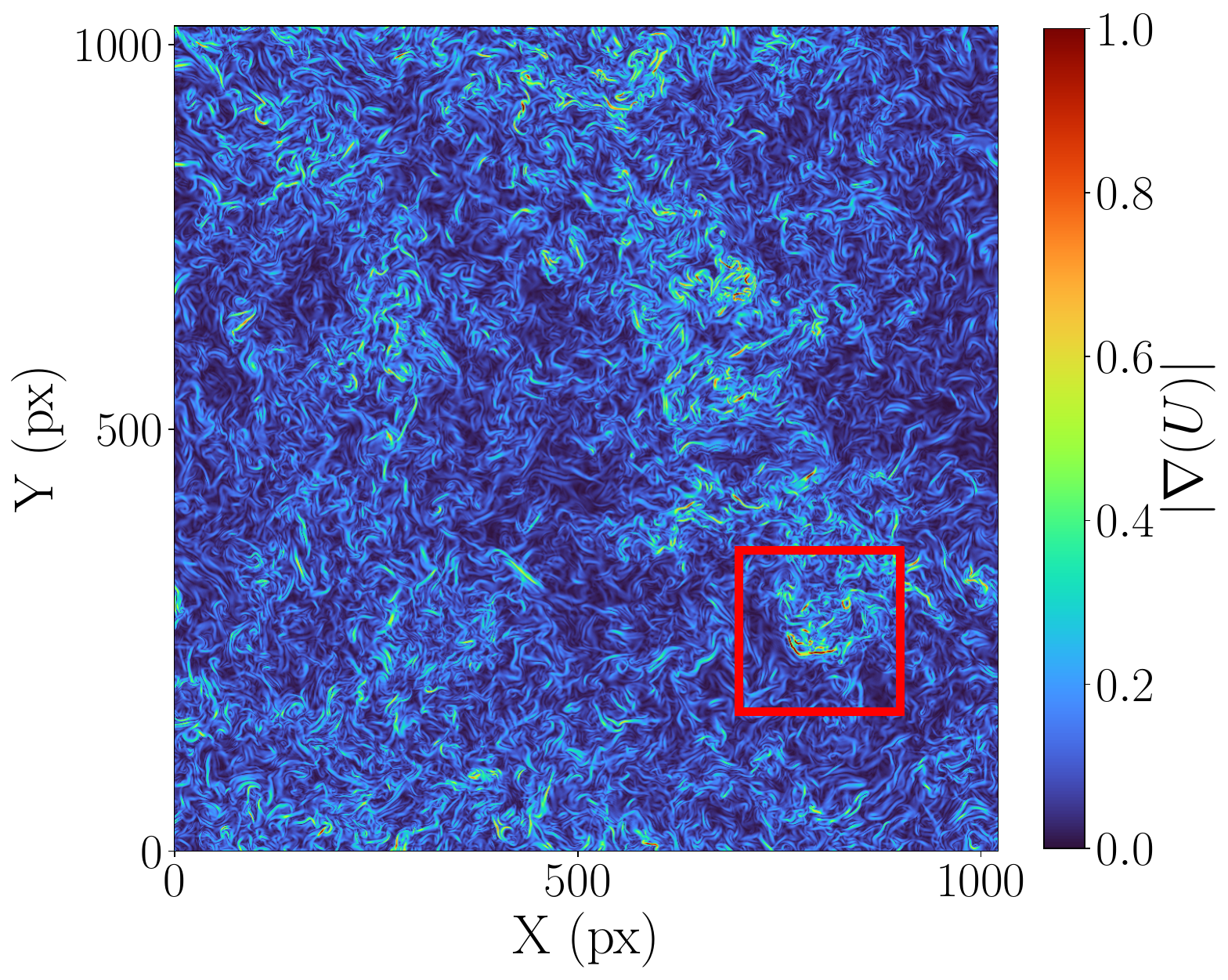}
    \caption{}
    \label{fig:grad_dns}
\end{subfigure}
\begin{subfigure}{0.48\linewidth}
    \centering
    \includegraphics[width=\linewidth]{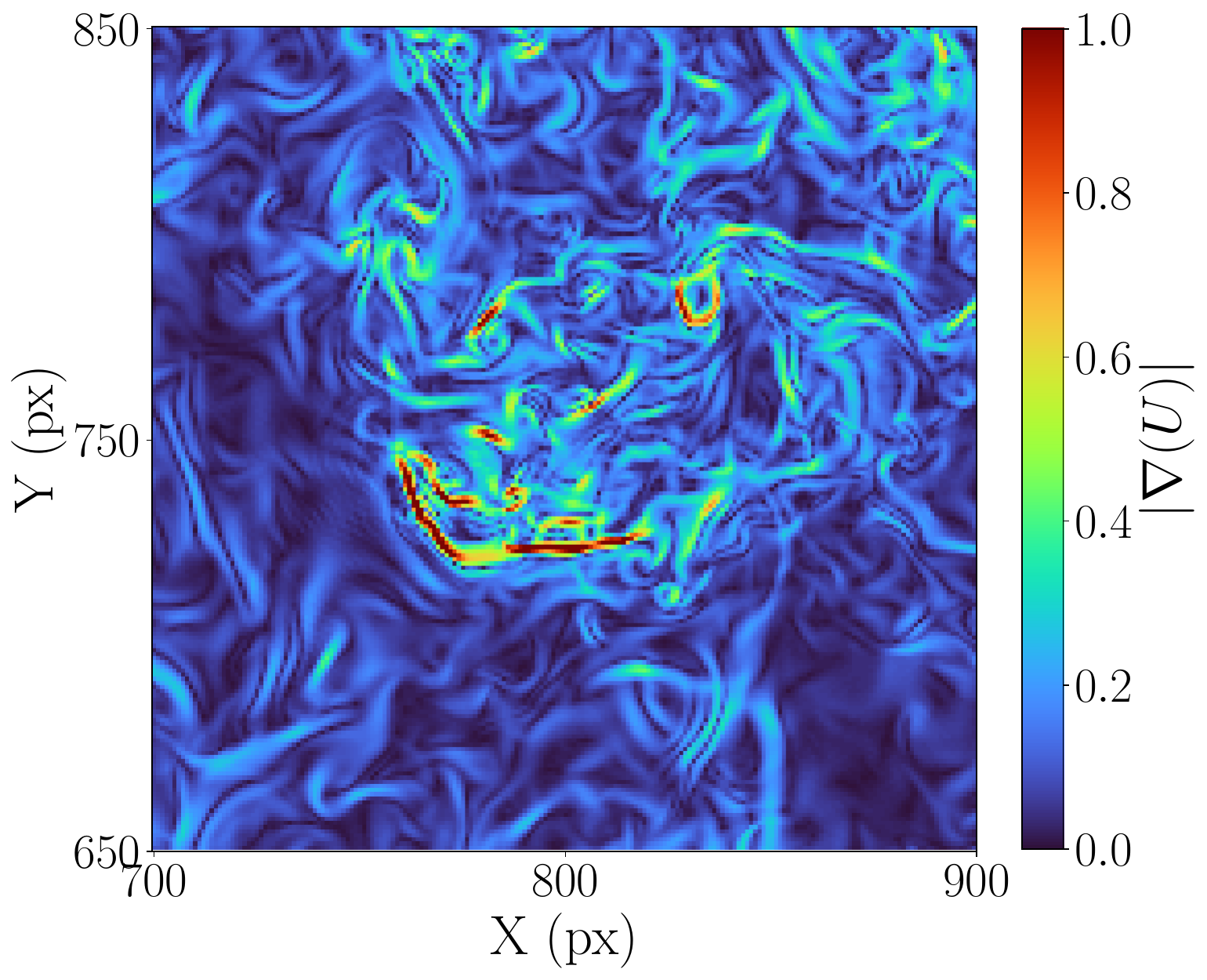}
    \caption{}
    \label{fig:grad_dns_zomm}
\end{subfigure}
    \caption{Structure detection — DNS. \textbf{(a)} Full field; red square marks the ROI. \textbf{(b)} Zoom into the ROI.}
    \label{fig:Gradient_dns}
\end{figure}

Finally, to \emph{visually} assess the ability to recover intense local gradients and small-–scale structures (as suggested by the spectra), we compute a gradient–-based structure map of the displacement magnitude $U(x,y)=\sqrt{u^2+v^2}$ following a standard edge-detection computer–vision approach \cite{Saif2016GradientBI}. The idea is simple: regions where $|\nabla U|$ is large mark sharp spatial variations (filaments, shear layers, vortex edges), i.e. precisely the features that challenge dense velocimetry methods. Using the discrete counterpart of Eq.~\ref{eq:gradient}:

\begin{equation}\label{eq:gradient}
    gU_x,\,gU_y = \nabla U(x,y), \qquad
    \left\lvert \nabla U(x,y)\right\rvert = \sqrt{gU_x^2 + gU_y^2},
\end{equation}

we form the gradient–-magnitude map $|\nabla U|$ for a given snapshot and compare the results obtained with DNS, OFV and CC–PIV.

\begin{figure}
\begin{subfigure}{0.48\linewidth}
    \centering
    \includegraphics[width=\linewidth]{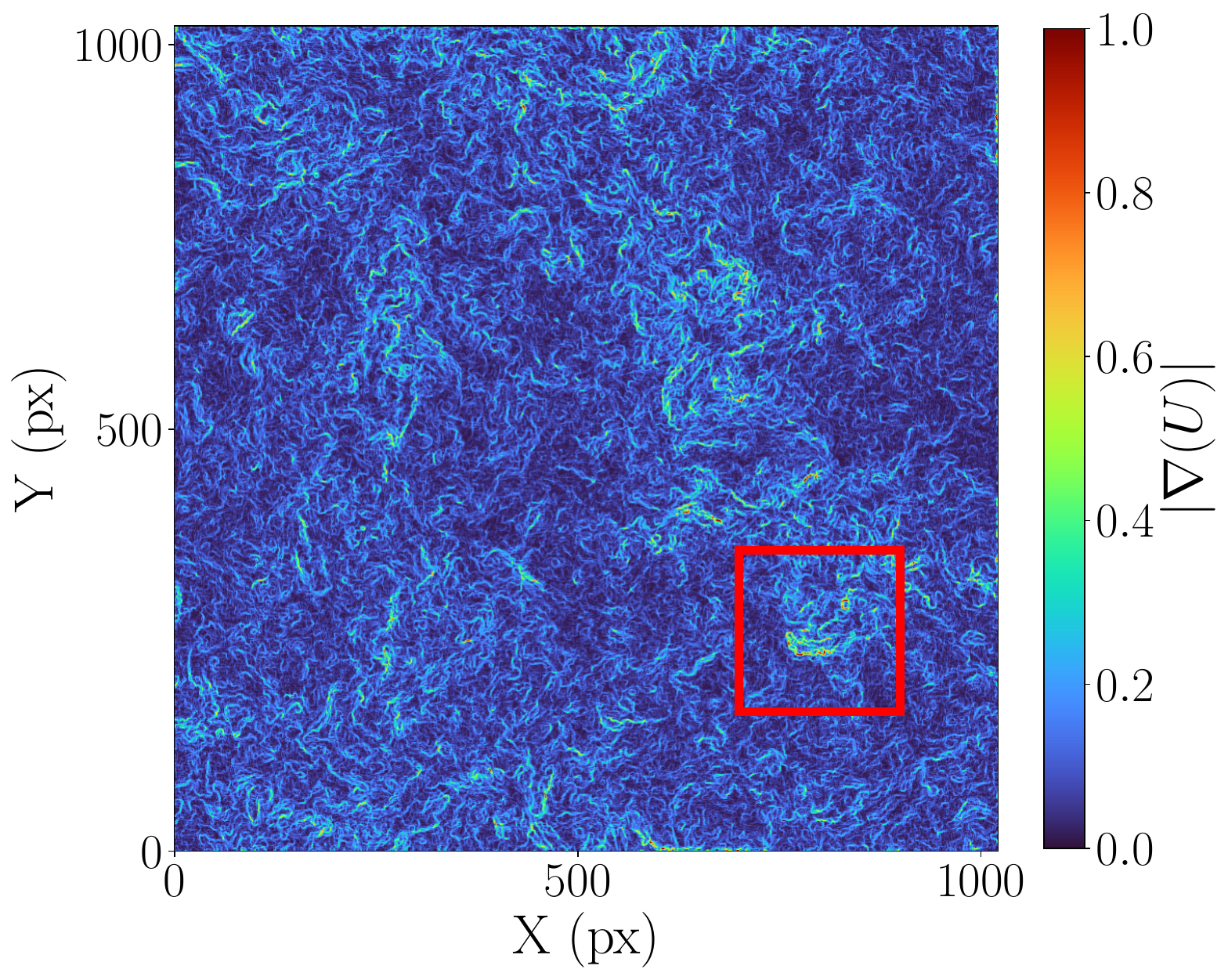}
    \caption{}
    \label{fig:grad_of}
\end{subfigure}
\begin{subfigure}{0.48\linewidth}
    \centering
    \includegraphics[width=\linewidth]{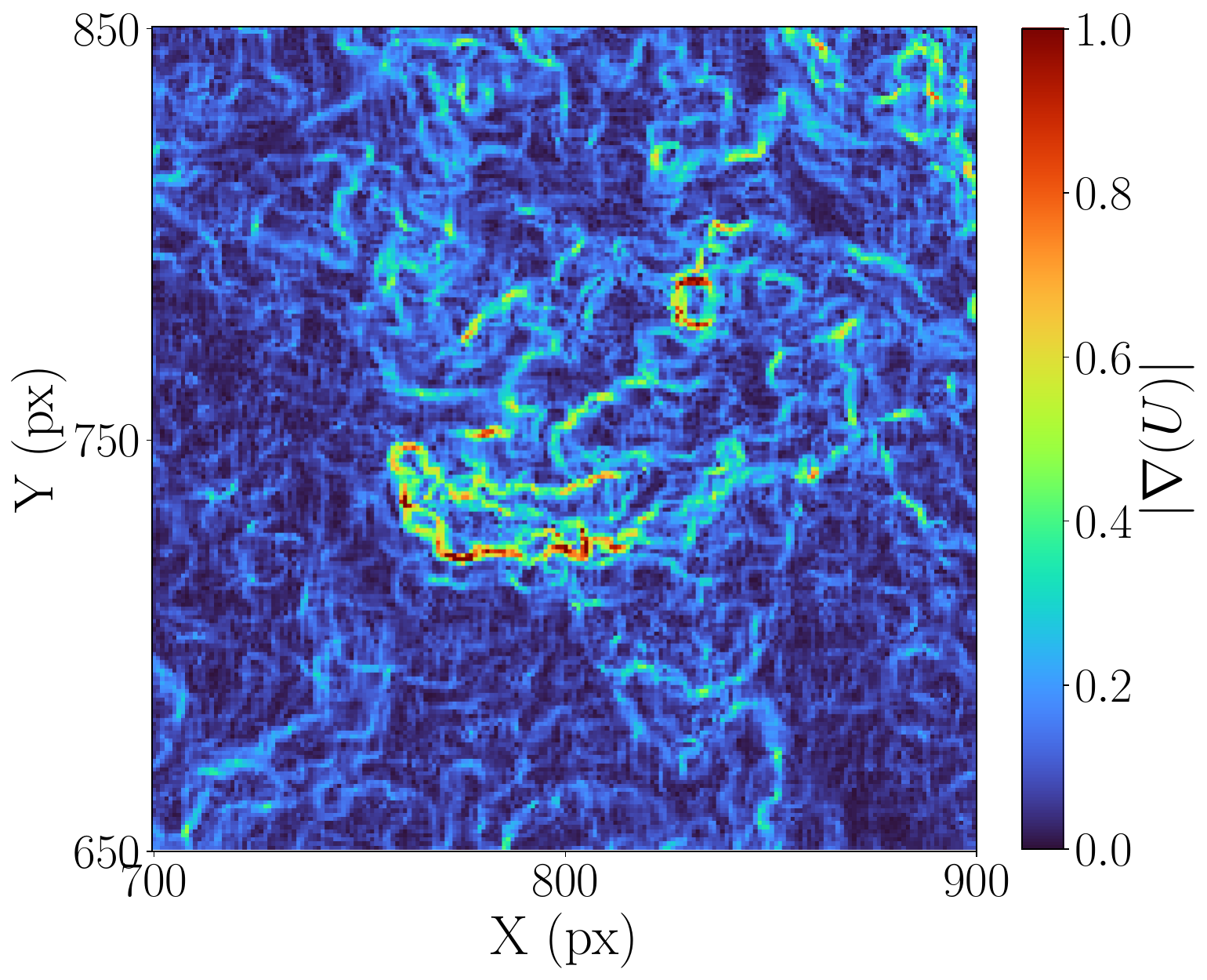}
    \caption{}
    \label{fig:grad_of_zomm}
\end{subfigure}
    \caption{Structure detection - OFV. a) Full field, red square marks a region of interest. b) Zoom-in over the region of interest.}
    \label{fig:Gradient_of}
\end{figure}

\begin{figure}[h!]
\begin{subfigure}{0.48\linewidth}
    \centering
    \includegraphics[width=\linewidth]{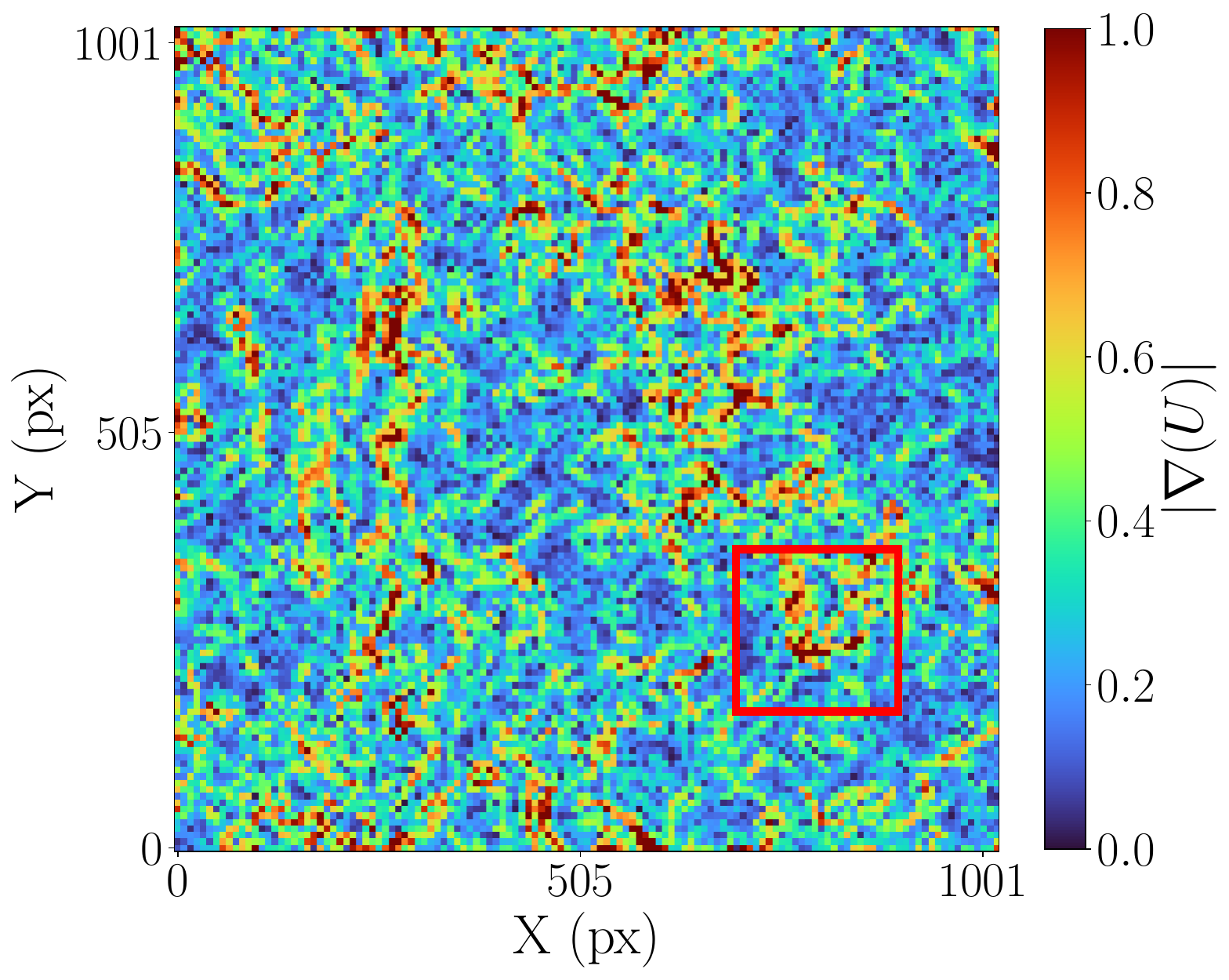}
    \caption{}
    \label{fig:grad_cc}
\end{subfigure}
\begin{subfigure}{0.48\linewidth}
    \centering
    \includegraphics[width=\linewidth]{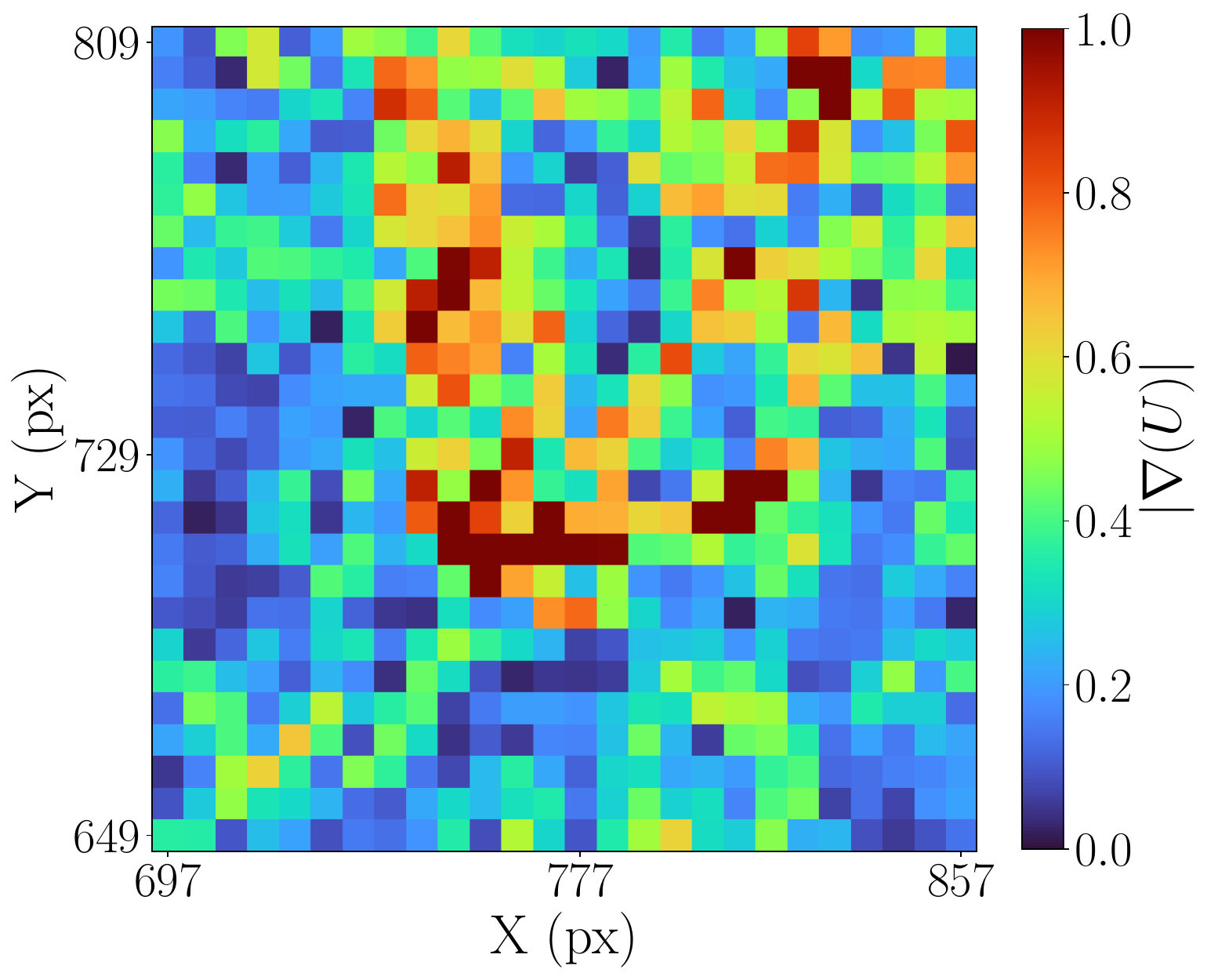}
    \caption{}
    \label{fig:grad_cc_zomm}
\end{subfigure}
    \caption{Structure detection - CCPIV. a) Full field, red square marks a region of interest. b) Zoom-in over the region of interest.}
    \label{fig:Gradient_cc}
\end{figure}

The maps \textcolor{black}{shown in Figs.~\ref{fig:Gradient_dns} to \ref{fig:Gradient_cc}} corroborate the spectral analysis: OFV reproduces the DNS organization at large and intermediate scales properly and retains much of the fine–-scale network of gradients, while the very thinnest features (those contributing near $k\eta\!\approx\!1$) are naturally most affected. In contrast, CC–PIV’s gradient maps reveal clear spatial under-–resolution as result of the IW. For these figures, we also show a $200\times200$~px region of interest (ROI) to better visualize the fine structures and strong gradients, \textcolor{black}{as it can be observed for the DNS in Fig.~\ref{fig:Gradient_dns}}. It reveals clearly the ability of OFV to capture relatively small structures, while, one the contrary, CC-PIV misses every small structures.

\subsection{\textcolor{black}{Uncertainty and sensitivity}\label{subsec:error_analysis}}

\textcolor{black}{In particle-image OFV, the dominant sources of uncertainty stem from (i) strong displacement gradients, which
challenge local estimation, and (ii) deviations from brightness constancy due to acquisition effects such as non-uniform or
time-varying illumination and reflections. In addition, the \emph{particle-image realization} itself (seeding density,
particle-image size/blur, and overlap) controls the richness and uniqueness of local intensity patterns. In our approach,
robustness is sought by emphasizing \emph{the texture of the images} (informative local intensity gradients) rather than tracking isolated particles:
dense particle fields of sufficiently small particle images provide richer intensity gradients within each kernel. To mitigate low-frequency
illumination variations, we apply local intensity normalization prior to OF estimation (Sec.~\ref{sec:LOFV}). Consistent
with this interpretation, the HIT/DNS benchmark shows that increasing particle density reduces the average error until saturation at high densities, suggesting a practical trade-off between texture richness and overlap/pattern non-uniqueness  Fig.~\ref{fig:error_dns}). Similarly, overly large or blurred particle images reduce intensity gradient sharpness and then increase the error by weakening the local texture.}

\begin{figure}[ht!]
    \begin{subfigure}{0.47\linewidth}
        \centering
        \includegraphics[width=\linewidth]{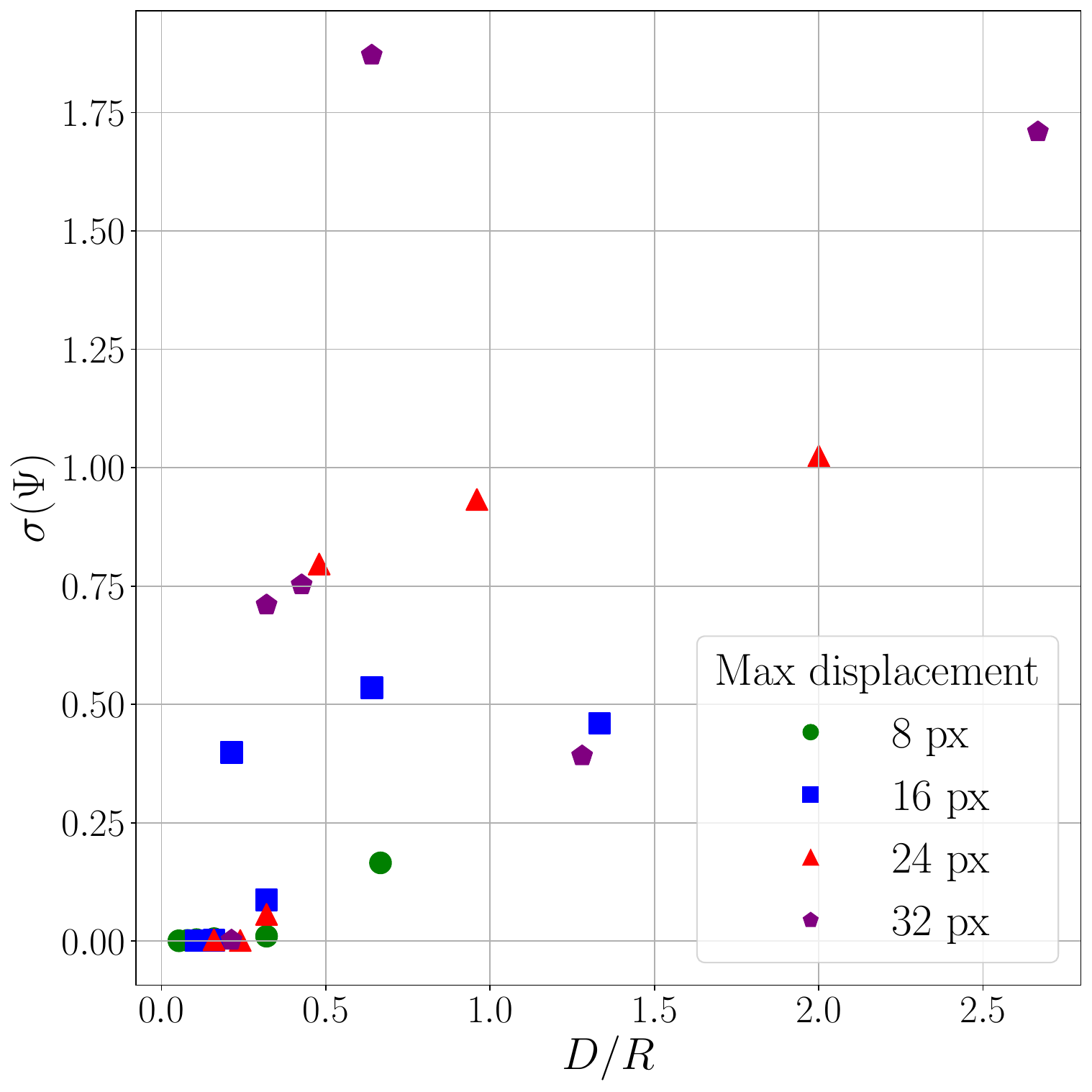}
        \caption{}
        \label{fig:psi_std}
    \end{subfigure}
    \begin{subfigure}{0.47\linewidth}
        \centering
        \includegraphics[width=\linewidth]{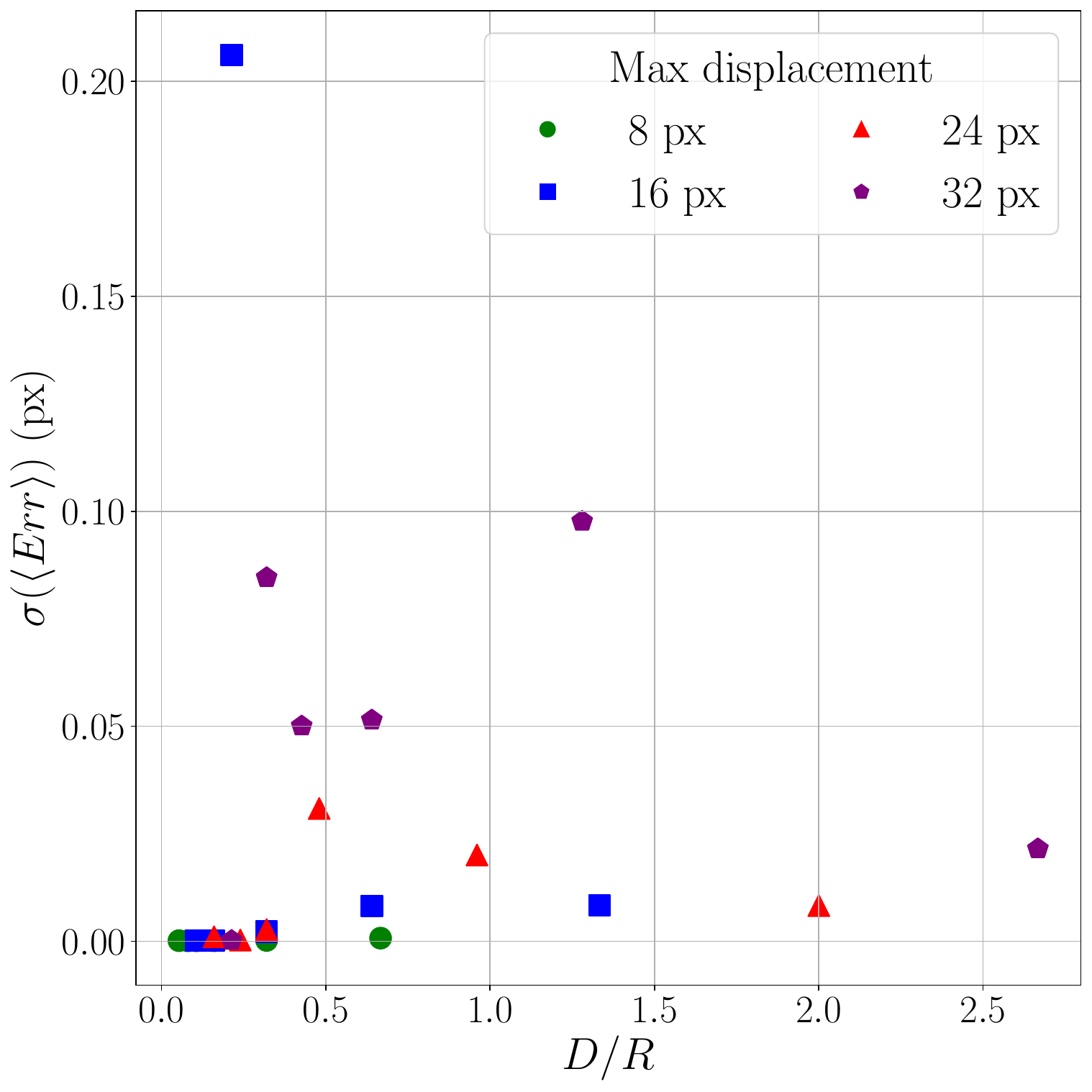}
        \caption{}
        \label{fig:mean_E_std}
    \end{subfigure}
    \caption{Standard deviation computed over 10 independent synthetic realizations for each $(D,R)$ case, using the OFV parameter set that yields the smallest error. (a) $\Psi$ metric. (b) Spatially averaged absolute displacement error $\langle Err \rangle$.}
    \label{fig:STD}
\end{figure}

\textcolor{black}{To quantify repeatability under identical nominal conditions, we report the standard deviation ($\sigma$) of both error metrics over 10 independent synthetic image realizations for each Rankine vortex core size and displacement combination.
Across these realizations, only the particle--image parameters vary (random initial particle positions and random
out-of-plane positions with standard deviation 0.025~mm), while the flow case and OFV settings are held fixed; the OFV
parameters correspond to the configuration that yields the smallest error (Sec.~\ref{subsec:RK_results}). Fig.~\ref{fig:STD}
shows that dispersion increases with the displacement gradient $D/R$. For mild gradients ($D/R \le 0.16$), the dispersion is small: $\sigma(\Psi)\sim O(10^{-3})$ and $\sigma(\langle Err\rangle)\sim O(10^{-4})$~px. For the steepest gradients
($D/R \ge 1.28$), the variability increases to $\sigma(\Psi)\sim O(1)$ (typically $0.6$--$1.6$) and
$\sigma(\langle Err\rangle)\sim O(10^{-2})$~px (typically $8\times10^{-3}$ to $2\times10^{-2}$~px), consistent with the
increased difficulty of estimating motion in the vortex core.}

\section{High frequency and high resolution Live computing and post-processing performance\label{sec:Comp}}

It has been shown in the previous sections that, with appropriate seeding, it is possible to obtain a dense (one vector per pixel) displacement field with a high effective spatial resolution. The following question is whether these high-quality measurements can be obtained in real-time and at high frequencies. For this purpose, the computational performances of the OFV algorithm were evaluated for both post-processing and real-time (live) measurements, using the dedicated workstation described in Sec.~\ref{sec:setup}.

For these benchmarks, a fixed set of OFV parameters was used:
\[
NR = 2~\text{px},\qquad KR = 10~\text{px},\qquad PSL \in \{0,1,2\},\qquad IT = 3.
\]
The performances were evaluated for image sizes ranging from 1 to 21~Mp, since the performance depends on the number of pixels in the images. The images do not depict a physical flow; they are used only to benchmark runtime under these conditions. A single NVIDIA RTX~5090 GPU was used.

\subsection{Post-processing (offline) performances\label{subsec:post}}

Post-processing speed was measured using a dedicated in--house benchmark tool that records the time taken to compute a prescribed set of image pairs, and thus the number of displacement fields per second. To determine the computation time, all image pairs were pre--loaded into memory before timing. For each image size, ten 8-bit image pairs were used and the computation was repeated 100 times, resulting in an average of over 1,000 image pairs.

For comparison, a CC-PIV implementation (DPIV\cite{AguilarCabello2022DPIVSoftOpenCLAM})  was also benchmarked, using an iterative multi-pass CC-FFT with three IW sizes
\(
32\times32,\;16\times16,\;8\times8~\text{px}
\)
at 50\% overlap and three iterations, chosen to mirror the iterative behavior of the OFV settings. For CC-PIV, one image pair per image size was timed to keep the total runtime reasonable at the largest resolutions. Images were also pre--loaded in the memory, so that only computing time was accounted for.

\begin{figure}[h]
    \begin{subfigure}{\linewidth}
        \centering
        \includegraphics[width=\linewidth]{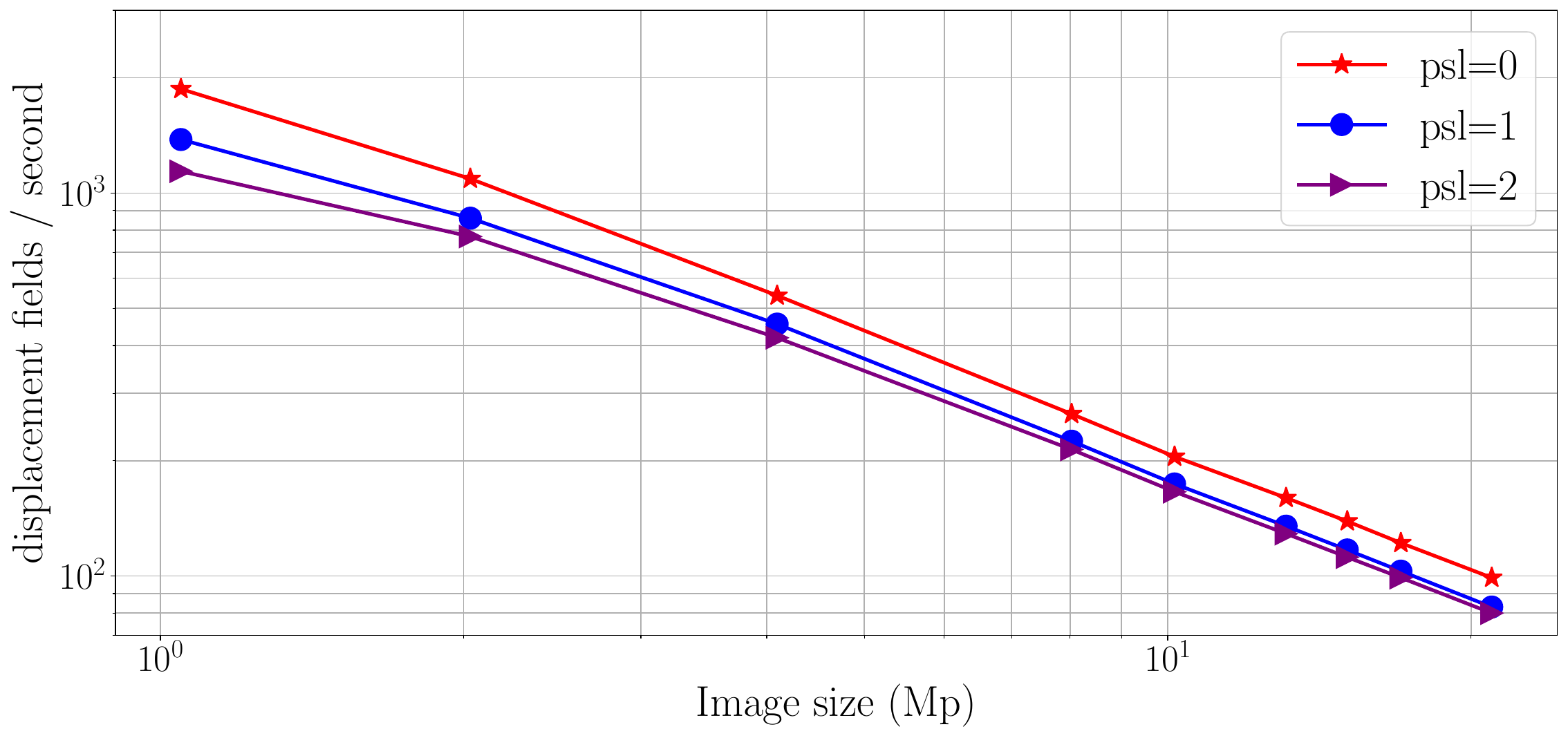}
        \caption{}
        \label{fig:perf_of_post}
    \end{subfigure}
    \begin{subfigure}{\linewidth}
        \centering
        \includegraphics[width=\linewidth]{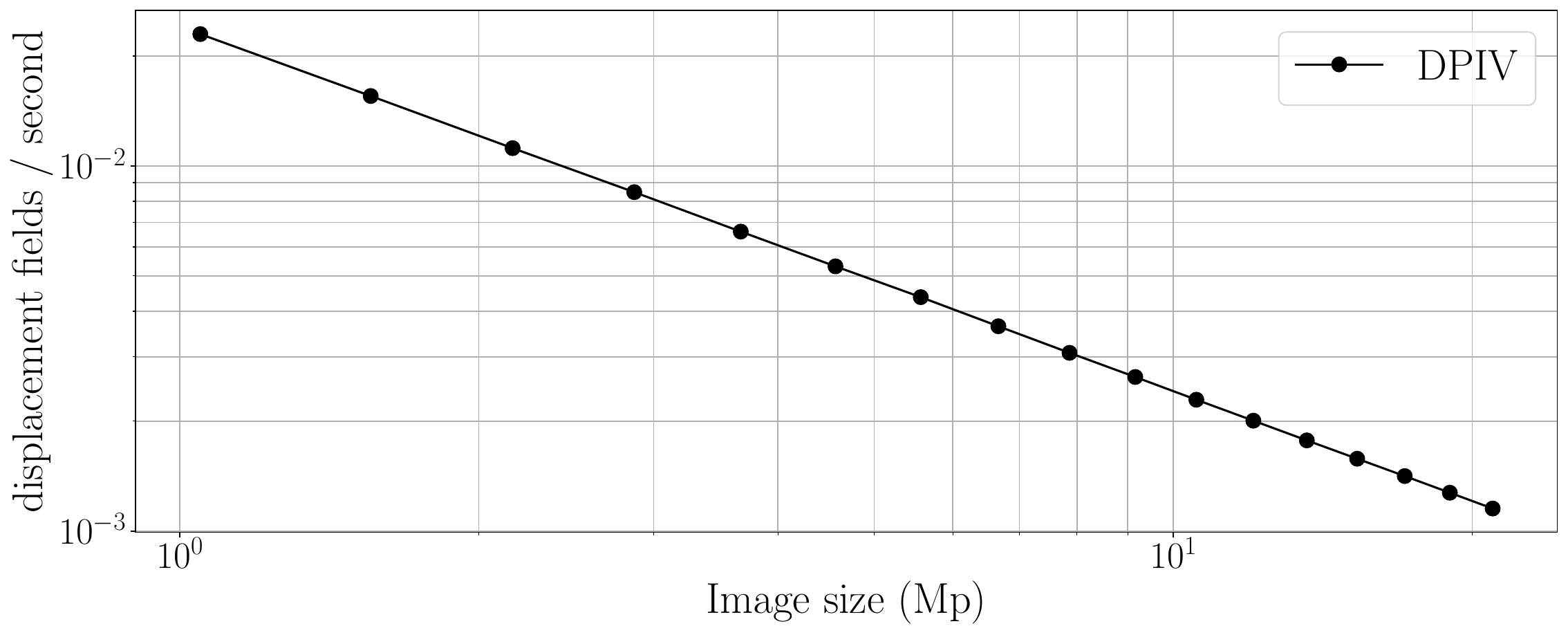}
        \caption{}
        \label{fig:perf_cc}
    \end{subfigure}
    
    \caption{Post-processing performance showing the number of displacement fields computed per second as a function of the size of images for \textbf{(a)} OFV and \textbf{(b)} DPIV (CC-PIV)).}
    \label{fig:perf_post}
\end{figure}

In Fig.~\ref{fig:perf_of_post}, the post-processing speed (number of displacement fields computed per second) is shown as a function of the size of the images, for 3 different pyramid sub-levels (PSL). This matters because PSL is tuned to the inter-frame displacement and has a non-negligible impact on runtime. In our implementation, $PSL=0$ means no pyramid level, meaning displacements of particles of order $\sim1$~px, while $PSL=1$ can deal with displacement up to $\sim4$~px and $PSL=2$ displacements up to $\sim9$~px.

First, one can see that the computing speed scales linearly with the image size: the smallest the image, the highest the number of velocity fields computed per second. 

For 1~Mp images, $1100$ to $1800$ fields are computed  per second, depending on the number of PSL. If there are mainly small displacements, one can compute up to $1800$ fields per second. Even with $PSL=2$ (displacements up to $\sim9$~px) the computing speed exceeds 1~kHz.
Interestingly, the performances are \textcolor{black}{also high} for 4~Mp images, which is the standard for PIV measurements, with $400$--$500$~fields computed per second. Even with large images, \textcolor{black}{performance remains relatively high}, with  $160$-–$200$~Hz for  10~Mp images and $80$–-$99$~Hz for 21~Mp images.
It is important to keep in mind that \textcolor{black}{this high computational efficiency is obtained with dense displacement fields} (1 vector per pixel), i.e. a high sampling rate that can lead to high effective spatial resolution.

Fig.~\ref{fig:perf_cc} shows the results obtained with CC-PIV (DPIV)  computed on CPU. The computing speed is approximately \emph{five} orders of magnitude lower than the one obtained with OFV, keeping in mind that the spatial resolution is also lower than the one obtained with OFV.

\textcolor{black}{Table \ref{tab:ml_compare} presents a brief comparison of some of the previously mentioned ML--based OF estimators with OFV. The datasets used in the different methods are not the same, but serves to present a general picture of accuracy and computational efficiency. FlowNet2 \cite{Ilg2016FlowNet2E} and LiteFlowNet3 \cite{hui2020liteflownet3resolvingcorrespondenceambiguity} show average endpoint errors (AEE) between $\sim 3$ and $\sim5$ px, and an inference time between 7--123 ms. This translates roughly to a range of 142--8 Hz for images of 0.446 Mpx. RAFT-PIV \cite{Lagemann2021DeepRO} shows better results in accuracy in both public and in--house synthetic PIV datasets, however it does not report on the inference time. Only a rough estimation can be made from \citet{Lagemann2022GeneralizationOD}, where the inference of 12000 image pairs took 12 hours, which would yield about 3.6 s of inference time per image pair, although the image size is not reported. This small comparison highlights OFV's advantage. If large models can generate good results, they still require massive amounts of training and are not suitable for Real-Time execution.}

\begin{table*}[t]
{
\centering
\scriptsize
\setlength{\tabcolsep}{4pt}
\renewcommand{\arraystretch}{1}
\begin{tabular}{|c|c|c|c|c|c|c|}
\hline
Method & Benchmark (metric) & Error (px) & \makecell[c]{Time\\(ms/pair)} & \makecell[c]{Input\\(Mpix)} & \makecell[c]{Throughput\\(Mp/s)} & Hardware \\
\hline
FlowNet2~\cite{Ilg2016FlowNet2E} &
MPI-Sintel (AEE final) &
 3.14--5.21 &
7--123 &
0.446 &
3.62--63.71 &
GTX 1080 \\
\hline
LiteFlowNet3~\cite{hui2020liteflownet3resolvingcorrespondenceambiguity} &
MPI-Sintel (AEE final) &
4.53 &
59 &
0.446 &
7.61 &
GTX 1080 \\
\hline
\makecell[c]{RAFT32-PIV\cite{Lagemann2021DeepRO}\\RAFT256-PIV\cite{Lagemann2021DeepRO}} &
\makecell[c]{Public PIV dataset\cite{Cai2019DenseME} (AEE)\\ PIV dataset\cite{Lagemann2021DeepRO}} & 
\makecell[c]{0.021 -- 0.004\\0.014--0.445} &   
-- &
0.0655 &
-- &
$2\times$ Quadro RTX 6000 \\
\hline
This work & 
\makecell[c]{Rankine ($\langle Err\rangle$)\\HIT ($\langle Err\rangle$)} &
\makecell[c]{0.048 -- 0.27\\0.163} & 
-- & 
1.05 & 
-- & 
RTX 5090 \\
This work & High-res throughput test & -- & 10.52 -- 0.56 & 1.15 -- 21.0 & 2064 -- 2079 & RTX 5090 \\
\hline
\end{tabular}

\caption{Indicative comparison with representative learning-based optical-flow / PIV methods.
Error is average endpoint error in pixels (AEE), i.e. the same mathematical definition as $\langle Err\rangle$ but evaluated on different datasets.
Throughput is computed as input megapixels divided by reported time per image pair.}
\label{tab:ml_compare}
}
\end{table*}

\subsection{Live (Real-Time) computing performances\label{Live}}

In the following, “Live” computing performances means that the velocity fields are computed in real-time, during the experiments. It means that the images are acquired by the camera, transferred through CoaXPress card to the workstation, then to the GPU. The OFV computation occurs on the GPU within the inter-frame interval. The details on the acquisition chain used, from camera to GPU, were presented in Sec.~\ref{sec:setup}.

Live performances were measured by streaming image pairs directly from the camera to the workstation and increasing the acquisition rate until system overruns (dropped frames, rejections, or errors) appeared. As in the offline tests, the images served purely to measure computational speed.

\begin{figure}[h]
    \centering
    \includegraphics[width=\linewidth]{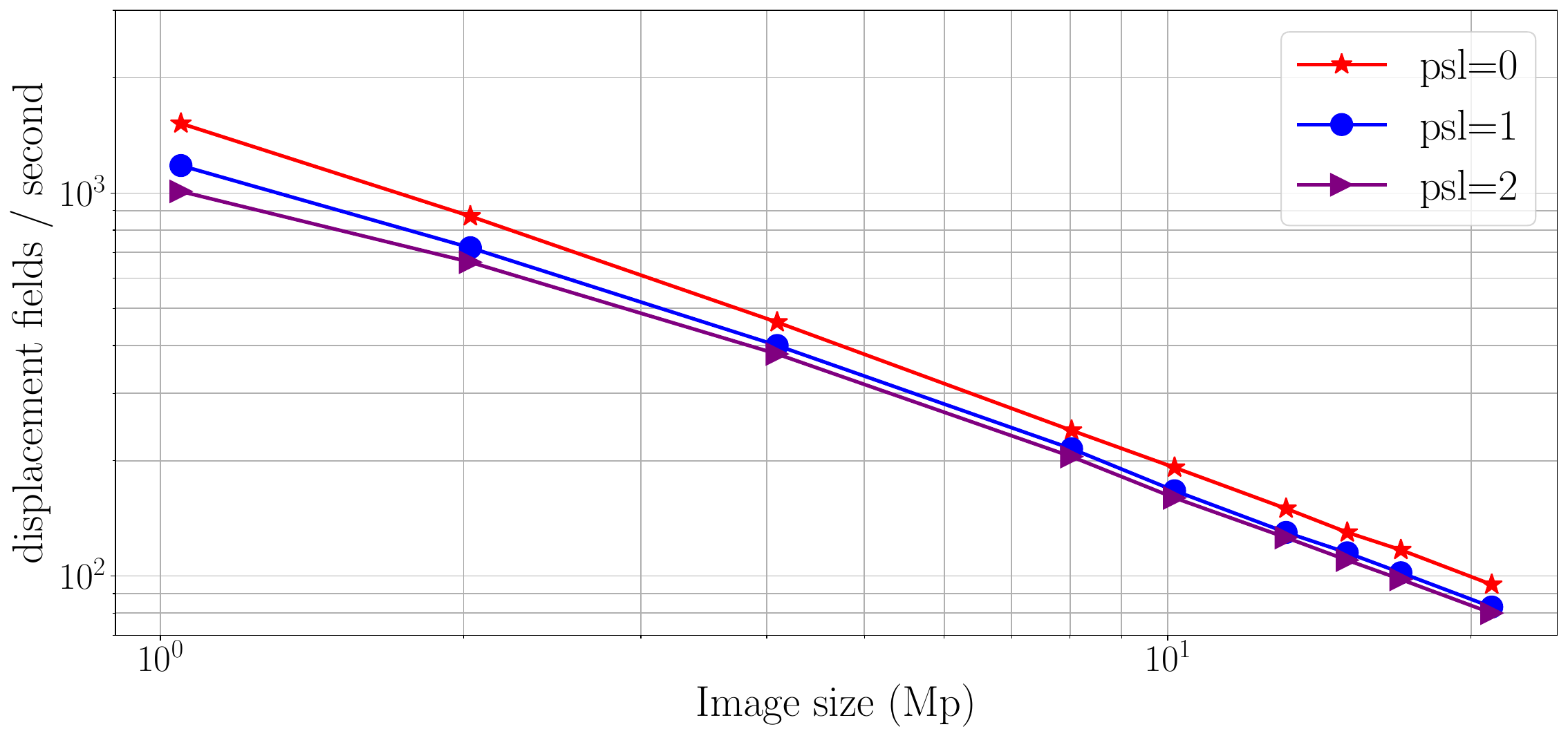}
    \caption{Live performance benchmark showing the number of displacement fields computed per second as a function of the size of images transferred by the camera directly to the GPU through a CoaxPress card. The influence of the PSL parameter is also shown, with a limited reduction of the computation speed for higher PSL. }
    \label{fig:perf_live}
\end{figure}

Figure~\ref{fig:perf_live} shows the number of displacement fields computed per second during the experiment as a function of the size of the images. The computing speed also scales linearly with the image size, similarly to the offline case.  Of course the performance also depends on the chosen OFV parameters, especially the number of PSL. Despite the impact of the data transfer from the camera to the GPU, the computing speed measured during live measurements remain notable, and we are not aware of higher published rates for comparable image sizes and dense displacement fields: from $\sim1000$--$\sim1400$ displacement fields can be computed per second for 1~Mp images, depending on the desired inter--frame displacements (i.e. number of PSL). For 4~Mp images (a common format for PIV), $380$--$460$ displacement fields per second can be computed, depending on PSL. For large images the system maintains high performance: $160$--$190$~Hz for 10~Mp images and $80$--$95$~Hz for 21~Mp images.

The difference between offline and Live performance decreases with increasing image size. It likely reflects driver/OS scheduling overheads that cap GPU task refresh for small problem sizes on a Windows system. As the image size (and thus work per frame) increases, this overhead becomes comparatively negligible, and offline and Live performances align.

\section{Experimental Application\label{sec:exp_application}}
\begin{figure}[!ht]
    \centering
    \includegraphics[width=\linewidth]{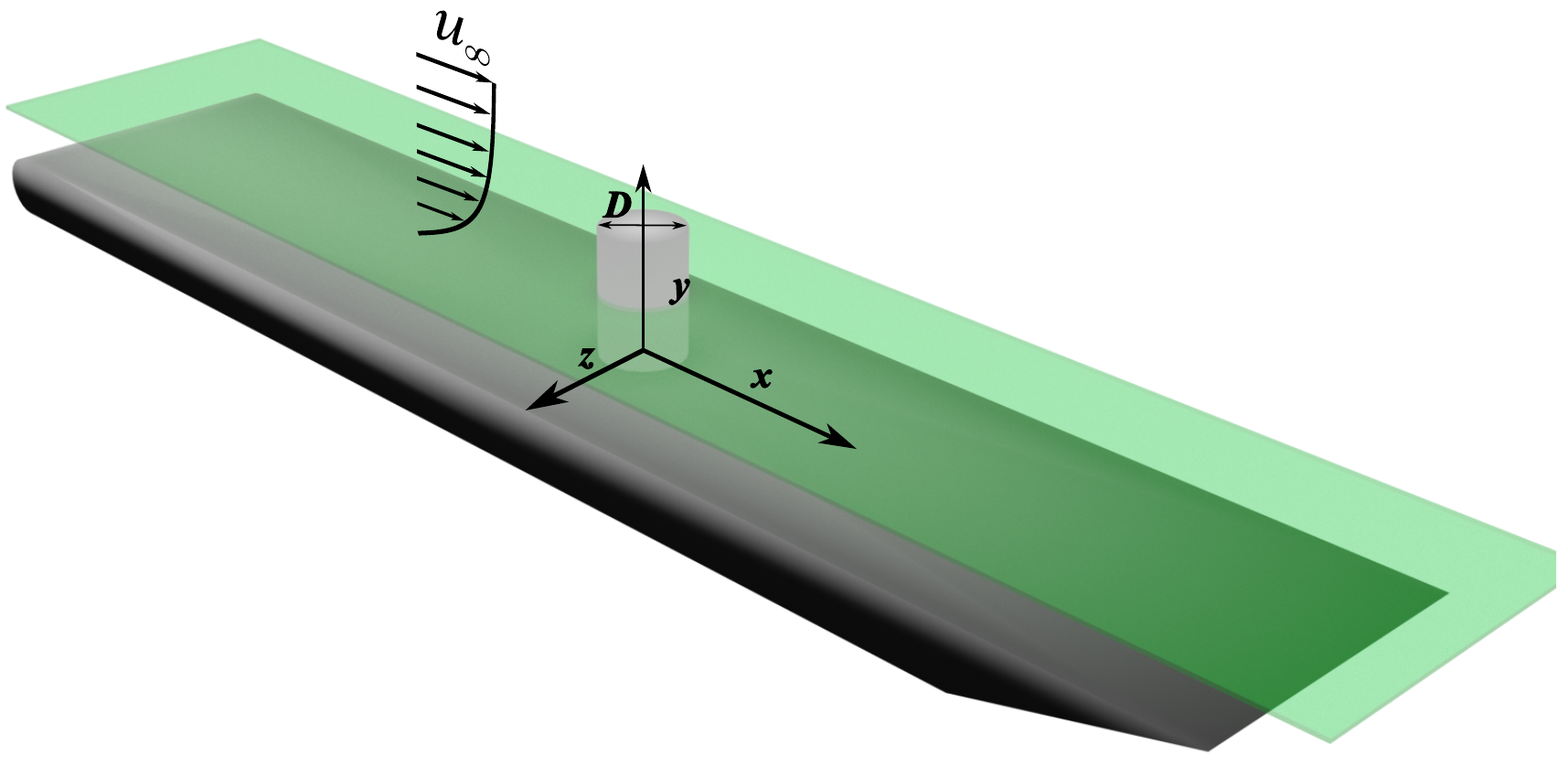}
    \caption{3D sketch of the flat plate with the cylinder located in the mid--plane of the flat plate, showing the laser plane where measurements are carried out at $y=H_c/2$.}
    \label{fig:exp_setup}
\end{figure}

\textcolor{black}{Finally, an example of OFV measurements of a real flow is shown in this section to illustrate the quality of the measured velocity fields, and possible new applications. For this experimental application, we focus on the canonical wake flow past a cylinder (Bénard - Von Karman flow) \cite{provansal1987benard,goujon1994downstream}. The experiment is carried out in our in--house recirculating hydrodynamic channel. The test section is composed of a rectangular cross section $W\times H=15\times7$~cm and spans over 80 cm in length \cite{cambonie2014seeding,gautier2015frequency,giannopoulos2020data}. A cylinder of diameter $D=4$~cm is fixed vertically on a flat plate centered at the mid--width and mid--length of the test section. Measurements are carried out on a horizontal plane at mid--height of the cylinder to avoid disturbances from the upper and lower walls. Fig.~\ref{fig:exp_setup} shows a 3D rendering of the model used for the measurements. The green plane corresponds to the horizontal laser sheet located at mid-height ($y=H_c/2$) of the cylinder.}

\textcolor{black}{The flow is illuminated with a Coherent Genesis$^{TM}$ continuous laser with a wavelength $\lambda=532$ nm at $1.5$ Watts. The beam passes through a 60$^o$ cylindrical Powell lens to form a light sheet of thickness $\sim1.5$~mm with a relatively homogeneous  illumination across the measurement plane. Water is seeded with polyamide tracer particles of 20 $\mu$m of diameter.} 

\textcolor{black}{Snapshots are acquired using the Mikrotron 21CXP12 camera described in Sec.~\ref{sec:setup} with a 50 mm f/1.2 Nikkor lens, fixed over the channel. The size of the images is $5120\times1600$ px and the exposure time is 1 ms. Image pairs are acquired at $f_{ac}=100$~Hz with an inter--frame time $\delta t = 5$ ms. The field of view is $X=31.33$~cm by $Z=10.67$~cm, with a pixel field of view $\Delta x=\Delta z \approx66.67 ~\mu\text{m}$. The resulting particle images have an average diameter $d_p=2.34$~px and a mean concentration $C_p=0.021$~ppp. Conventional post-processing and Live (L-OFV) measurements are performed. For the post processing a series of 1000 continuous images are recorded and saved. The Live measurements on the other hand are carried out for over \textit{four hours} to show L-OFV's capabilities of computing and extracting data from velocity fields in Real--Time. The Live measurements are illustrated by one of the two online movies associated to Fig.~\ref{fig:cyl_flow} in the supplementary material.}

\textcolor{black}{Fig.~\ref{fig:cyl_flow} shows a snapshot of the instantaneous velocity fields of the flow past the cylinder at $Re_D=U_\infty D/\nu=6482$. A great level of detail can be appreciated in the streamwise and flow--normal velocity components (Fig.~\ref{fig:cyl_u} and~\ref{fig:cyl_w}) showing the complexity of the flow, and refinement of small scales. }

\begin{figure}[h!]
    \begin{subfigure}{\linewidth}
        \centering
        \includegraphics[width=13cm,height=4.5cm]{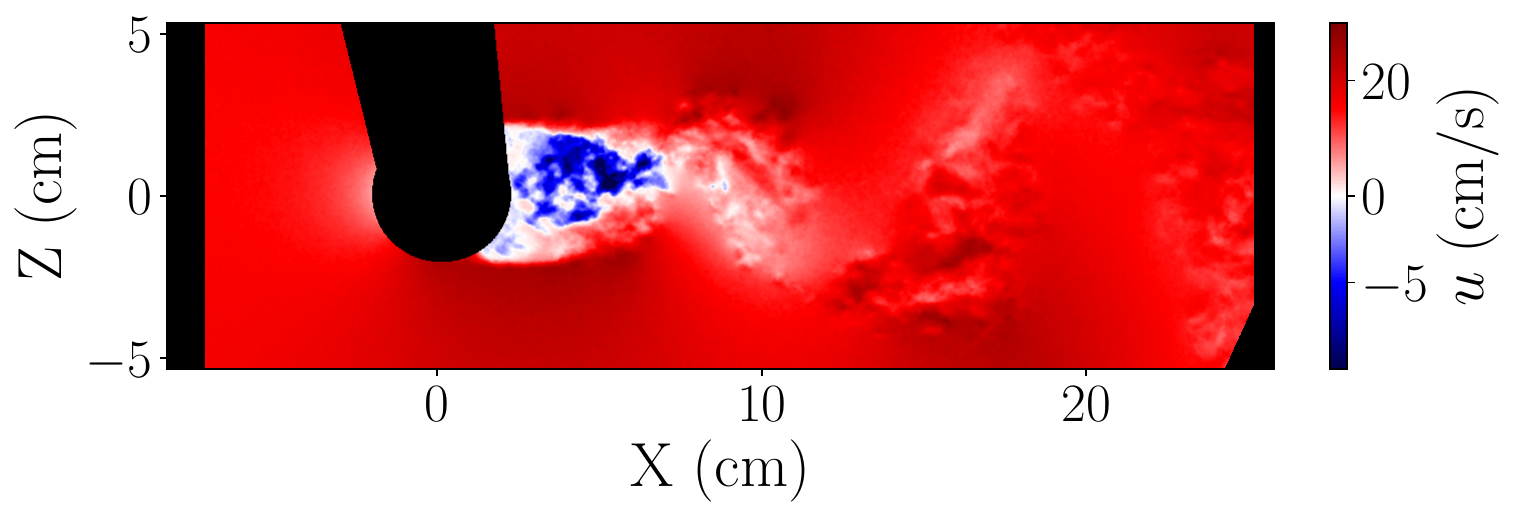}
        \caption{}
        \label{fig:cyl_u}
    \end{subfigure}
    \begin{subfigure}{\linewidth}
        \centering
        \includegraphics[width=13cm,height=4.5cm]{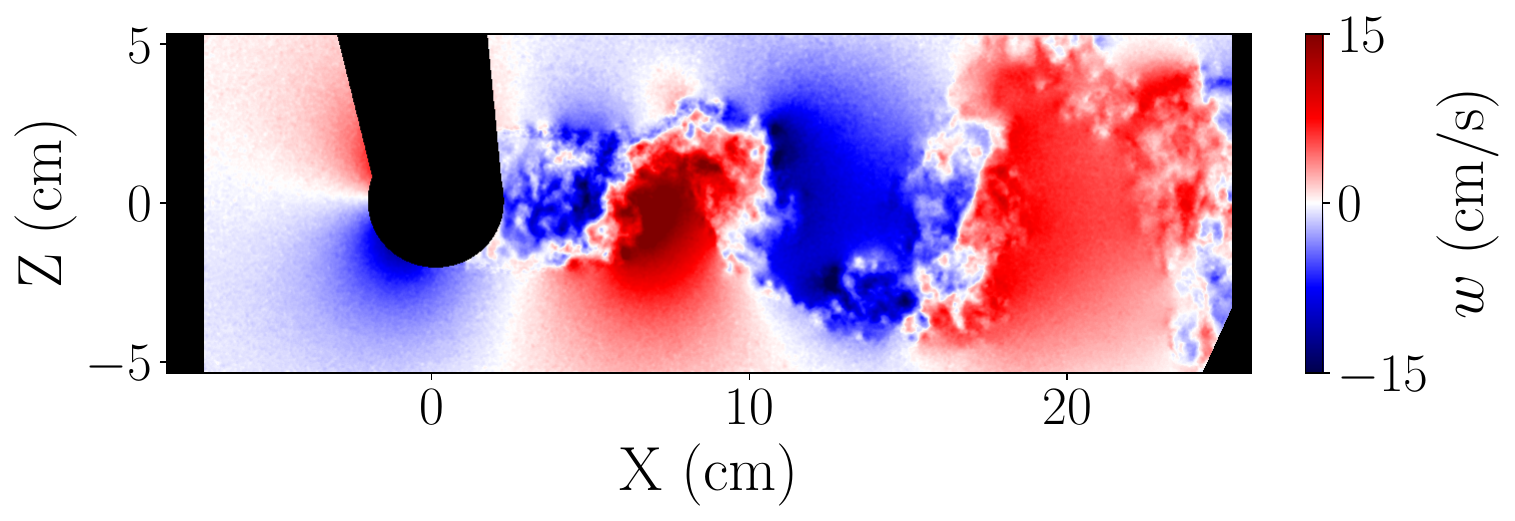}
        \caption{}
        \label{fig:cyl_w}
    \end{subfigure}
    \begin{subfigure}{\linewidth}
        \includegraphics[width=13cm,height=4.5cm]{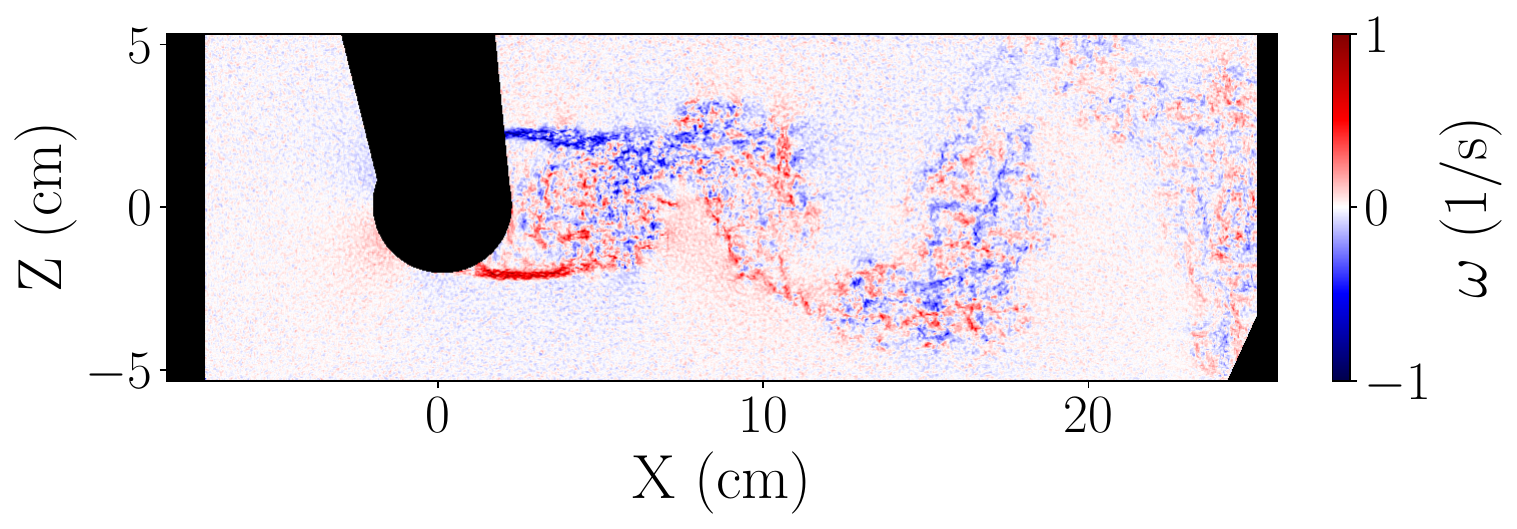}
        \caption{}
        \label{fig:cyl_vort}
    \end{subfigure}
    
    \caption{\textcolor{black}{Instantaneous velocity fields of the wake flow past a Cylinder of diameter $D=4~cm$  at $Re_D=6482$ measured in a horizontal plane at $y=H_c/2$. The black area corresponds to the masked regions (e.g. the cylinder and its shadow). (a) streamwise velocity component $u(x, z, t_0)$. (b) Cross-stream velocity component $w(x, z, t_0)$. (c) Out of plane vorticity $\omega (x, z, t_0)$ computed also in real-time using the instantaneous ($u, w$) 2D velocity field. The first movie online corresponds to the time series of 2D vorticity field. The second movie online illustrate the live measurements during an experiment. }}
    \label{fig:cyl_flow}
\end{figure}

\textcolor{black}{The ability to capture fine details of the flow is also appreciated in the vorticity field (Fig.~\ref{fig:cyl_vort}), showing rich and complex flow with many small vortices generated in the shear layers in the wake of the cylinder. No additional spatial smoothing or interpolation was applied for visualization. The apparent \textit{smoothness} in the fields results from dense per--pixel estimation combined with sufficiently rich seeding texture. It is also important to note that the vorticity field can be computed in Real--Time using the 2D2C velocity field measured at each time step. This is another example of the great advantage of this approach which gives immediate access to complex quantities derived in Real--Time from the measured 2D2C velocity fields. }

\begin{figure}
    \begin{subfigure}{\linewidth}
        \centering
        \includegraphics[width=\linewidth]{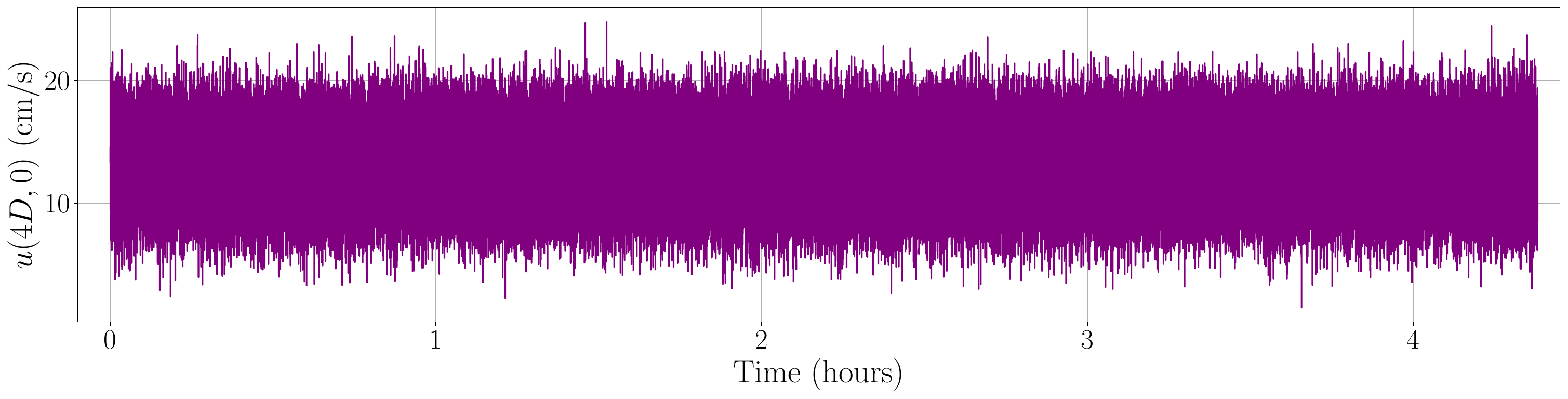}
        \caption{}
        \label{fig:u_ts}    
    \end{subfigure}
    \begin{subfigure}{\linewidth}
        \centering
        \includegraphics[width=\linewidth]{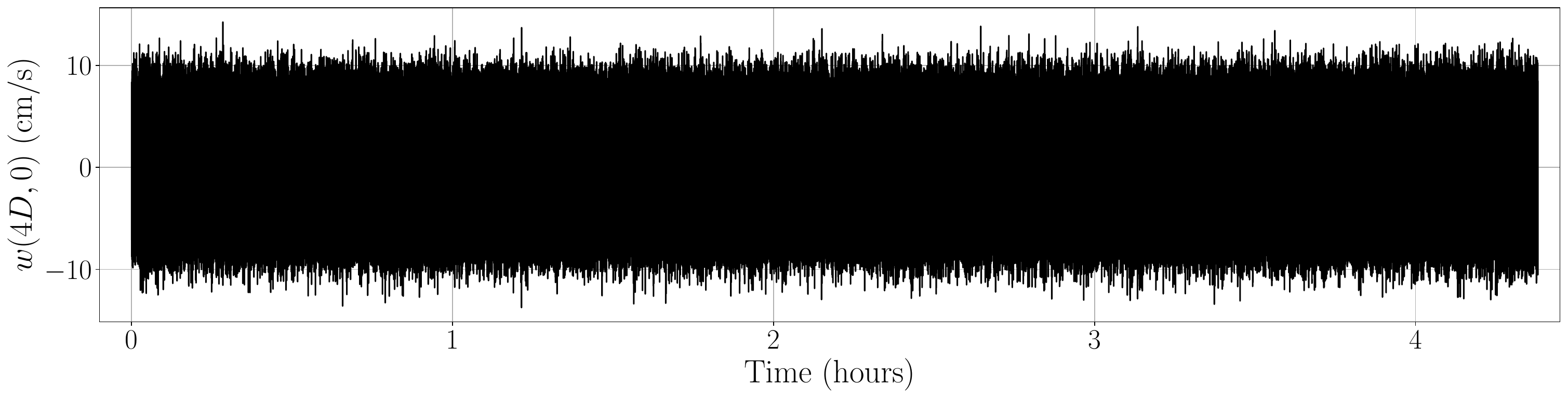}
        \caption{}
        \label{fig:w_ts}    
    \end{subfigure}
    \caption{Time series of velocity probes $u$ and $w$ located at $(x,z)=(4D,0)$ for over four hours at $f_{ac}=100$~Hz. (a) Streamwise velocity $u$. (b) Cross--stream velocity $w$.}
    \label{fig:probes}
\end{figure}

\textcolor{black}{Beyond the appreciated quality of the instantaneous velocity fields, live data acquisition enables long-duration monitoring, unlocking new experimental possibilities, e.g. on-the-fly statistical convergence assessment. To illustrate this capability, we report on two types of observables: first, local velocity probes extracted from the velocity fields and stored at each time step, second, the recirculation area, a global quantity extracted at each time steps from the 2D instantaneous velocity fields.}

\textcolor{black}{Fig.~\ref{fig:probes} shows the time series of the two components of the velocity field measured in a velocity probe located at $(x,z)=(4D,0)=(16,0)$~cm, acquired live and \textit{recorded for over four hours} at high frequency ($f_{ac}=100$~Hz). Such measurements are impractical with standard workflows at comparable resolution due to storage and post--processing burden ($5120\times1600$~px at 100 Hz for 4.38 hours corresponds to 1,576,800 image pairs; with 8-bit dynamic range this corresponds to $\sim22$ TB only in images). Not only does L-OFV provide access to data that is generally inaccessible experimentally, but it also allows for significant savings in time and energy. }

\begin{figure}
    \begin{subfigure}{\linewidth}
        \centering
        \includegraphics[width=\linewidth]{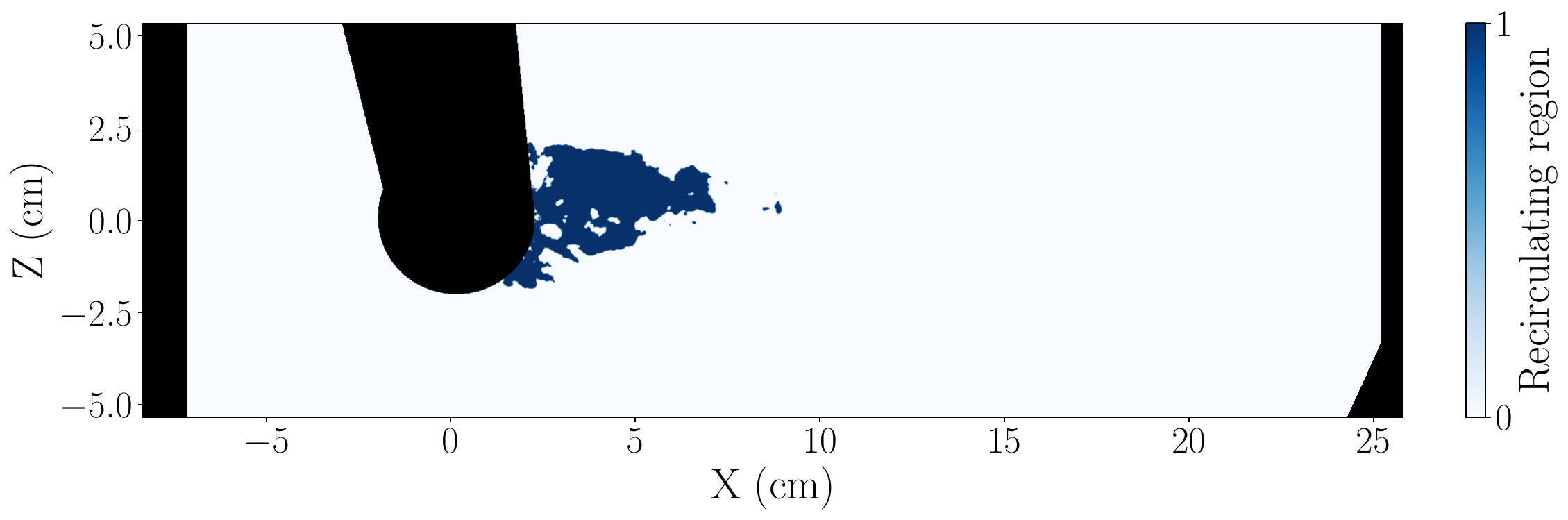}
        \caption{}
        \label{fig:RA_cyl}    
    \end{subfigure}
    \begin{subfigure}{\linewidth}
        \centering
        \includegraphics[width=\linewidth]{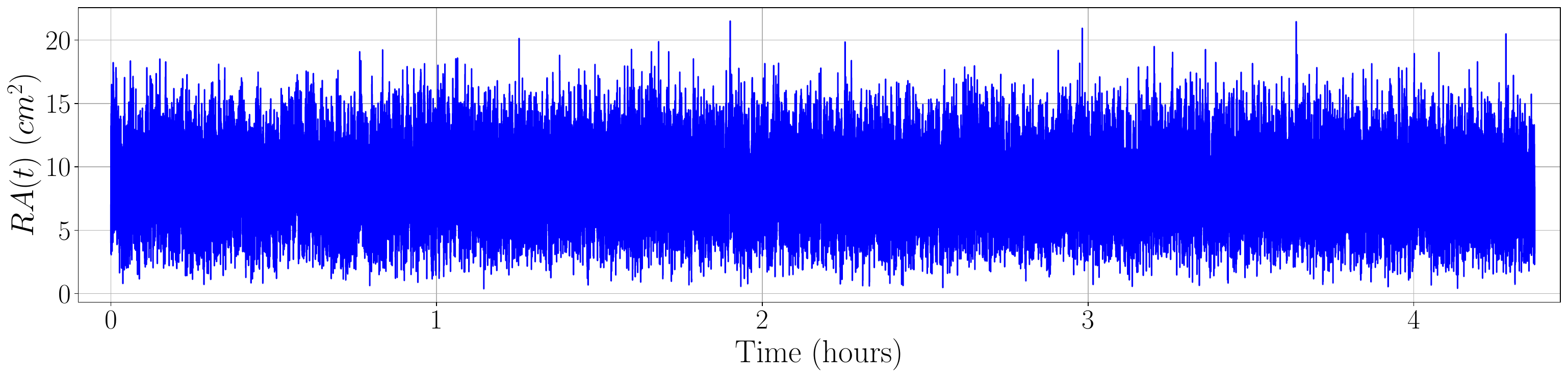}
        \caption{}
        \label{fig:RA_ts}    
    \end{subfigure}
    \caption{\textcolor{black}{Recirculation bubble in the wake of a cylinder at $Re_D=6482$ in a horizontal plane at $y=H_c/2$. (a) Example of the instantaneous recirculation bubble, as defined in Eq.~\ref{eq:RA}, downstream the cylinder in the measurement plane. (b) Time series of the recirculation area $RA(t)$, monitored over more than four hours.}}
    \label{fig:Recirculation}
\end{figure}

\textcolor{black}{Fig.~\ref{fig:RA_cyl} highlights the ability to extract meaningful global information from the instantaneous 2D velocity fields. Indeed, we can define the recirculation bubble generated downstream of the cylinder as the region where the direction of the streamwise velocity is opposite to the free-stream direction \cite{gautier2013control}. This binary field is converted into a scalar by summing the corresponding pixel areas. The instantaneous recirculation area $RA(t)$ can be written as:
\begin{equation}
    RA(t)=\sum_{i,j}H(-u_{i,j}(t))\,\Delta x \Delta z,
    \label{eq:RA}
\end{equation}
where $H$ is the Heaviside function, defined as $H(-u)=1$ if $u<0$, and $0$ otherwise. With one vector per pixel and using a calibration of $\Delta x, \Delta z$, $RA(t)$ is expressed in cm$^2$. The resulting $RA(t)$ time series is shown in Fig.~\ref{fig:RA_ts}  over more than \textit{four hours}. As with the velocity time series measured in a local probe, it represents 1,576,800 velocity fields. Such data can then be used to investigate low-frequency dynamics (very low frequencies), long-time drifts, or establish rigorous convergence criteria for time-averaged observables. A compact video illustrating the experimental setup and the live processing workflow (real-time field computation and Live extraction of derived quantities) is provided as Supplementary Material.}

\begin{figure}[h!]
    \begin{subfigure}{\linewidth}
        \centering
        \includegraphics[width=\linewidth]{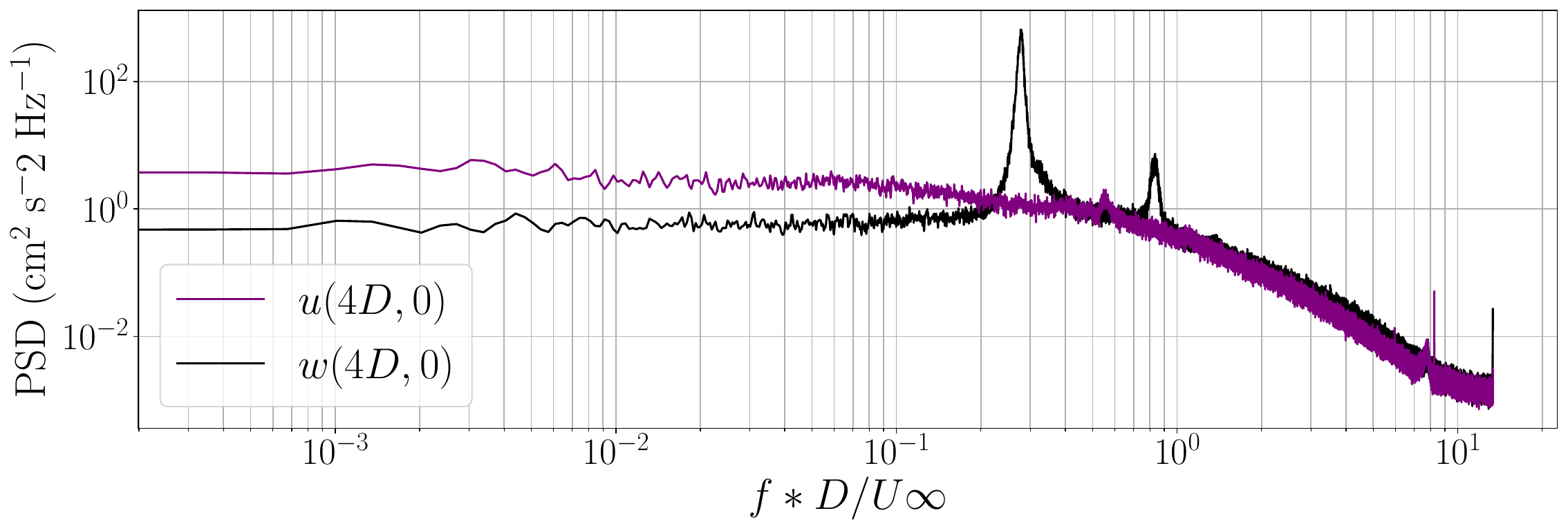}
        \caption{}
        \label{fig:spectra_vw}
    \end{subfigure}
    \begin{subfigure}{\linewidth}
        \centering
        \includegraphics[width=\linewidth]{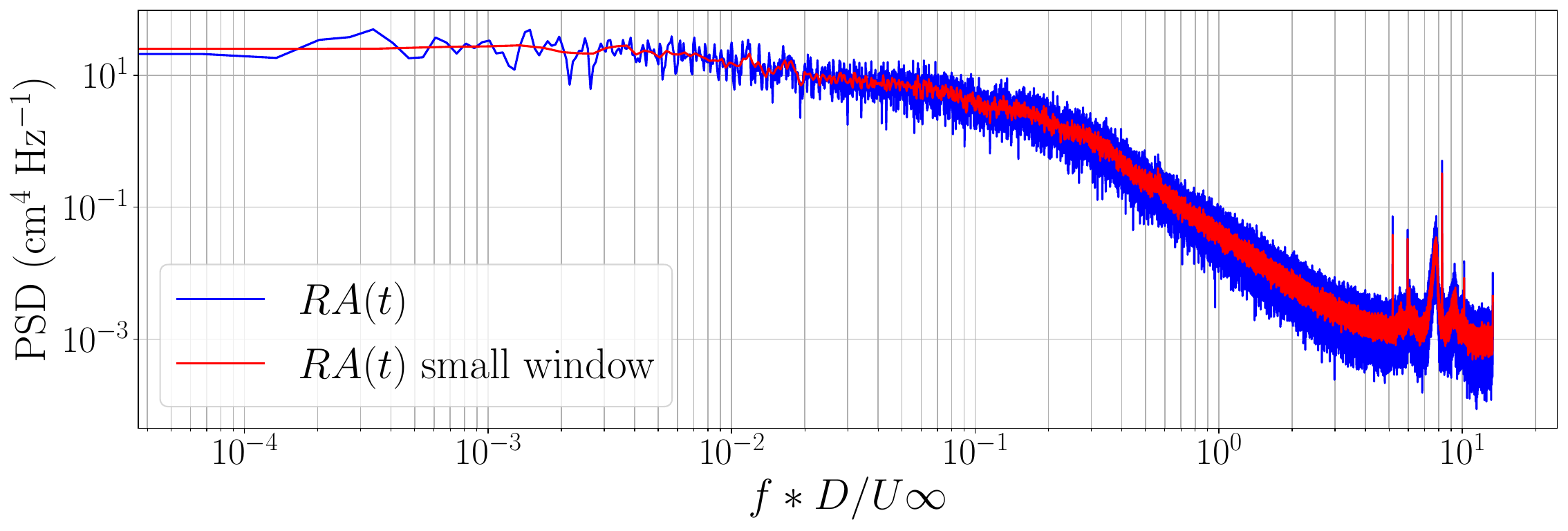}
        \caption{}
        \label{fig:spectra_RA}
    \end{subfigure}
    \caption{Power spectral density of observables captured live for over four hours of monitoring at $f_{ac}=100$~Hz. (a) Velocity probes $u$ and $w$ located at $(x,w)=(4D,0)$. (b) Recirculation area scalar.}
    \label{fig:spectra_exp}
\end{figure}

\textcolor{black}{The advantage of live data acquisition is evident: it gives an immediate access to the data of interest with virtually no limit of observation. For instance, the monitoring of both the velocity probes and the recirculation area scalar allows for the study of characteristic frequencies of the fluid flow. Fig.~\ref{fig:spectra_exp} presents the Power Spectral Density (PSD) for the time series. The PSD is computed using Welch's \cite{Welch1967TheUO} method, with a \textit{Hann} window type, and a window length of $2\%$ of the signal's length for the velocity probes and the recirculation area marked as small window. To explore the lower frequencies from the global signal, the recirculation area spectrum was also computed with a \textit{Hann} window of $25\%$ of the signal's length.}

\textcolor{black}{In Fig.~\ref{fig:spectra_vw} the shedding frequency is clearly recovered from the $w$ component with a Strouhal number $St=f\cdot D/U_\infty\approx0.27$ ($U_\infty$ being the streamwise velocity upstream of the cylinder), higher than the standard Bénard-Von-Karman shedding frequency ($St_s\approx0.205$) \cite{Fey1998ANS} for this regime. 
Since the cylinder occupies a significant fraction of the channel width ($\beta=D/W=0.267$) we correct the characteristic velocity using a continuity--based gap velocity $U_{gap}=U_\infty/(1-\beta)$ to account for the blockage effect, as commonly used under confinement conditions \cite{Awbi1983EffectOB,King1977ARO}. Using $U_{gap}$, our dominant shedding peak corresponds to $St_{s,gap}=f\cdot D/U_{gap}=0.204$, consistent with the standard circular--cylinder wake Strouhal number in the subcritical regime. This agreement provides an experimental validation of the L-OFV measurements.} 

\textcolor{black}{The second narrow--band peak is attributed to a harmonic of the shedding frequency. In comparison, the streamwise component shows higher energy in the lower frequencies with wide--band peaks at $St\approx2\times10^{-3}$. 
The lower frequencies dominate the spectrum of the recirculation area (Fig.~\ref{fig:spectra_RA}), which does not clearly display the shedding frequency but rather suggests that the stronger variations of the recirculation region are caused by very low frequency \textit{breathing} modes. }

\textcolor{black}{This analysis illustrates the advantages of Live data acquisition and Live data extraction. It not only allows to recover the expected physical results, e.g. $St_s\approx0.2$, but also to deepen our understanding of fluid flows with possible novel experiments and analysis (e.g. Recirculation area spectra, access to very low frequencies together with high frequencies, among other examples).}

\section{Conclusions}

In this study we presented a variational Optical Flow Velocimetry (OFV) approach that produces dense velocity fields (one vector per pixel) with high accuracy, high spatial resolution and high processing rates (up to kHz), both live and offline, on commodity hardware.

Accuracy was benchmarked precisely using synthetic particle images with two types of datasets. First, an isolated Rankine vortex demonstrated that increasing particle density in the images (i.e., richer \emph{image texture}) reduces error and enables resolution of small structures with steep velocity gradients. This analysis also suggested a practical parameter guideline: the optimal kernel radius tends to \emph{decrease} as displacement gradients increase. Using $D/R$ (maximum displacement over core radius) as a gradient indicator proved useful; results quality degraded as $D/R$ approached $\sim 0.7$, consistent with the challenge of small, fast vortices.

Second, velocity fields from a DNS HIT dataset confirmed these trends. OFV benefited from higher seeding, resolved small scales down to near the dissipation range, and outperformed CC-PIV in reconstructing sharp gradients. Mid-height profiles, kinetic-energy comparisons, and time-averaged spectra showed close agreement with DNS at large and intermediate scales, with divergence emerging near $k\eta\!\approx\!0.5$ and an effective cutoff close to $k\eta\!\approx\!1$.

For both benchmarks, the results yielding higher accuracy levels were produced with higher particle seeding. In particular, the DNS HIT dataset showed a specific preference of OFV for high density with small particle size. 

\textcolor{black}{We also characterized computational performance in both post-processing and Live (real-time) modes on a single RTX~5090 Nvidia GPU. Computing speeds reached up to $1800$~Hz offline and  $1400$~Hz online (live) for $1$~Mp images, $\sim 400$–-$500$~Hz for $4$~Mp images and $\sim 80$–-$99$~Hz for $21$~Mp images. These results are obtained while keeping the dense displacement fields, i.e. one vector per pixel. Live results were comparable to offline results for large image sizes. To our knowledge, these performances are among the fastest reported for dense, per-pixel velocimetry. Moreover, these performances are obtained using a single GPU RTX~5090. One can expect scalability of the results with multiple GPU. It means that the computation frequency could be significantly increased if the computation is properly distributed over the GPUs. Doubling the computation frequency with 2 or 3 GPUs appears a reasonable objective for future developments of both the software and hardware. } 

\textcolor{black}{Finally, the OFV is used to characterize the Bénard - Von Karman flow downstream a cylinder in an hydrodynamic tunnel. The instantaneous velocity fields demonstrate that indeed OFV allows the measurements of small scale structures even during live measurements at high frequencies. It was also shown that it is possible to measure in real-time complex observables derived from the 2D instantaneous velocity fields, like the vorticity field or the recirculation area. Moreover these quantities can be measured and stored as long as needed, allowing nearly unlimited data acquisitions (4 hours measurements at 100 Hz in the present application). } 

\textcolor{black}{In summary, the proposed OFV setup and algorithm enable high--frequencies Live PIV experiments without excessive data buffering. It is now possible to extract the data of interest (either local probes or global quantities) directly during the experiments, without having to store the raw data. It also opens new opportunities for closed-loop flow-control, rare events detection or rapid in-situ diagnostics. It additionally allows for significant time and power consumption savings, either through fast post-processing or direct live measurements of the quantities of interest, avoiding data storage and post-processing. Practical guidance emerges from the benchmarks: favor rich image texture (high seeding that we understand can be a limitation depending on the application), use smaller kernels for higher gradients, tune pyramid levels to inter-frame displacement and monitor $D/R$ as a simple gradient proxy. A companion application paper will extend the validation to laboratory flows and explore a novel experimental application.}

\begin{acknowledgments}
This work has been supported by the Association National Recherche Techonologie (ANRT) and Photon Lines SAS. 
\end{acknowledgments}

\clearpage
\bibliography{LOFV}

@article{provansal1987benard,
  title={B{\'e}nard-von K{\'a}rm{\'a}n instability: transient and forced regimes},
  author={Provansal, M and Mathis, C and Boyer, L},
  journal={Journal of Fluid Mechanics},
  volume={182},
  pages={1--22},
  year={1987},
  publisher={Cambridge University Press}
}

@article{goujon1994downstream,
  title={Downstream evolution of the B{\'e}nard--von K{\'a}rm{\'a}n instability},
  author={Goujon-Durand, S and Jenffer, P and Wesfreid, JE},
  journal={Physical Review E},
  volume={50},
  number={1},
  pages={308},
  year={1994},
  publisher={APS}
}

@article{giannopoulos2020data,
  title={Data-driven order reduction and velocity field reconstruction using neural networks: The case of a turbulent boundary layer},
  author={Giannopoulos, Antonios and Aider, Jean-Luc},
  journal={Physics of Fluids},
  volume={32},
  number={9},
  year={2020},
  publisher={AIP Publishing}
}

@article{gautier2015frequency,
  title={Frequency-lock reactive control of a separated flow enabled by visual sensors},
  author={Gautier, Nicolas and Aider, J-L},
  journal={Experiments in Fluids},
  volume={56},
  number={1},
  pages={16},
  year={2015},
  publisher={Springer}
}

@article{cambonie2014seeding,
  title={Seeding optimization for instantaneous volumetric velocimetry. Application to a jet in crossflow},
  author={Cambonie, Tristan and Aider, Jean-Luc},
  journal={Optics and Lasers in Engineering},
  volume={56},
  pages={99--112},
  year={2014},
  publisher={Elsevier}
}

@article{gautier2013control,
  title={Control of the separated flow downstream of a backward-facing step using visual feedback},
  author={Gautier, N and Aider, J-L},
  journal={Proceedings of the Royal Society A: Mathematical, Physical and Engineering Sciences},
  volume={469},
  number={2160},
  pages={20130404},
  year={2013},
  publisher={The Royal Society Publishing}
}

@article{romera2023optical,
  title={Optical flow algorithms optimized for speed, energy and accuracy on embedded GPUs},
  author={Romera, Thomas and Petreto, Andrea and Lemaitre, Florian and Bouyer, Manuel and Meunier, Quentin and Lacassagne, Lionel and Etiemble, Daniel},
  journal={Journal of Real-Time Image Processing},
  volume={20},
  number={2},
  pages={32},
  year={2023},
  publisher={Springer}
}

@article{westerweel2013particle,
  title={Particle image velocimetry for complex and turbulent flows},
  author={Westerweel, Jerry and Elsinga, Gerrit E and Adrian, Ronald J},
  journal={Annual Review of Fluid Mechanics},
  volume={45},
  number={1},
  pages={409--436},
  year={2013},
  publisher={Annual Reviews}
}

@article{Jassal2025ARO,
  title={A review of optical flow velocimetry in fluid mechanics},
  author={Gauresh Raj Jassal and Bryan E. Schmidt},
  journal={Measurement Science and Technology},
  year={2025},
  volume={36},
  url={https://api.semanticscholar.org/CorpusID:276003332}
}

@article{theunissen2006adaptive,
  title={An adaptive sampling and windowing interrogation method in PIV},
  author={Theunissen, Raf and Scarano, Fulvio and Riethmuller, Michel L},
  journal={Measurement Science and Technology},
  volume={18},
  number={1},
  pages={275},
  year={2006},
  publisher={IOP Publishing}
}

@book{raffel2018particle,
  title={Particle image velocimetry: a practical guide},
  author={Raffel, Markus and Willert, Christian E and Scarano, Fulvio and K{\"a}hler, Christian J and Wereley, Steve T and Kompenhans, J{\"u}rgen},
  year={2018},
  publisher={springer}
}

@article{scarano2000advances,
  title={Advances in iterative multigrid PIV image processing},
  author={Scarano, Fulvio and Riethmuller, Michel L},
  journal={Experiments in fluids},
  volume={29},
  number={Suppl 1},
  pages={S051--S060},
  year={2000},
  publisher={Springer}
}

@article{keane1995super,
  title={Super-resolution particle imaging velocimetry},
  author={Keane, RD and Adrian, RJa and Zhang, Yuanhui},
  journal={Measurement Science and Technology},
  volume={6},
  number={6},
  pages={754},
  year={1995},
  publisher={IOP Publishing}
}

@article{susset2006novel,
  title={A novel architecture for a super-resolution PIV algorithm developed for the improvement of the resolution of large velocity gradient measurements},
  author={Susset, A and Most, JM and Honor{\'e}, D},
  journal={Experiments in Fluids},
  volume={40},
  number={1},
  pages={70--79},
  year={2006},
  publisher={Springer}
}

@article{choi2023deep,
  title={Deep learning-based spatial refinement method for robust high-resolution PIV analysis},
  author={Choi, Jun Sung and Kim, Eung Soo and Seong, Jee Hyun},
  journal={Experiments in Fluids},
  volume={64},
  number={3},
  pages={45},
  year={2023},
  publisher={Springer}
}

@inproceedings{ilg2017flownet,
  title={Flownet 2.0: Evolution of optical flow estimation with deep networks},
  author={Ilg, Eddy and Mayer, Nikolaus and Saikia, Tonmoy and Keuper, Margret and Dosovitskiy, Alexey and Brox, Thomas},
  booktitle={Proceedings of the IEEE conference on computer vision and pattern recognition},
  pages={2462--2470},
  year={2017}
}

@article{stitou2001extension,
  title={Extension of PIV to super resolution using PTV},
  author={Stitou, Adel and Riethmuller, ML},
  journal={Measurement Science and Technology},
  volume={12},
  number={9},
  pages={1398},
  year={2001},
  publisher={IOP Publishing}
}

@inproceedings{zhang2020unsupervised,
  title={Unsupervised learning of particle image velocimetry},
  author={Zhang, Mingrui and Piggott, Matthew D},
  booktitle={High Performance Computing: ISC High Performance 2020 International Workshops, Frankfurt, Germany, June 21--25, 2020, Revised Selected Papers 35},
  pages={102--115},
  year={2020},
  organization={Springer}
}

@article{HORN1981185,
title = {Determining optical flow},
journal = {Artificial Intelligence},
volume = {17},
number = {1},
pages = {185-203},
year = {1981},
issn = {0004-3702},
doi = {https://doi.org/10.1016/0004-3702(81)90024-2},
url = {https://www.sciencedirect.com/science/article/pii/0004370281900242},
author = {Berthold K.P. Horn and Brian G. Schunck}
}

@inproceedings{Lucas1981AnII,
  title={An Iterative Image Registration Technique with an Application to Stereo Vision},
  author={Bruce D. Lucas and Takeo Kanade},
  booktitle={International Joint Conference on Artificial Intelligence},
  year={1981},
  url={https://api.semanticscholar.org/CorpusID:2121536}
}

@article{Champagnat2011FastAA,
  title={Fast and accurate \textsc{PIV} computation using highly parallel iterative correlation maximization},
  author={Fr{\'e}d{\'e}ric Champagnat and Aur{\'e}lien Plyer and Guy Le Besnerais and Benjamin Leclaire and Samuel Davoust and Yves Le Sant},
  journal={Experiments in Fluids},
  year={2011},
  volume={50},
  pages={1169-1182},
  url={https://api.semanticscholar.org/CorpusID:120766529}
}

@article{Gautier2013RealtimePF,
  title={Real-time planar flow velocity measurements using an optical flow algorithm implemented on \textsc{GPU}s},
  author={N. Gautier and J. L. Aider},
  journal={Journal of Visualization},
  year={2015},
  volume={18},
  pages={277-286},
  url={https://api.semanticscholar.org/CorpusID:9799419}
}

@article{Gautier2013ControlOT,
  title={Control of the separated flow downstream of a backward-facing step using visual feedback},
  author={N. Gautier and J. L. Aider},
  journal={Proceedings of the Royal Society A: Mathematical, Physical and Engineering Sciences},
  year={2013},
  volume={469},
  url={https://api.semanticscholar.org/CorpusID:118040396}
}

@article{Gautier2015FrequencylockRC,
  title={Frequency-lock reactive control of a separated flow enabled by visual sensors},
  author={N. Gautier and J. L. Aider},
  journal={Experiments in Fluids},
  year={2015},
  volume={56},
  pages={1-10},
  url={https://api.semanticscholar.org/CorpusID:119779600}
}

@article{MENDES_piv_image_gen,
title = {\textsc{PIV}-image-generator: An image generating software package for planar \textsc{PIV} and Optical Flow benchmarking},
journal = {SoftwareX},
volume = {12},
pages = {100537},
year = {2020},
issn = {2352-7110},
doi = {https://doi.org/10.1016/j.softx.2020.100537},
url = {https://www.sciencedirect.com/science/article/pii/S2352711020300339},
author = {Luís Mendes and Alexandre Bernardino and Rui M.L. Ferreira}
}

@article{Beresh_velocity_gradient_PIV_turbulence,
author = {Steven Beresh},
title = {The Influence of Velocity Gradients on \textsc{PIV} Measurements of Turbulence Statistics: A Preliminary Study},
journal = {26th AIAA Aerodynamic Measurement Technology and Ground Testing Conference},
year = {2008},
pages = {},
doi = {10.2514/6.2008-3710},
URL = {https://arc.aiaa.org/doi/abs/10.2514/6.2008-3710},
eprint = {https://arc.aiaa.org/doi/pdf/10.2514/6.2008-3710}
}

@article{westerweel_piv_gradients,
    author = {Westerweel, J}, 
    title =  {On velocity gradients in \textsc{PIV} interrogation},
    journal = {Exp Fluids 44, 831–842},
    year = {2008},
    doi = {https://doi.org/10.1007/s00348-007-0439-3}
}

@article{ANDALIBI20151,
title = {Effects of texture addition on optical flow performance in images with poor texture},
journal = {Image and Vision Computing},
volume = {40},
pages = {1-15},
year = {2015},
issn = {0262-8856},
doi = {https://doi.org/10.1016/j.imavis.2015.04.008},
url = {https://www.sciencedirect.com/science/article/pii/S0262885615000554},
author = {Mehran Andalibi and Lawrence.L. Hoberock and Hossein Mohamadipanah},
}

@article{Yu2023DeepDR,
  title={Deep dual recurrence optical flow learning for time-resolved particle image velocimetry},
  author={Changdong Yu and Yiwei Fan and Xiaojun Bi and Yun-fei Kuai and Yongpeng Chang},
  journal={Physics of Fluids},
  year={2023},
  url={https://api.semanticscholar.org/CorpusID:257576268}
}

@inproceedings{Teed2020RAFTRA,
  title={RAFT: Recurrent All-Pairs Field Transforms for Optical Flow},
  author={Zachary Teed and Jia Deng},
  booktitle={European Conference on Computer Vision},
  year={2020},
  url={https://api.semanticscholar.org/CorpusID:214667893}
}

@article{Lagemann2021DeepRO,
  title={Deep recurrent optical flow learning for particle image velocimetry data},
  author={Christian Lagemann and Kai Lagemann and Sach Mukherjee and Wolfgang Schr{\"o}der},
  journal={Nature Machine Intelligence},
  year={2021},
  volume={3},
  pages={641 - 651},
  url={https://api.semanticscholar.org/CorpusID:237869288}
}

@article{Cai2019DenseME,
  title={Dense motion estimation of particle images via a convolutional neural network},
  author={Shengze Cai and Shichao Zhou and Chao Xu and Qi Gao},
  journal={Experiments in Fluids},
  year={2019},
  volume={60},
  url={https://api.semanticscholar.org/CorpusID:86818733}
}

@article{Lagemann2022GeneralizationOD,
  title={Generalization of deep recurrent optical flow estimation for particle-image velocimetry data},
  author={Christian Lagemann and Kai Lagemann and Swarnava Mukherjee and Wolfgang Schroeder},
  journal={Measurement Science and Technology},
  year={2022},
  volume={33},
  url={https://api.semanticscholar.org/CorpusID:249109179}
}

@article{Saif2016GradientBI,
  title={Gradient Based Image Edge Detection},
  author={Jamil Saif and Mahgoub H. Hammad and Ibrahim A. A. Alqubati},
  journal={International journal of engineering and technology},
  year={2016},
  volume={8},
  pages={153-156},
  url={https://api.semanticscholar.org/CorpusID:18572027}
}

@article{AguilarCabello2022DPIVSoftOpenCLAM,
  title={DPIVSoft-OpenCL: A multicore CPU-GPU accelerated open-source code for 2D Particle Image Velocimetry},
  author={Jorge Aguilar-Cabello and Luis Parras and Carlos del Pino},
  journal={SoftwareX},
  year={2022},
  volume={20},
  pages={101256},
  url={https://api.semanticscholar.org/CorpusID:249170272}
}

@book{gibson2002theory,
  title={A theory of direct visual perception},
  author={Gibson, James J},
  year={2002}
}

@article{Burt1983TheLP,
  title={The Laplacian Pyramid as a Compact Image Code},
  author={Peter J. Burt and Edward H. Adelson},
  journal={IEEE Trans. Commun.},
  year={1983},
  volume={31},
  pages={532-540},
  url={https://api.semanticscholar.org/CorpusID:8018433}
}

@inproceedings{Bouguet1999PyramidalIO,
  title={Pyramidal implementation of the lucas kanade feature tracker},
  author={J.-Y. Bouguet},
  year={1999},
  url={https://api.semanticscholar.org/CorpusID:9350588}
}

@inproceedings{Varon2017ReactiveCO,
  title={Reactive control of large-scale structures in a turbulent wake},
  author={Eliott Varon and J. L. Aider and Yoann Eulalie and Stéphie Edwige and Philippe Gilotte},
  year={2017},
  url={https://api.semanticscholar.org/CorpusID:59567954}
}

@phdthesis{varonthesis,
  TITLE = {{Closed-loop control separated flows using real-time PIV}},
  AUTHOR = {Varon, Eliott},
  URL = {https://pastel.hal.science/tel-01865213},
  NUMBER = {2017PSLET008},
  SCHOOL = {{Universit{\'e} Paris sciences et lettres}},
  YEAR = {2017},
  MONTH = Oct,
  TYPE = {Theses},
  HAL_ID = {tel-01865213},
  HAL_VERSION = {v1},
}

@article{Pan2015EvaluatingTA,
  title={Evaluating the accuracy performance of Lucas-Kanade algorithm in the circumstance of PIV application},
  author={Chong Pan and Dong Xue and Yang Xu and Jinjun Wang and Runjie Wei},
  journal={Science China Physics, Mechanics \& Astronomy},
  year={2015},
  volume={58},
  pages={1-16},
  url={https://api.semanticscholar.org/CorpusID:124955763}
}

@article{Thielicke_2021, doi = {10.5334/jors.334}, url = {https://doi.org/10.5334%2Fjors.334}, year = 2021, month = {may}, publisher = {Ubiquity Press, Ltd.}, volume = {9}, number = {1}, pages = {12}, author = {William Thielicke and Ren{\'{e}} Sonntag}, title = {Particle Image Velocimetry for {MATLAB}: Accuracy and enhanced algorithms in {PIVlab}}, journal = {Journal of Open Research Software} }

@inproceedings{Gibson1979TheEA,
  title={The Ecological Approach to Visual Perception},
  author={James Jerome Gibson},
  year={1979},
  url={https://api.semanticscholar.org/CorpusID:33656271}
}

@inproceedings{Gibson1967TheSC,
  title={The Senses Considered As Perceptual Systems},
  author={James Jerome Gibson},
  year={1967},
  url={https://api.semanticscholar.org/CorpusID:144676485}
}

@book{Gibon1950,
    author ={James Jerome Gibson},
    title = {The perception of th visual world},
    publisher ={Houghton Mifflin} ,
    year = {1950}
}

@article{Ilg2016FlowNet2E,
  title={FlowNet 2.0: Evolution of Optical Flow Estimation with Deep Networks},
  author={Eddy Ilg and Nikolaus Mayer and Tonmoy Saikia and Margret Keuper and Alexey Dosovitskiy and Thomas Brox},
  journal={2017 IEEE Conference on Computer Vision and Pattern Recognition (CVPR)},
  year={2016},
  pages={1647-1655},
  url={https://api.semanticscholar.org/CorpusID:3759573}
}

@article{Piga2021ROFTRO,
  title={ROFT: Real-Time Optical Flow-Aided 6D Object Pose and Velocity Tracking},
  author={Nicola A. Piga and Yuriy Onyshchuk and Giulia Pasquale and U. Pattacini and Lorenzo Natale},
  journal={IEEE Robotics and Automation Letters},
  year={2021},
  volume={7},
  pages={159-166},
  url={https://api.semanticscholar.org/CorpusID:241606466}
}

@article{Brebion2021RealTimeOF,
  title={Real-Time Optical Flow for Vehicular Perception With Low- and High-Resolution Event Cameras},
  author={Vincent Brebion and Julien Moreau and Franck Davoine},
  journal={IEEE Transactions on Intelligent Transportation Systems},
  year={2021},
  volume={23},
  pages={15066-15078},
  url={https://api.semanticscholar.org/CorpusID:245334808}
}

@inproceedings{Wang2011RealTimeVS,
  title={Real-Time Video Stabilization for Unmanned Aerial Vehicles},
  author={Yue Wang and Zujun Hou and Karianto Leman and Richard Chang},
  booktitle={IAPR International Workshop on Machine Vision Applications},
  year={2011},
  url={https://api.semanticscholar.org/CorpusID:1292575}
}

@article{Lim2017RealtimeOF,
  title={Real-time optical flow-based video stabilization for unmanned aerial vehicles},
  author={Anli Lim and Bharath Ramesh and Yue Yang and Cheng Xiang and Zhi Gao and Feng Lin},
  journal={Journal of Real-Time Image Processing},
  year={2017},
  volume={16},
  pages={1975 - 1985},
  url={https://api.semanticscholar.org/CorpusID:13479503}
}

@article{Mendes2021ACS,
  title={A comparative study of optical flow methods for fluid mechanics},
  author={Lu{\'i}s P. N. Mendes and Ana M. Ricardo and Alexandre J. M. Bernardino and Rui M. L. Ferreira},
  journal={Experiments in Fluids},
  year={2021},
  volume={63},
  url={https://api.semanticscholar.org/CorpusID:245228313}
}

@misc{fischer2015flownetlearningopticalflow,
      title={FlowNet: Learning Optical Flow with Convolutional Networks}, 
      author={Philipp Fischer and Alexey Dosovitskiy and Eddy Ilg and Philip Häusser and Caner Hazırbaş and Vladimir Golkov and Patrick van der Smagt and Daniel Cremers and Thomas Brox},
      year={2015},
      eprint={1504.06852},
      archivePrefix={arXiv},
      primaryClass={cs.CV},
      url={https://arxiv.org/abs/1504.06852}, 
}

@misc{hui2020liteflownet3resolvingcorrespondenceambiguity,
      title={LiteFlowNet3: Resolving Correspondence Ambiguity for More Accurate Optical Flow Estimation}, 
      author={Tak-Wai Hui and Chen Change Loy},
      year={2020},
      eprint={2007.09319},
      archivePrefix={arXiv},
      primaryClass={cs.CV},
      url={https://arxiv.org/abs/2007.09319}, 
}

@article{Cardesa2017TheTC,
  title={The turbulent cascade in five dimensions},
  author={Jos{\'e} I. Cardesa and Alberto Vela-Mart{\'i}n and Javier Jim{\'e}nez},
  journal={Science},
  year={2017},
  volume={357},
  pages={782 - 784},
  url={https://api.semanticscholar.org/CorpusID:206661410}
}

@inproceedings{Maruyama2001AnAT,
  title={An Approach to Real-Time Visualization of PIV Method with FPGA},
  author={Tsutomu Maruyama and Yoshiki Yamaguchi and Atsushi Kawase},
  booktitle={International Conference on Field-Programmable Logic and Applications},
  year={2001},
  url={https://api.semanticscholar.org/CorpusID:2016090}
}

@inproceedings{Fujiwara2003ARV,
  title={A Real-Time Visualization System for PIV},
  author={Toshihito Fujiwara and Kenji Fujimoto and Tsutomu Maruyama},
  booktitle={International Conference on Field-Programmable Logic and Applications},
  year={2003},
  url={https://api.semanticscholar.org/CorpusID:42101709}
}

@inproceedings{Siegel2003RealTimePI,
  title={Real-Time Particle Image Velocimetry for Closed-Loop Flow Control Studies},
  author={Stefan Siegel and Kelly Cohen and Thomas McLaughlin and James H. Myatt},
  year={2003},
  url={https://api.semanticscholar.org/CorpusID:119879877}
}

@inproceedings{Willert2010RealtimePI,
  title={Real-time particle image velocimetry for closed-loop flow control applications},
  author={Christian E. Willert and Matthew Munson and Morteza Gharib},
  year={2010},
  url={https://api.semanticscholar.org/CorpusID:9761905}
}

@article{McCormick2024ReactiveCO,
  title={Reactive control of velocity fluctuations using an active deformable surface and real-time PIV},
  author={Findlay McCormick and Bradley Gibeau and Sina Ghaemi},
  journal={Journal of Fluid Mechanics},
  year={2024},
  volume={985},
  url={https://api.semanticscholar.org/CorpusID:269202879}
}

@inproceedings{Shchapov2018SupercomputerRE,
  title={Supercomputer Real-Time Experimental Data Processing: Technology and Applications},
  author={Vladislav A. Shchapov and Alexander Pavlinov and Elena N. Popova and Andrei Sukhanovskii and Stanislav L. Kalyulin and Vladimir Ya. Modorskii},
  booktitle={Russian Supercomputing Days},
  year={2018},
  url={https://api.semanticscholar.org/CorpusID:59264452}
}

@phdthesis{pimientaalvernia:tel-05105277,
  TITLE = {{High frequency and high resolution Live Optical Flow velocimetry. Application to the monitoring and control of fluid flows.}},
  AUTHOR = {Pimienta Alvernia, Juan},
  URL = {https://pastel.hal.science/tel-05105277},
  NUMBER = {2024UPSLS077},
  SCHOOL = {{Universit{\'e} Paris sciences et lettres}},
  YEAR = {2024},
  MONTH = Dec,
  TYPE = {Theses},
  HAL_ID = {tel-05105277},
  HAL_VERSION = {v1},
}

@phdthesis{giannopoulos:tel-03364421,
  TITLE = {{Optical Flow Velocimetry : optimization, benchmarking and application to system identification, modelling and control of shear flows}},
  AUTHOR = {Giannopoulos, Antonios},
  URL = {https://theses.hal.science/tel-03364421},
  NUMBER = {2021SORUS097},
  SCHOOL = {{Sorbonne Universit{\'e}}},
  YEAR = {2021},
  MONTH = Jul,
  TYPE = {Theses},
  HAL_ID = {tel-03364421},
  HAL_VERSION = {v1},
}

@article{Welch1967TheUO,
  title={The use of fast Fourier transform for the estimation of power spectra: A method based on time averaging over short, modified periodograms},
  author={Peter D. Welch},
  journal={IEEE Transactions on Audio and Electroacoustics},
  year={1967},
  volume={15},
  pages={70-73},
  url={https://api.semanticscholar.org/CorpusID:13900622}
}

@article{Fey1998ANS,
  title={A new Strouhal–Reynolds-number relationship for the circular cylinder in the range 47},
  author={Uwe Fey and Michael K{\"o}nig and Helmut Eckelmann},
  journal={Physics of Fluids},
  year={1998},
  volume={10},
  pages={1547-1549},
  url={https://api.semanticscholar.org/CorpusID:119694512}
}

@article{Awbi1983EffectOB,
  title={Effect of blockage on the strouhal number of two-dimensional bluff bodies},
  author={Hazim Awbi},
  journal={Journal of Wind Engineering and Industrial Aerodynamics},
  year={1983},
  volume={12},
  pages={353-362},
  url={https://api.semanticscholar.org/CorpusID:109529355}
}

@article{King1977ARO,
  title={A review of vortex shedding research and its application},
  author={Roger King},
  journal={Ocean Engineering},
  year={1977},
  volume={4},
  pages={141-171},
  url={https://api.semanticscholar.org/CorpusID:110111629}
}

\end{document}